\documentclass{aa}
\usepackage{graphicx}
\usepackage{txfonts}
\usepackage[]{hyperref}
\hypersetup{colorlinks=true, urlcolor=blue, citecolor=cyan, pdfborder={0 0 0},}

\begin{document}

\title{Sixteen overlooked open clusters in the fourth Galactic quadrant}
\subtitle{A combined analysis of UBVI photometry and Gaia DR2 with 
\texttt{ASteCA}}

\author{G. I. Perren\inst{1}
      \and
      E. E. Giorgi\inst{2}
      \and
      A. Moitinho\inst{3}
      \and
      G. Carraro\inst{4}
      \and
      M. S. Pera\inst{1}
      \and
      R. A. Vázquez\inst{2}
}

\institute{Instituto de Astrof\'isica de La Plata (IALP-CONICET), La Plata,
Argentina\\
\email{gabrielperren@gmail.com}
\and
Facultad de Ciencias Astronómicas y Geofísicas (UNLP-IALP-CONICET), 1900 La
Plata, Argentina
\and
CENTRA, Faculdade de Ci\^encias, Universidade de Lisboa, Ed. C8, Campo Grande,
1749-016 Lisboa, Portugal
\and
Dipartimento di Fisica Astronomia Galileo Galilei, Vicolo Osservatorio 3,
Padova, I-35122, Italy
}

\date{Received November xx, 2019; accepted xxxx xx, 2019}

\abstract
{}
{This paper has two main objectives:
(1) To determine the intrinsic properties of sixteen faint and mostly
unstudied open clusters in the poorly known sector
of the Galaxy at $270^\circ-300^\circ$, to probe the Milky Way
structure in future investigations.
(2) To address previously reported systematics in Gaia DR2 parallaxes by
comparing the cluster distances derived from photometry with
those derived from parallaxes.
}
%
{Deep \textit{UBVI} photometry of 16 open clusters was carried out.
Observations were reduced and analyzed in an automatic way using the 
\texttt{ASteCA} package to get individual distances, reddening,
masses, ages and metallicities. Photometric distances were compared to
those obtained from a Bayesian analysis of Gaia DR2 parallaxes.
}
%
{Ten out of the sixteen clusters are true or highly probable open
clusters. Two of them are quite young and follow the trace of the Carina Arm
and the already detected warp.
The rest of the clusters are placed in the interarm zone between the
Perseus and Carina Arms as expected for older objects.
We found that the cluster van den Berg-Hagen 85 is $7.5\times10^9$ yrs
old becoming then one of the oldest open cluster detected in our Galaxy so
far.
The relationship of these ten clusters with the Galaxy structure in
the solar neighborhood is discussed.
The comparison of distances from photometry and parallaxes data, in turn,
reveals a variable level of disagreement.
}
%
{
Various zero point corrections for Gaia DR2 parallax data recently
reported were considered for a comparison between photometric and
parallax based distances.
The results tend to improve with some of these corrections. 
Photometric distance analysis suggest an average correction of
$\sim$+0.026 mas (to be added to the parallaxes).
The correction may have a more intricate distance dependency, but addressing
that level of detail will require a larger cluster sample.
}


\maketitle

\section{Introduction}

Galactic open clusters are routinely used as probes of the structure and
evolution of the Milky Way disk. Their fundamental parameters, like age,
distance, and metallicity allow us to define the large scale structure of the
disk and to cast light on its origin and assembly \citep{Janes_1982,
Moitinho_2010,2018A&A...618A..93C}. Young open clusters can be used to trace
spiral arms and star forming regions \citep{Moitinho_2006,Vazquez2008}, while
older clusters are better probes of the chemical evolution of the thin disk
\citep{2009yCat..35120063M}. The recent second release of Gaia satellite data
\citep{GaiaDR2_2018} is producing a tremendous advance in the study of the
Galactic disk and its star cluster population.

Basic parameters for a large number of clusters are now available with
unprecedented accuracy \citep{2018A&A...618A..93C,Soubiran_2018,Bossini_2019,
Monteiro_2019}. Proper motions may be employed to select cluster members and
parallaxes can be used to derive distances. However, in some cases Gaia
parallax distances disagree with distances derived from other methods 
(i.e., photometric or spectrophotometric). It may occur that for
short distances the photometric and parallax distances yield similar results
within the uncertainties \citep{2018A&A...618A..93C}. But the situation is
complex regarding the existence of a bias correction to be applied to
Gaia parallaxes. The analysis of quasar measurements in Gaia DR2 by
\cite{Lindegren_2018} led to the determination of a global zero point
correction to parallaxes of approximately 0.03 mas, with variations of a
comparable size depending on magnitude, colour, and position.
More recently, by analyzing a sample of stars \cite{Schonrich2019} have put in
evidence that not only a parallax offset must be applied to Gaia data but also
that exists a quasi linear dependence with distances. The study presented by 
\cite{Xu_2019}, who compared distances of a variety of astronomical objects
between Gaia and VLBI parallaxes, also report a zero point parallax correction
of $\sim$0.075 mas.
It is difficult to establish the critical distance at which Gaia parallax
distances start diverging from values from other methods and the dependence of
the bias on position, parallax or other measurements. The  task of
establishing distances and other essential parameters for open clusters using
the Gaia data looks arduous, since other factors such as the interstellar
absorption and the level of crowding of a given star cluster also play a
role.

In this article we present a sample of sixteen catalogued star clusters
\citep{Dias_2002} previously unstudied, located in a poorly known Galactic sector
at approximately $270^\circ<l<300^\circ$ along the galactic plane.
With one exception, this is
the first systematic study carried out for the clusters in our sample. In this
sense, we provide CCD \textit{UBVI} photometry complemented with data available
from Gaia DR2. The purpose of this investigation is twofold. First, we look
for a reliable estimation of the true nature of these objects. Gaia DR2 offers
us a long sought opportunity since we can make our analysis more robust by
combining  ground-based \textit{UBVI} CCD data with space-based astrometry 
(parallax, and proper motions) and photometry. Second, since distance is the
main derived parameter for mapping the Galaxy’s structure, we seek to
understand and take
into account the corresponding biases in Gaia DR2 parallaxes.
In following studies we aim at investigating the structure of the
Galactic disk in this region. Traces of the Perseus Arm coming from of the
third Galactic quadrant  would be expected, despite this arm being only
prominent in the second quadrant. However, one must keep in mind 
that some of these clusters may be associated to the Carina Arm.

Analyzing this sector of the Galaxy has proved to be quite a challenging
task since the extinction is particularly strong and variable. This
makes it not only difficult to derive accurate basic parameters of a cluster
but, even worse, to establish if a visual stellar aggregate is a physical
cluster or simply a random enhancement of field stars produced by patchy
extinction. 
For achieving these two purposes, we employed the Automated Stellar
Cluster Analysis code \citep[\texttt{ASteCA};][]{Perren_2015} to derive
clusters' fundamental parameters from \textit{G-UBVI} data, and two Bayesian
techniques to extract membership probabilities and distances from Gaia DR2. The
sample of clusters studied in this paper is shown in Table 
\ref{tab:clust_list} together with their galactic coordinates and their
equatorial coordinates referred to the J2000.0 equinox.

The layout of the paper is as follows: in Section \ref{sec:clust_sample} we
present the cluster sample.
Section \ref{sec:photo_obs} is devoted to explain the observations and the
reduction process of photometry. In Section \ref{sec:photom_analysis} we
describe the tools to analyze the photometric data and the method to connect
Gaia DR2 with photometric results. A cluster by cluster report of the results
obtained is presented in Section \ref{sec:cluster_discuss}. In Section
\ref{sec:gaia_distances} three different corrections to Gaia DR2 parallax
data are applied and discussed. Conclusions of the paper are given in Section 
\ref{sec:results_concl}.

\section{The cluster sample}
\label{sec:clust_sample}

Table \ref{tab:clust_list} lists the equatorial coordinates ($\alpha$,
$\delta$) and galactic coordinates (\textit{l}, \textit{b}) of the 16 cluster
fields studied here, ordered by increasing right ascension $\alpha$.
Equatorial coordinates refer to the J2000.0 equinox.

\begin{table}[ht]
    \centering
\caption{List of objects surveyed in the present article. Note: van den
Bergh-Hagen clusters \citep{vdBH1975} are indicated by vdBH. In a similar   
way Ruprecht \citep{Ruprecht_1996} and Trumpler \citep{Trumpler_1930} clusters
are mentioned as RUP and TR followed by the respective numbers.}
    \begin{tabular}{lcccc}
    \hline \hline 
        Cluster name & $\alpha_{2000}$ & $\delta_{2000}$ & \emph{l} & \emph{b}\\
         & hh:mm:ss & dd:mm:ss & $^\circ$ & $^\circ$\\
       \hline
        vdBH 73 & 09:31:56 & -50:13:00 & 273.634 & 0.951\\
        vdBH 85 & 10:01:52 & -49:34:00 & 276.914 & 4.544\\
        RUP 87 & 10:15:32 & -50:43:00 & 279.372 & 4.883\\
        RUP 85 & 10:01:33 & -55:01:12 & 280.15 & 0.160\\
        vdBH 87 & 10:04:18 & -55:26:00 & 280.719 & 0.059\\
        vdHB 92 & 10:19:07 & -56:25:00 & 282.984 & 0.438\\
        TR 12 & 10:06:29 & -60:18:00 & 283.828 & -3.698\\
        vdBH 91 & 10:17:16 & -58:42:00 & 284.03 & -1.600\\
        TR 13 & 10:23:48 & -60:08:00 & 285.515 & -2.353\\
        vdBH 106 & 10:52:42 & -54:14:00 & 286.048 & 4.700\\
        RUP 88 & 10:18:55 & -63:08:00 & 286.661 & -5.186\\
        RUP 162 & 10:52:54 & -62:19:00 & 289.638 & -2.545\\
        Lynga 15 & 11:42:24 & -62:29:00 & 295.053 & -0.672\\
        Loden 565 & 12:08:06 & -60:43:12 & 297.65 & 1.710\\
        NGC 4230 & 12:17:20 & -55:06:06 & 298.025 & 7.445\\
        NGC 4349 & 12:24:08 & -61:52:18 & 299.719 & 0.830\\
        \hline
    \end{tabular}
    \label{tab:clust_list}
\end{table}

These objects form part of a long term joint effort aimed at studying the
complicated structure of the Galaxy in the solar neighborhood. With this
motivation, during the last decade we have been collecting and producing
homogeneous $UBVI$ observations of open clusters in the third Galactic quadrant
(3GQ: $180^\circ\leq l \leq270^\circ$) of the Milky Way. We understand that for
a better interpretation of the galaxy structure from an optical point of view
is essential to increase the number of these  objects with well estimated
parameters. In this fashion, we have contributed significantly to the present
understanding of the spiral structure in this Galactic region
\citep{Carraro_2005,Moitinho_2006,Vazquez2008,Carraro_2010}.
In this article we decided to focus in unknown open clusters placed
between the end of the 3GQ and 300 in galactic longitude, aimed at a similar
purpose.

\begin{figure*}[ht]
    \centering
    \includegraphics[width=\hsize]{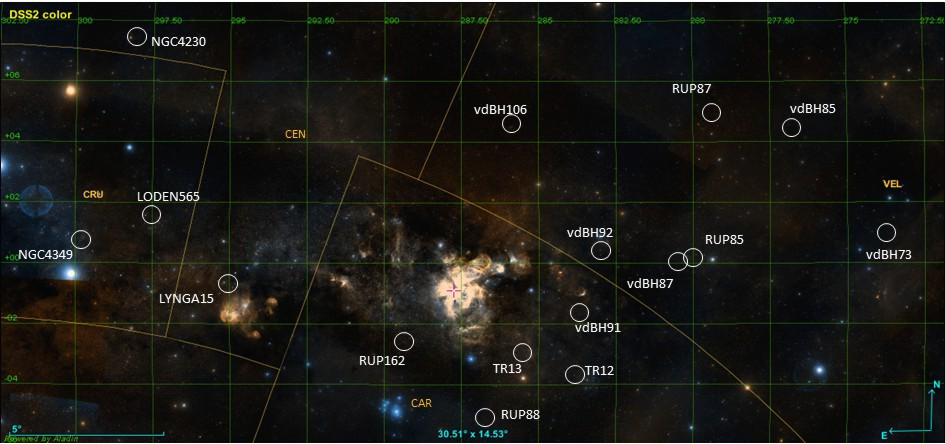}
    \caption{DSS2 color Aladin image showing with white circles the position of
    the clusters surveyed in the present sample. The galactic coordinates $l$
    and $b$ are depicted by a green grid while constellation limits for Carina,
    Vela, Centaurus and Crux appear in yellow lines.}
    \label{fig1}
\end{figure*}

Positions of the clusters in the Galaxy are shown in Fig. \ref{fig1} superposed
onto the Aladin Sky Atlas DSS2 color image. Our sampling covers essentially the
first 30 degrees of the fourth Galactic quadrant, from latitudes
$l\sim$273$^\circ$ to $l\sim$300$^\circ$, encompassing the region around the
Carina OB association and the south-east part of Vela with some objects in Crux
and Centaurus.

\section{Photometric observations}
\label{sec:photo_obs}

A first series of CCD \emph{UBVI} photometry was carried out in 13 open clusters
placed in the galactic region going from 270$^\circ$ to 300$^\circ$ in galactic
longitude and from 7$^\circ$ to -5$^\circ$ in galactic latitude. This region
covers the Carina Arm, the inter-arm region between the Perseus and Carina arms
and also a part of the Local Arm.
The observations were made on 9 nights in April and May 2002, using the YALO
(Yale, AURA, Lisbon, OSU)
\footnote{http://www.astronomy.ohio-state.edu/YALO/}
facilities at Cerro Tololo Inter-american
Observatory (CTIO). The images were taken with a $2048\times2048$ px CCD
attached to the 1.0 m telescope and the set of \textit{UBVI} filters.
The field of view is $10^\prime\times10^\prime$ given the
$0.3^{\prime\prime}$/px plate scale. All images were acquired using the
ANDICAM\footnote{\url{http://www.astronomy.ohio-state.edu/~depoy/research/instrumentation/andicam/andicam.html}}
which was moved to the 1.3 m CTIO telescope in 2003.

A second series of CCD photometry was implemented during on March 2010 at CTIO
to get \textit{UBVI} photometry in two other clusters,
NGC 4349 and Lynga 15; both at a slightly larger galactic longitude 
(298$^\circ$). Images in a first run were taken with the
SMARTS 0.9 m telescope\footnote{
\url{http://www.ctio.noao.edu/noao/content/SMARTS-09-m-Telescope}}
using a $2048\times2046$ px Tek2K
detector\footnote{\url{http://www.ctio.noao.edu/noao/content/Tek2K}} with a
scale $0.401^{\prime\prime}$/px covering thus $13.6^{\prime}$ on a side. A
second run of images taken at the SMARTS 1.0 m telescope\footnote{
\url{http://www.ctio.noao.edu/noao/content/SMARTS-10-m-Telescope}}
of the same clusters was carried out with a $4064\times4064$ px
Y4KCam\footnote{\url{http://www.ctio.noao.edu/noao/content/y4kcam}}
CCD with a scale of $0.289^{\prime\prime}$/px thus covering
$20^\prime\times20^\prime$ on a side.
The first run (at the 0.9 m) was not photometric, and therefore we tied all
the images to the second run (at the 1.0 m), which was photometric. During
this second run, we took multiple images of the standard star fields PG 1047
and SA98 \citep{1992AJ....104..340L}.

Finally, in the year 2015 the open cluster vdBH 73, located at a smaller
longitude ($\sim 273^\circ$) was observed in the \textit{UBVI} filters with the
1.0 m Swope telescope\footnote{\url{http://www.lco.cl/telescopes-information/henrietta-swope/telescope-control-system/telescopes-information/henrietta-swope/instruments/}}
at Las Campanas Observatory, Chile. On this occasion,
direct images were acquired with the 4kx4k E2V CCD with a scale of
$0.435^{\prime\prime}$/px covering $29.7^\prime\times29.8^\prime$.\\

Short exposures were always obtained to avoid bright star saturation in the
frame. Notwithstanding, sometimes we could not help to lose very bright stars.
Details of air masses, seeing values, and exposure times per filter and
telescope can be seen in Table \ref{tab:log_yalo} for all the observations.

\begin{table*}[ht]
    \centering
    \caption{Log of observations at YALO (CTIO) and Las Campanas.
    Reference for the telescopes used: 1 (1.0 m YALO), 2 (0.9 m, 1.0 m SMARTS),
    3 (1.0 m Swope). Air masses and seeing are averaged values for the short and
    long exposures.}
    \begin{tabular}{lcccccc}
    \hline \hline 
        Cluster & Date & Telescope & U & B & V &  I\\
                &      &           &
        \multicolumn{4}{c}{(airmass, seeing [''], short exp/long exp [sec])}\\
       \hline
        vdBH 73   & 06/2015 & 3 & 1.2, 2.8, 50/150 & 1.2, 2.8, 20/60 &
        1.17, 2.0, 15/45  & 1.16, 2.43, 15/45\\
        vdBH 85   & 04/2002 & 1 & 1.09, 1.7, 30/300 & 1.07, 1.7, 5/200 &
        1.07, 1.5, 3/160 & 1.14, 1.6, 1/120\\
        RUP 87    & 04/2002 & 1 & 1.14, 1.9, 30/300 & 1.11, 1.7, 5/200 &
        1.09, 2.0, 3/160 & 1.07, 1.6, 1/120\\
        RUP 85    & 04/2002 & 1 & 1.11, 2.5, 30/300 & 1.11, 2.1, 5/200 &
        1.11, 1.9, 3/160 & 1.13, 1.7, 1/120\\
        vdBH 87   & 04/2002 & 1 & 1.11, 2.2, 30/300 & 1.11, 2.5, 5/200 &
        1.12, 2.0, 3/160 & 1.14, 1.7, 1/120\\
        vdBH 92   & 05/2002 & 1 & 1.12, 1.9, 60/300 & 1.12, 1.9, 20/200 &
        1.12, 2.0, 10/160 & 1.12, 1.8, 10/120\\
        TR 12     & 04/2002 & 1 & 1.19, 1.7, 30/300 & 1.17, 1.8, 5/200 &
        1.16, 1.6, 3/160 & 1.16, 1.5, 1/120\\
        vdBH 91   & 05/2002 & 1 & 1.14, 2.1, 60/300 & 1.14, 2.0, 20/200 &
        1.15, 2.0, 10/160 & 1.17, 1.8, 10/120\\
        TR 13     & 05/2002 & 1 & 1.17, 1.8, 60/300 & 1.16, 1.6, 20/200 &
        1.16, 1.6, 10/160 & 1.16, 1.4, 10/120\\
        vdBH 106  & 05/2002 & 1 & 1.10, 2.3, 60/300 & 1.11, 2.3, 20/200 &
        1.13, 2.1, 10/160 & 1.15, 2.1, 10/120\\
        RUP 88    & 05/2002 & 1 & 1.19, 2.2, 60/300 & 1.19, 2.1, 20/200 &
        1.2, 2.0, 10/160 & 1.21, 1.8, 10/120\\
        RUP 162   & 05/2002 & 1 & 1.18, 1.6, 60/300 & 1.19, 1.6, 20/200 &
        1.0, 1.5, 10/160 & 1.2, 1.4, 10/120\\
        Lynga 15  & 03/2010 & 2 & 1.19, 1.9, 5/2400 & 1.25, 1.9, 3/1800 &
        1.28, 1.19, 3/1100 & 1.27, 1.19, 3/1100\\
        Loden 565 & 05/2002 & 1 & 1.16, 1.9, 60/300 & 1.17, 1.7, 20/200 &
        1.17, 1.7, 10/160 & 1.19, 1.6, 10/120\\
        NGC 4230  & 05/2002 & 1 & 1.11, 2.1, 60/300 & 1.12, 1.8, 20/200 &
        1.13, 1.8, 10/160 & 1.16, 1.6, 10/120\\
        NGC 4349  & 03/2010 & 2 & 1.18, 1.8, 5/2400 & 1.18, 1.6, 3/1800 &
        1.18, 1.5, 3/1100 & 1.18, 1.4, 3/1100\\
        \hline
    \end{tabular}
    \label{tab:log_yalo}
\end{table*}

\subsection{Photometric reduction process}
\label{ssec:photom_reduc}

The basic reduction of the CCD science frames has been done in the standard way
using the IRAF 4 package \texttt{ccdred}. The photometry was performed using
IRAF's DAOPHOT \citep{Stetson_1987,Stetson_1990} and \texttt{photcal} packages.
Aperture photometry was performed to obtain the instrumental magnitudes of
standard stars and some bright cluster stars. Profile-fitting photometry was
performed in each program frame by constructing the corresponding point spread
function. The zero-point of the instrumental magnitudes for each image was
determined with aperture photometry and growth curves.

The transformation equations to convert instrumental magnitudes into the
standard system were always of the form:

\begin{equation}
\begin{aligned}
  u  &=  U+u_1+u_2xX+u_3x(U-B), \\
  b  &=  B+b_1+b_2xX+b_3x(B-V), \\
  v  &=  V+v_1+v_2xX+v_3x(B-V), \\
  i  &=  I+i_1+i_2xX+i_3x(V-I)
\end{aligned}
\end{equation}

\noindent where $u_2, b_2, v_2, i_2$ are the extinction coefficients computed
for each night and $X$ is the air-mass. No color dependence of higher
order was found for either filter.

In each case detector coordinates were cross-matched with Gaia astrometry to
convert pixels into equatorial $\alpha$ and $\delta$ for the equinox J2000.0,
thus providing Gaia-based positions for the entire cluster catalog.
This process was performed in three steps. First the
Astrometry.net\footnote{\url{http://astrometry.net/}} service was used to
assign ($\alpha$, $\delta$) coordinates to the brightest stars in our observed
frames. The second step involves employing our own code called 
\texttt{astrometry}\footnote{\url{https://github.com/Gabriel-p/astrometry}} to
apply a transformation from pixel to equatorial coordinates to all the observed
stars, using the coordinates already assigned to the brightest stars matched in
the previous step. The algorithm in this code applies the affine transformation
method developed by
J Elonen\footnote{\url{https://elonen.iki.fi/code/misc-notes/affine-fit/}}
based on the work by \cite{Spath2004}. The transformation equations are of
the form $\alpha=c_0+c_1x+c_2y$ where $\alpha$ is the right ascension,
$(x, y)$ are the pixel coordinates, and the $c_X$ coefficients are fitted
(similarly for $\delta$, more details in the code's site).
Finally, in the third step we use another one of our open source codes called
\texttt{CatalogMatch}\footnote{\url{https://github.com/Gabriel-p/catalog_match}}
to cross match our frames (which by now have equatorial coordinates assigned)
with Gaia DR2\footnote{\url{https://www.cosmos.esa.int/web/gaia/dr2}} data. The
matching tolerance used here ranges from 2 to 4 arcsec, with mean
minimum/maximum differences in the matches of 0.3 and 0.9 arcsec
respectively (for all the observed frames).\\

With the exception of the cluster NGC 4349, the rest of the objects in our
sample have no dedicated photometric studies. Notwithstanding we could perform
a comparison of our photometry in $V$, $B$, and $(B-V)$ with available
photometry from APASS DR10 (The AAVSO Photometric All-Sky
Survey\footnote{\url{https://www.aavso.org/apass}}) that has a
magnitude limit near 18 mag (enough to identify the presence of RGB stars), and
Gaia DR2.
In this comparison we have put special care in those clusters belonging to the
observing runs in 2002 since they are mostly very faint.

For APASS data, we downloaded a region centered on each
observed frame and cross-matched it with our data, taking care of removing
bad matches by enforcing a tolerance of 0.7 arcsec in the matches for all the
frames (this value was selected because it gave a reasonable number of
matches with a minimum of bad-match contamination).

We also compared our photometry with that from Gaia DR2 using the
Carrasco photometric
relationships\footnote{\url{https://gea.esac.esa.int/archive/documentation/GDR2/Data_processing/chap_cu5pho/sec_cu5pho_calibr/ssec_cu5pho_PhotTransf.html}}
between the Johnson-Cousins system and Gaia passbands. The process
requires to transform the $G$ magnitude into $V$ and $B$ magnitudes through the
transformation equations provided there. For the $V$ filter we employed the
$(G-V)$ vs ($BP-RP$) polynomial. For the $B$ filter there is no similar
polynomial presented, so we fitted our own using the same list of
cross-matched Landolt standards used by Carrasco.\footnote{This list was kindly
provided by Carrasco upon our request. We thank Dr Carrasco very much for
sharing this data.} This third degree polynomial is:

\begin{equation}
\begin{aligned}
G - B = {} & 0.003[0.009]-0.64[0.02]\,(BP-RP)- \\
        & 0.42[0.03]\,(BP-RP)^2+0.067[0.007]\,(BP-RP)^3
\end{aligned}
\end{equation}

\noindent
where the values in brackets are the standard deviations of each
coefficient, and the RMS of the residuals is $\sigma\sim0.066$.
As a result of applying these two polynomials we obtain transformed $G$
magnitude values into $V_{Gaia}$ and $B_{Gaia}$ magnitudes, which we can use to
compare with our own $V$ and $B$ magnitudes directly.

\begin{figure*}[ht]
    \centering
     \includegraphics[width=\hsize]{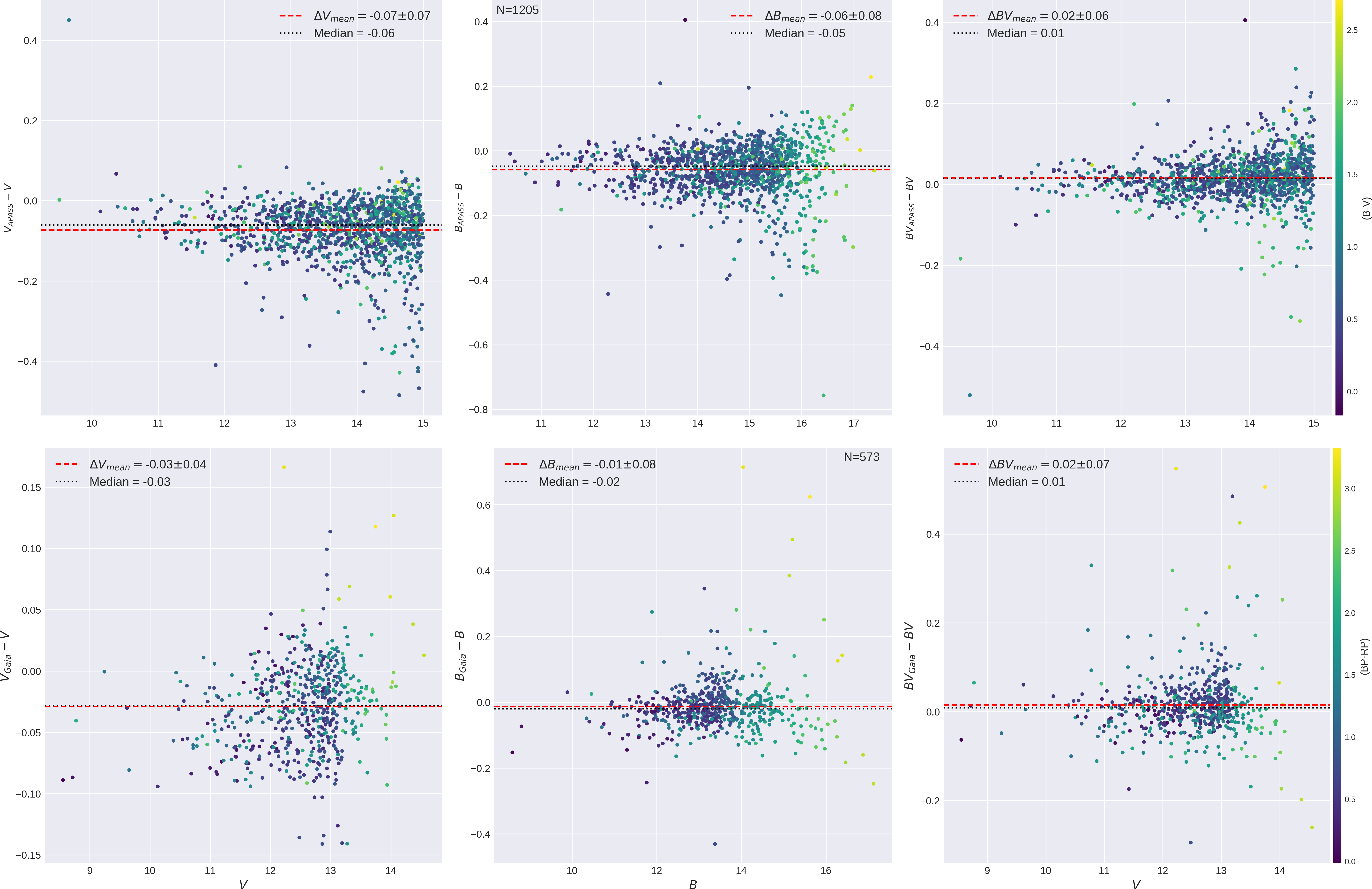}   
\caption{Top row: differences between the APASS DR10 data for the $V$
(left), $B$ (center) magnitudes and $(B-V)$ color (right) and our own
photometry. Bottom row: same for Gaia DR2 data versus our photometry. Details
in the text.}
    \label{fig:gaia_transf}
\end{figure*}

The results are shown in Table \ref{tab:phot_diffs} where the $\Delta V$,
$\Delta B$ and $\Delta (B-V)$ columns display the mean differences
between our photometry and APASS DR10/Gaia DR2 data for all the observed
regions.
In each frame the groups of stars to compare were selected according
to the filter criteria imposed by Carrasco: $G<13$, $\sigma_{G}<0.01$.
The mean differences for $V$, $B$ and $(B-V)$ combining all
the frames are shown in Fig \ref{fig:gaia_transf}. Although there are no
visible trends, there are offsets in the $V$ and $B$ magnitudes between our
photometry and APASS of ($\Delta V=-0.07\pm0.07$, $\Delta B=0.06\pm0.08$) and
between our photometry and Gaia of ($\Delta V=-0.03\pm0.04$,
$\Delta B=-0.01\pm0.08$).
The reason for the differences found for the offsets between our data and
APASS/Gaia arises from the fact that APASS DR10 has itself an offset with Gaia
DR2 of ($\Delta V=0.04\pm0.07$, $\Delta B=0.05\pm0.10$), in the sense (Gaia -
APASS).
These values were found cross-matching APASS data (for the regions where
our 16 frames are located) directly with Gaia data, and applying the mentioned
transformations for the $G$ magnitude into $V,B$.
In any case, these offsets are not relevant because we only use the $(B-V)$
color in the analysis so that the offsets tend to compensate each other
and result in a smaller value of $\sim$0.015 mag.
The effect that this $(B-V)$ offset in our photometry has on the estimated
photometric distances will be addressed in Sect~\ref{sec:gaia_distances}.\\

\begin{table*}[ht]
    \centering
\caption{Mean differences between APASS and the Carrasco transformation
polynomials and our own photometry. The columns named N show the
number of stars used to estimate these values for each cluster.}
    \begin{tabular}{lcccc|cccc}
    \hline \hline 
Cluster & \multicolumn{3}{c}{APASS} & \multicolumn{4}{c}{Gaia}\\
 & $\Delta V$ & $\Delta B$ & $\Delta (B-V)$ & N & $\Delta V$ &
$\Delta B$ & $\Delta (B-V)$ & N\\
    \hline
vdBH 73   & -0.07$\pm$0.05 & -0.04$\pm$0.05 & 0.03$\pm$0.03 & 301 &
-0.03$\pm$0.03 & -0.01$\pm$0.07 & 0.01$\pm$0.07 & 95\\
vdBH 85   & 0.01$\pm$0.04 & 0.03$\pm$0.05 & 0.03$\pm$0.04 & 32 &
0.01$\pm$0.02 & 0.02$\pm$0.07 & 0.00$\pm$0.07 & 11\\
RUP 87    & -0.02$\pm$0.05 & 0.01$\pm$0.09 & 0.02$\pm$0.07 & 41 &
0.00$\pm$0.02 & 0.00$\pm$0.03 & 0.00$\pm$0.04 & 17\\
RUP 85    & -0.04$\pm$0.05 & -0.02$\pm$0.10 & 0.02$\pm$0.08 & 36 &
-0.01$\pm$0.02 & 0.02$\pm$0.03 & 0.03$\pm$0.03 & 22\\
vdBH 87   & -0.03$\pm$0.05 & -0.02$\pm$0.06 & 0.01$\pm$0.04 & 37 &
-0.02$\pm$0.03 & 0.02$\pm$0.06 & 0.04$\pm$0.08 & 18\\
vdBH 92   & -0.06$\pm$0.05 & -0.05$\pm$0.06 & 0.01$\pm$0.04 & 34 &
-0.02$\pm$0.04 & 0.02$\pm$0.07 & 0.03$\pm$0.04 & 20\\
TR 12     & -0.07$\pm$0.07 & -0.07$\pm$0.07 & 0.00$\pm$0.05 & 37 &
-0.01$\pm$0.04 & -0.03$\pm$0.09 & -0.02$\pm$0.07 & 29\\
vdBH 91   & -0.06$\pm$0.06 & -0.04$\pm$0.09 & 0.02$\pm$0.05 & 81 &
-0.01$\pm$0.02 & 0.00$\pm$0.04 & 0.01$\pm$0.05 & 33\\
TR 13     & -0.13$\pm$0.10 & -0.08$\pm$0.07 & 0.05$\pm$0.05 & 38 &
-0.04$\pm$0.03 & 0.01$\pm$0.10 & 0.04$\pm$0.10 & 42\\
vdBH 106  & -0.07$\pm$0.08 & -0.07$\pm$0.08 & -0.01$\pm$0.06 & 44 &
-0.01$\pm$0.01 & -0.04$\pm$0.04 & -0.03$\pm$0.04 & 12\\
RUP 88    & -0.06$\pm$0.05 & -0.04$\pm$0.07 & 0.02$\pm$0.04 & 44 &
-0.01$\pm$0.01 & -0.02$\pm$0.06 & -0.01$\pm$0.06 & 29\\
RUP 162   & -0.16$\pm$0.14 & -0.13$\pm$0.19 & 0.04$\pm$0.10 & 20 &
-0.02$\pm$0.05 & 0.02$\pm$0.14 & 0.04$\pm$0.11 & 28\\
Lynga15   & -0.08$\pm$0.08 & -0.09$\pm$0.06 & -0.01$\pm$0.07 & 98 &
-0.06$\pm$0.04 & -0.06$\pm$0.09 & 0.00$\pm$0.07 & 53\\
Loden 565 & -0.03$\pm$0.04 & -0.02$\pm$0.07 & 0.00$\pm$0.04 & 43 &
-0.01$\pm$0.03 & 0.01$\pm$0.04 & 0.02$\pm$0.04 & 23\\
NGC 4230  & -0.03$\pm$0.04 & 0.00$\pm$0.06 & 0.03$\pm$0.04 & 23 &
-0.03$\pm$0.02 & 0.02$\pm$0.10 & 0.05$\pm$0.10 & 11\\
NGC 4349  & -0.11$\pm$0.08 & -0.10$\pm$0.09 & 0.01$\pm$0.07 & 296 &
-0.05$\pm$0.04 & -0.03$\pm$0.09 & 0.02$\pm$0.08 & 131\\
    \hline
    \end{tabular}
    \label{tab:phot_diffs}
\end{table*}

Figure \ref{fig:Vim} shows the CCD $V$ images of the clusters areas where we
have carried out the photometric surveys. The series of panels shown from upper
left to the lower right are ordered by increasing longitude and labeled
with the cluster name inserted in every panel. Equatorial decimal
coordinates, $\alpha$ and $\delta$, for the J2000.0 equinox are shown in each
panel as reference for the reader.\\

Final tables containing star number, x,y detector coordinates and $\alpha$,
$\delta$ equatorial coordinates together with magnitude and colors are
accessible in separate form for each cluster at
Vizier\footnote{\url{http://vizier.u-strasbg.fr/viz-bin/VizieR?-source=XXX}}.

\begin{figure*}[htp]
    \centering
     \includegraphics[width=1\hsize]{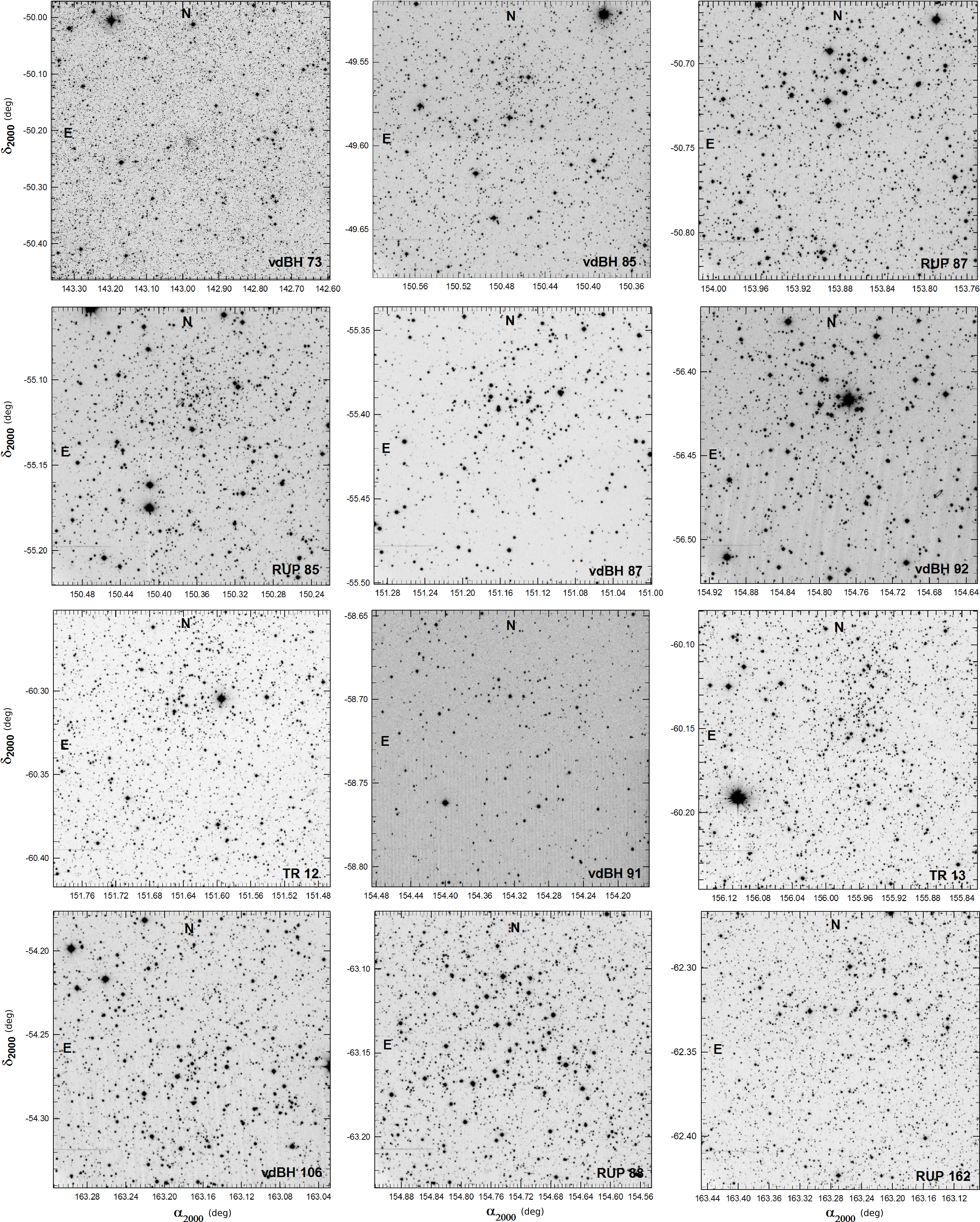}   
\caption{The V images (charts) of the observed clusters (names inserted)
ordered from top to bottom and from left to right by increasing 
longitude. Decimal $\alpha$ and $\delta$ coordinates for the 2000 equinox are
indicated. North and East are also shown.}
    \label{fig:Vim}
\end{figure*}

\begin{figure*}[htp]
    \addtocounter{figure}{-1}
    \centering
    \includegraphics[width=1\hsize]{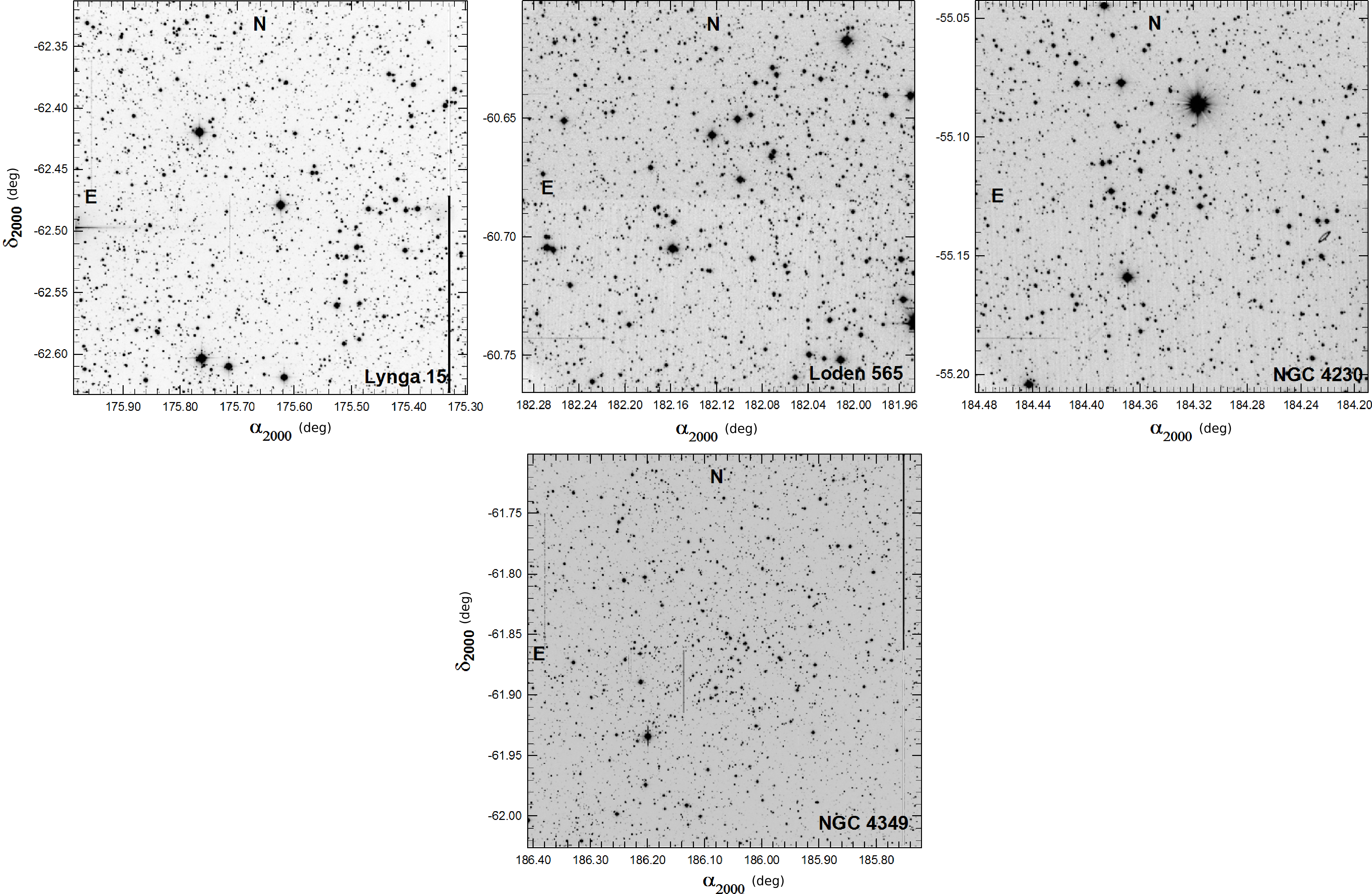}
    \caption{Continued}
    \label{fig:Vim2}
\end{figure*}

\section{Photometric data analysis process: Gaia data and the \texttt{ASteCA}
code}
\label{sec:photom_analysis}
For analyzing the large number of objects studied in this paper in a
systematic, reproducible and homogeneous way, we have used the \texttt{ASteCA}
code\footnote{\url{http://asteca.github.io/}}. The main goal of this code is to
put the user apart, as much as possible, from the analysis of a stellar cluster
to derive its fundamental parameters. We shall limit ourselves to give a brief
summary about the way the positional and photometric data are employed by the
code. A complete description of the analysis carried out by
\texttt{ASteCA} can be found in \cite{Perren_2015} and \cite{Perren_2017}.
The basic hypothesis of any stellar cluster analysis is that the
region occupied by a real cluster and the surrounding field should show ``a
priori'' different properties.
This is, we should see an increase in the star density  (not always true) where
a cluster is supposed to exist; the kinematic properties of cluster members
should differ from similar ones for the surrounding region; members of a
cluster must be at a same distance while non members may show all kind of
distances; the photometric diagrams composed by members of a cluster should
follow a well defined star sequence while field stars should not.

\subsection{Gaia data}
\label{ssec:gaia_data}

The second data release for the Gaia mission \citep{GaiaDR2_2018} was presented
on April 2018 with improved coverage, particularly for the five-parameter
astrometric solution.
We crossed-match our complete set of photometric data with those of Gaia DR2
and employed Gaia's $G$ magnitude, parallax, and proper motions in our
analysis as described in Sect~\ref{ssec:asteca_works}.

No uncertainty-based cut-off has been done on Gaia DR2 parallax or proper
motion data following the advice given in \cite{Luri_2018}, where the
authors explain that even parallaxes with negative values or large
uncertainties carry important information. Negative values in the parallax
data were thus kept during the processing. The parallax values
were processed with a Bayesian approach to get an independent estimate of the
distance to each cluster. In this approach, the model for the cluster is
taken from the accompanying tutorial by Bailer-Jones on inferring the distance
to a cluster via astrometry
data\footnote{
\url{https://github.com/agabrown/astrometry-inference-tutorials}}.
The full model (i.e., the likelihood in the Bayesian approach) can be
written as:

\begin{equation}
\begin{aligned}
P\left(\{\varpi\} | r_{c}\right)
= & {} \prod_{i=1}^{N} \int \int \frac{1}{2 \pi \sigma_{\varpi_{i}} s_{c}} \\
& \exp \left[-\frac{1}{2} \left( \frac{\left(\varpi_{i}-1 / r_{i}\right)^{2}}{
\sigma_{\varpi}^{2}} + \frac{\left(r_{i}-r_{c}\right)^{2}}{s_{c}^
{2}}\right)\right]
d r_{i}\,d s_{c}
\label{eq:plx_lkl}
\end{aligned}
\end{equation}

\noindent where $\{\varpi\}$ is the set of all parallax values (our
data), $N$ is the number of processed stars in the cluster,
$\varpi_{i}$ and $\sigma_{\varpi_{i}}$ are the parallax value and its
uncertainty for star $i$, $r_{i}$ is the distance to that star in parsec,
$s_{c}$ is a shape parameter that describes the size of the cluster, and
$r_{c}$ the distance to the cluster (the parameter we want to estimate).
Our model marginalizes not only over the individual distances ($r_i$;
as done in the original model by Bailer-Jones) but also
over the shape parameter ($s_c$), estimating only the overall cluster distance
$r_c$ using the parallax value and its uncertainty for each star in the
decontaminated cluster region (the membership probabilities process is
described with more detail in Sect.~\ref{ssec:asteca_works}).
The prior for the distance in the Bayesian model is a Gaussian centered at a
maximum likelihood estimate of the distance to the cluster region, with a large
standard deviation (1 kpc). This maximum likelihood was
obtained through a Differential Evolution algorithm built into
\texttt{scipy}\footnote{\url{https://docs.scipy.org/doc/scipy/reference/generated/scipy.optimize.differential_evolution.html}},
applied on Eq. \ref{eq:plx_lkl}, i.e, the model.
The results of this analysis will be shown in Sect~\ref{sec:cluster_discuss}
and discussed in Sect.~\ref{sec:gaia_distances}.

We include in our analysis a two-sample Anderson-Darling
test,\footnote{\url{https://docs.scipy.org/doc/scipy/reference/generated/scipy.stats.anderson_ksamp.html}}
comparing the distribution of Gaia parallax and proper motions,
between the cluster and the estimated stellar field regions, to quantify how
``similar'' these two regions are among each other.
The results of the test in each case are indicated with AD and the
corresponding p-value\footnote{The null hypothesis ($H_{0}$) is the
hypothesis that the distributions of the two samples are drawn from the same
population. The significance level ($\alpha$) is the probability of mistakenly
rejecting the null hypothesis when it is true, also known as Type I error. The
p-value indicates the $\alpha$ with which we can reject $H_{0}$. The usual 5\%
significance level corresponds to an AD test value of 1.961, for the case of two
samples.} in Fig \ref{fig:plx_bys_vdBH85} and the similar figures for
the remaining clusters.
The p-value indicates at what significance level the null hypothesis
can be rejected. This is, the smaller the p-value, the larger the probability
for the cluster region of being a true physical entity rather than a random
clustering of field stars. When using parallax and proper motions three
p-values are generated that are combined into a single p-value using Fisher's
combined probability
test\footnote{\url{https://docs.scipy.org/doc/scipy/reference/generated/scipy.stats.combine_pvalues.html}}.

\subsection{The way \texttt{ASteCA} works}
\label{ssec:asteca_works}

Since the first release of \texttt{ASteCA} the code has grown
considerably. The purpose of the tool and the core set of analysis it is able
to perform are still properly described in \cite{Perren_2015}, although
several modifications have been implemented since. The most relevant changes
include the ability to combine parallax and proper motion data in the
membership analysis algorithm, which was initially purely photometric. This
means one can currently use up to 7 dimensions of data in this process:
magnitude, three colors, parallax, and proper motions.

The several tasks performed by \texttt{ASteCA} can be roughly divided into three
main, independent analysis blocks: structural study including the
determination of a cluster region identified primarily by an overdensity,
individual membership probability estimation for stars inside the overdensity,
and the search for the best fit parameters.\\

The first block estimates center and radius values that define in each case the
cluster region. Robust estimations of these two quantities can only be achieved
when a clear overdensity and a large number of members are detected.
If a cluster is not clearly defined as an overdensity on the observed frame and
if its boundaries are weakly established, \texttt{ASteCA} allows center and
radius to be manually fixed since the automatic procedure may return incorrect
values. We have chosen to fix all radii values manually since many of
our observed frames are structurally sparse and with a low number of members,
and display very noisy radial density profiles (hereafter RDP).\footnote{
The radii values are estimated using the frames in pixels coordinates,
and then converted to arcmins.
}
Every point of the RDP was obtained by generating rings around the
center defined for the potential cluster, i.e., the comparison field.
In the present case the comparison field may contain between 1 and 10 regions
of equal area to that of the cluster, depending on the cluster area and the
available size of the remaining of the frame. In each ring the found number of
stars (with no magnitude cut applied) is divided by the respective
area to get a value of the radial density.
For the computation of the density level of the field (foreground/background),
outliers in the RDP are iteratively discarded to avoid biasing the final value.
This procedure is repeated until converging to an equilibrium value, equivalent
to the density of the star field at a given distance from the potential
cluster center.

King profile \citep{King_1962} fittings have been performed in those cases when
a fit could be generated. No formal core or tidal radius are given because
their values, due mainly to the shape of the RDP, were not within reasonable
estimates (the process to fit the King profiles to the RDP returned either
large and unrealistic values, or values with very large uncertainties).
This could due to the non-spheric geometry of sparse open clusters
combined with the field contamination within the cluster region. Although
photometric incompleteness is not taken into account in the generation of the
RDP, these are not clusters largely affected by crowding; thus we do not
expect this to have a major effect on the estimated radii.\\

The second block assigns membership probabilities to the defined
cluster region, an often disregarded process in simpler cluster studies, 
and removes the most probable field stars that contaminate this region.
By itself, an over-density does not guarantee the presence of a real cluster;
many times an overdensity is generated by random fluctuations in the field star
density. To avoid such a mistake a comparison of the properties for cluster and
field stars must be done. Ideally, we look for firm evidence of the presence of
a cluster sequence at some evolutionary stage. \texttt{ASteCA} employs a
Bayesian algorithm to compare the photometric, parallax, and proper motions
distribution of the stars in the cluster region with a similar distribution in
the surrounding field areas \citep{Perren_2015}. Initially the analysis
was carried out in an N-dimensional data space that combined the
$G$ magnitude, parallax, and proper motions from Gaia, with colors from our own
photometry: $(V-I)$, $(B-V)$, $(U-B)$. In this case thus, the data space where
the algorithm works is characterized by N=7.
Combining all the available data is though not always optimal. A
data dimension can sometimes introduce noise in the analysis instead of helping
disentangle members from field stars. In our case we found that using
parallax and proper motions, i.e., N=3, resulted in more clearly defined
cluster sequences than if we included photometric dimensions (with N=7 as
mentioned above).

Briefly, the algorithm compares the properties of this N-dimensional data space,
for stars inside (cluster region) and outside (field region)
the adopted cluster limits. All the data dimensions are previously normalized 
(to prevent any dimension from out-weighting others) and 4 sigma outliers are
rejected.
The position of every star inside the cluster in this data space is compared
against each star in all the defined equivalent-area field regions,
assuming a Gaussian probability density (centered at the given values for each
data dimension, with standard deviations given by the respective
uncertainties). This procedure is repeated hundreds or thousands of times 
(defined by the user) each time selecting different stars to construct an
approximation of the clean cluster region. The outcoming result of this
algorithm are thousands of probability values that are averaged to a final
single membership probability value for each star within the cluster region.

This block ends with the cleaning of the photometric diagrams in the cluster
region. Each cluster region photometric diagram is divided into cells, and
the same is done for the equivalent diagram of the field regions. The star
density number found in the field is then subtracted from the cluster
photometric diagram, cell by cell, starting with stars that have low
membership probabilities. Therefore, the final cluster photometric diagrams
contain not only star membership assignation but it is also cleaned from the
expected field star contamination. This two-step process is of the utmost
importance to ensure that the fundamental parameters analysis that
follows is performed on the best possible approximation to the cluster
sequence (particularly when the cluster contains few members).\\

Finally, the third block performs the cluster's parameters estimation through
the minimization of a likelihood function \citep{Dolphin_2002} employing a
genetic algorithm numerical optimization \citep{Charbonneau_1995}. This last
stage includes the assignment of uncertainties for each fitted parameter via a
standard bootstrap method \citep{efron1986}. Again, all of these processes are
described in much more detail in \cite{Perren_2015} and \cite{Perren_2017}.

It is worth noting that, unlike other tools \citep[e.g.:][]{Yen_2018},
\texttt{ASteCA} does not fit isochrones to cluster sequences in photometric
diagrams. Instead, it fits synthetic clusters generated from a set of
theoretical isochrones, a given initial mass function, and completeness and
uncertainties functions estimated directly from the observations.
These synthetic clusters are represented as two or three-dimensional
color-magnitude diagrams, depending on the number of photometric colors
available in our observations. The ``best fit'' isochrones shown in green in
the photometric diagrams shown in Fig~\ref{fig:fundpars_vdBH85} for vdBH85 (and
similar figures for the rest of the clusters) are there for convenience
purposes only as a way to guide the eye.

The code makes use of the PARSEC v1.2S \citep{Bressan_2012} theoretical
isochrones (obtained from the CMD
service\footnote{\url{http://stev.oapd.inaf.it/cgi-bin/cmd}}), and the
\cite{Kroupa_2002} form for the initial mass function. A dense grid of
isochrones with fixed $z$ and $\log(age)$ values is requested to the CMD
service\footnote{Grid values: $z$ range [0.0005, 0.0295] with a step of
0.0005; $\log(age)$ range [7, 9.985] with a step of 0.015}, which are later on
used in the fundamental parameters estimation
process. The full processing yields five parameters: metallicity, age,
extinction, distance, and mass, along with their respective uncertainties. The
binary fraction was always fixed to 0.3, a reasonable
estimate for open clusters \citep{Sollima_2010}. As for the final mass of each
cluster, although the values are corrected by the effects of star loss
due to photometric incompleteness at large magnitudes and the
percentage of rejected stars with large photometric uncertainties, it is not
corrected by the dynamical mass loss due to the cluster's orbiting through the
Galaxy. Hence, it should be regarded as a lower limit on the actual initial
mass value.

From a practical point of view the code proceeds as follows to estimate the
cluster's parameters. Firstly, individual three dimensional $G$ vs $(B-V)$ vs $
(U-B)$ photometric diagrams are analyzed fixing the metallicity to a solar
value ($z = 0.0152$) in order to reduce the dimensionality of the parameter
space, and thus its complexity. Although several of the aforementioned diagrams
contain, in the present case, a rather small number of stars due to the
presence of the $U$ filter, they are very useful to get reddening and thus
extinction via the inspections of the $(U-B)$ vs $(B-V)$ diagrams
\citep[e.g.,][]{Vazquez2008} .

The individual $E(B-V)$ values in each region were always checked against
maximum values given by the \cite{Schlafly_2011} maps\footnote{Through the
NASA/IPAC service \url{https://irsa.ipac.caltech.edu/applications/DUST/}}.
The only information extracted from this first step, and in particular
by inspection of the $(U-B)$ vs $(B-V)$ diagram, is thus a reasonable range for
the $E(B-V)$ parameter.
Secondly, the analysis of the $G$ vs $(B-V)$ vs $(V-I)$ diagram is carried out
restricting now the reddening space to the $E(B-V)$ range obtained previously,
while still fixing the metallicity to solar value. We get from this process
estimates for the age, distance, and cluster mass.
Finally, in a third stage the parameter ranges derived above are applied
including now the metallicity as a free parameter.
As a result of the entire procedure, we obtain a five parameter best model fit
for each observed cluster, along with the associated one sigma uncertainties
for each one. In all the cases we have adopted $R=A_v/E(B-V) = 3.1$ to produce 
absorption-free distance moduli.

During the maximum likelihood and bootstrap processes,
each observed cluster is compared to $\sim2\times10^7$
synthetic clusters. This number is obtained combining those synthetic
clusters generated in the maximum likelihood and bootstrap processes, by
varying the fundamental parameters values.

\section{Cluster-by-cluster discussion on structural and intrinsic parameters
provided by \texttt{ASteCA}}
\label{sec:cluster_discuss}

We now present the results from the spatial and photometric analysis
carried out with \texttt{ASteCA}, together with the outcome of the application
of the Anderson-Darling test that compares parallax and proper motion
distributions in cluster regions with their respective field regions.
It is important to emphasize that the code will always fit the
best possible synthetic cluster to a given star distribution, no matter we
face a true open cluster or not.

Our sample contains clusters with a large variety of properties: some are
robust, bright, well detached from the cluster background and therefore with a
clearly defined main sequence (TR 13, TR 12, NGC 4349, vdBH 87, vdBH 92),
others are clusters which are faint, with sparse star population and easy to
confuse with the background (vdBH 73, vdBH 85, vdBH 106, RUP 162, RUP 85).
Therefore, given the amount of figures to be shown in this paper we decided to
add them to an Appendix, and limited ourselves here to present
the case of three extreme types of cluster according to the statement
above: a poorly defined (vdBH 85), a well-defined (NGC 4349) and a not
cluster (RUP 87).

\subsection{van den Bergh-Hagen 85}
\label{ssec:vdbh85}

\begin{figure*}[ht]
    \centering
    \includegraphics[width=\hsize]{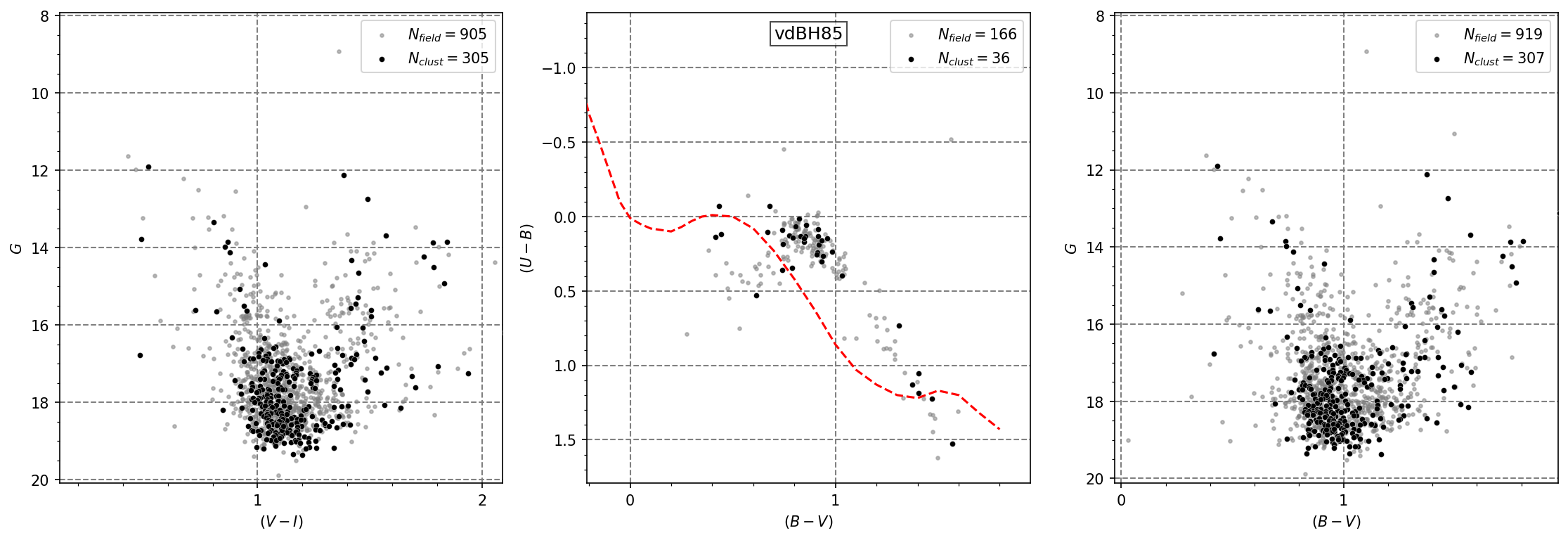}
\caption{From left to right: The $G$ vs $(V-I)$, $(B-V)$ vs $(U-B)$, and
$V$ vs $(B-V)$ diagrams for all the stars observed in the region of van
den Bergh-Hagen 85.
The red dashed line in the two color diagram gives the position of the ZAMS
\citep{Aller1982}. Insets in each diagram contain the number of stars 
in the cluster region ($N_{clust}$, black circles) and the surrounding
field ($N_{field}$, grey circles)
}
    \label{fig:photom_vdBH85} 
\end{figure*}

\begin{figure*}[ht]
    \centering
    \includegraphics[width=\hsize]{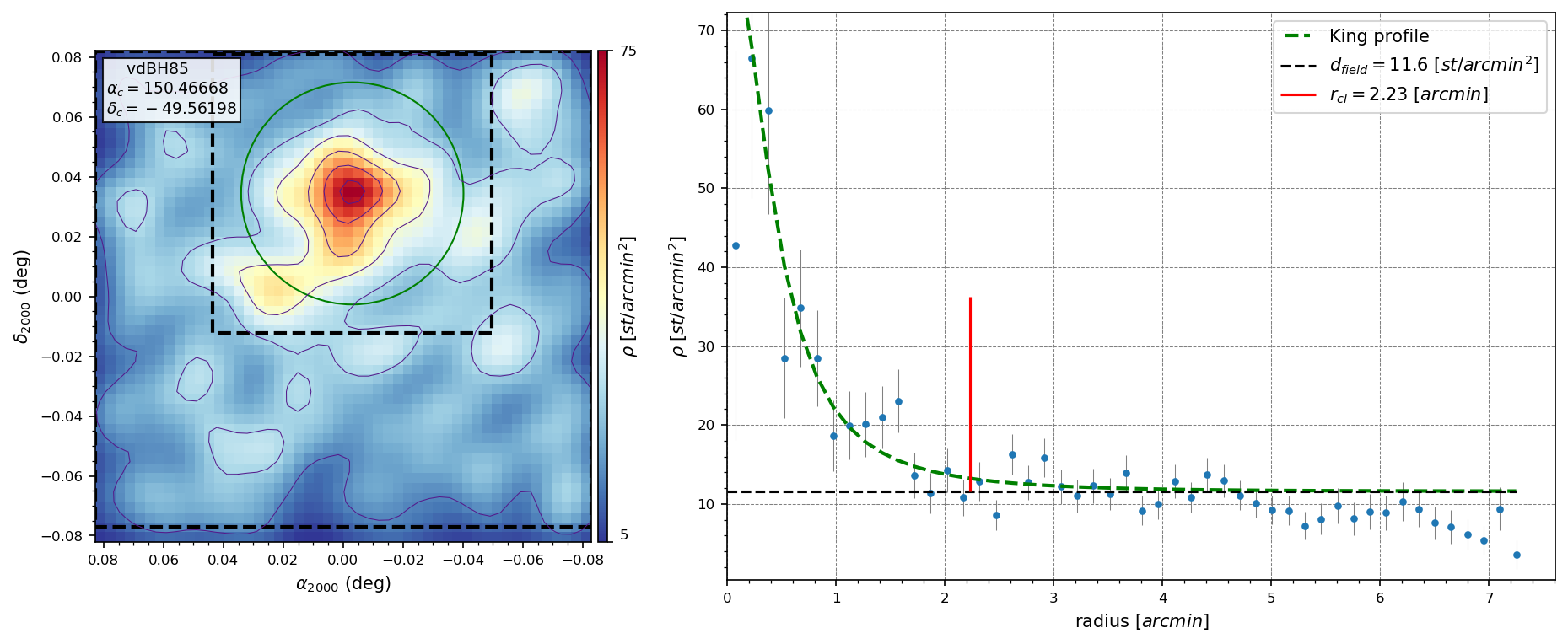}
\caption{From left to right. First panel:
Contour plot showing the position of the overdensity associated to vdBH 85.
Green inner circle gives the cluster size while the two black dashed lines
squares enclose the region used for \texttt{ASteCA} to estimate the field stars
properties. The lower density values at the frame's borders are an
artifact of the kernel density estimate method employed to generate the density
maps.
Equatorial coordinates in decimal format are indicated.
The colorbar denotes the star number per square arcmin (linear scale).
These values are slightly different from those in the panel to the right
because they are obtained with a different method (nearest neighbors).
Second panel: The RDP is shown as blue dots with standard deviations shown as
vertical black lines. King profile is shown in dashed green line. The
horizontal black line is the mean field star density. Vertical red line is the
adopted cluster radius.
}
    \label{fig:struct_vdBH85}
\end{figure*}

\begin{figure*}[ht]
    \centering
    \includegraphics[width=\hsize]{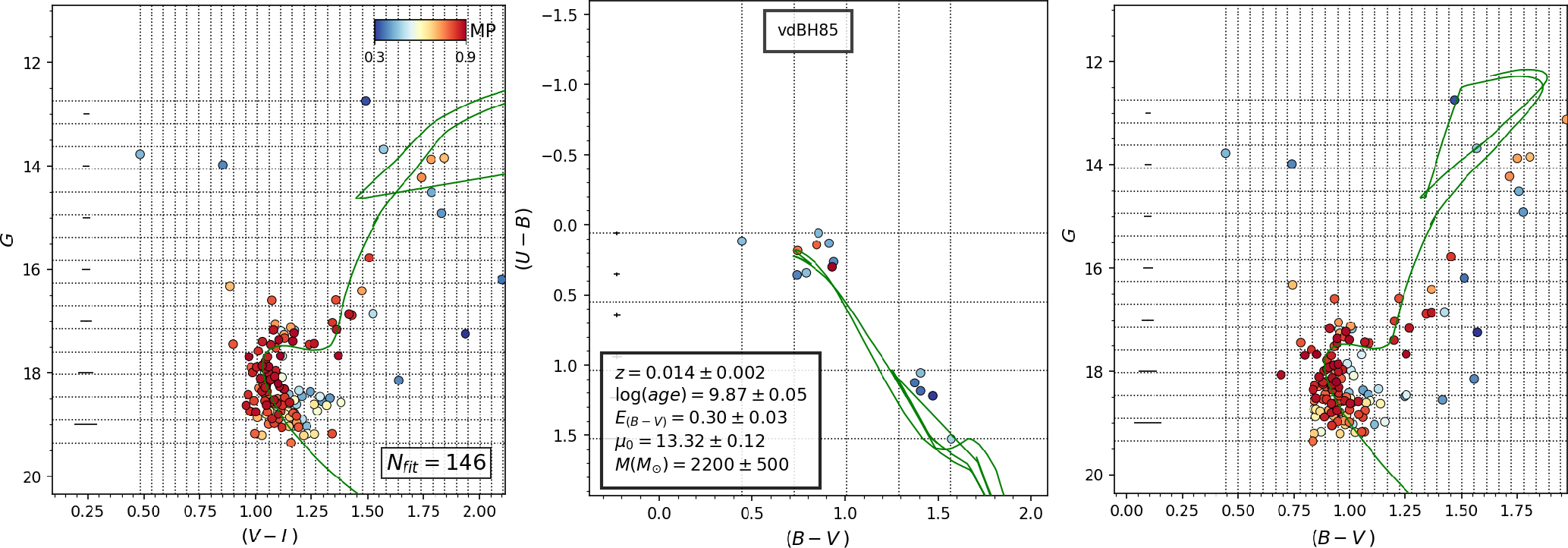}
\caption{From left to right: The $G$ vs $(V-I)$ , $(B-V)$ vs $(U-B)$,
and $G$ vs $(B-V)$ clean diagrams after the removal by field interlopers
made by \texttt{ASteCA} over vdBH 85. The color of each star reflects
its membership probability. Corresponding values are in the color bar at
the upper right corner in the $G$ vs $(V-I)$ diagram (left) labeled
MP. The CCD in the middle will always show fewer stars due to the
use of the $U$ filter.
Inset at the lower right corner in the $G$ vs $(V-I)$ diagram shows
the number of stars used by \texttt{ASteCA} to compare with synthetic clusters.
Inset in the mid panel includes the final
results for metallicity, $\log(age)$, $E(B-V)$, the corrected distance
modulus, and the cluster total mass provided by \texttt{ASteCA}. The green
continuous line in the three diagrams is a reference isochrone. In
particular, the green line in the color-color diagram, mid panel, shows the
most probable $E(B-V)$ value fitting found by \texttt{ASteCA}.
}
    \label{fig:fundpars_vdBH85}
\end{figure*}

\begin{figure*}[ht]
    \centering
    \includegraphics[width=\hsize]{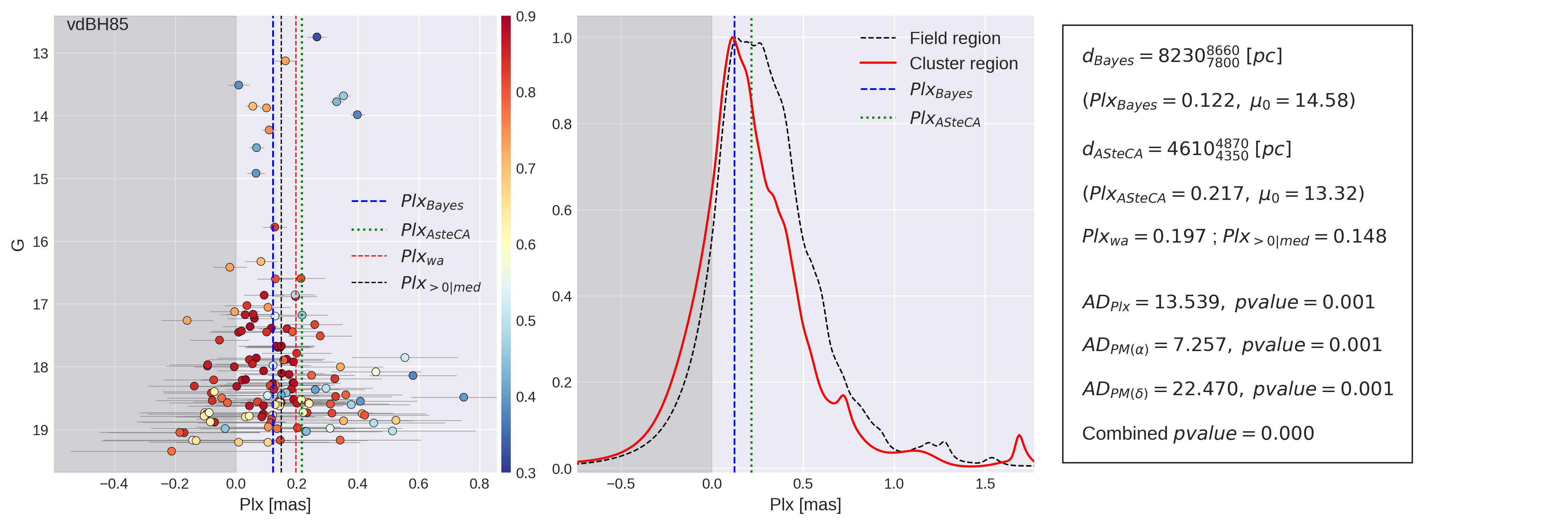}
\caption{Left panel: distribution of parallax for all stars with membership
probabilities in the cleaned cluster region, as a function of the apparent
magnitude $G$ (vertical color scale is for the star membership
probability) in vdBH 85.
Horizontal bars are the parallax errors as given by Gaia. The different
parallax value fittings are shown by dashed lines of different colors:
blue is the Bayesian parallax estimate, green is \texttt{ASteCA}'s
photometric distance, red is the weighted average, and black is the median 
(without negative values).
The mid panel is a normalized comparison between the parallax
distributions inside the cluster region (red line) and outside it (dashed
black line). The frame at the right summarizes the distances in parsecs
according to the Bayesian analysis ($d_{Bayes}$) and \texttt{ASteCA} ($d_
{ASteCA}$) followed by the parallax corresponding value, $Plx$, and
corrected distance modulus ($\mu_0$). Both fittings are indicated by the
vertical blue and green dashed lines. The last four text lines in the right
panel are the AD values for $Plx$, $PM(\alpha)$, $PM(\delta)$ followed by
the corresponding $p$-values and, finally the combined $p$-value.
}
    \label{fig:plx_bys_vdBH85}
\end{figure*}

The open cluster vdBH 85 appears in the sky slightly east of the center of the
Vela constellation. The $V$ chart in Fig. \ref{fig:Vim} shows a weak star
concentration near the north side of the observed field extending a little bit
to the south east. 
The color-color and color-magnitude diagrams (from now on CCD and CMDs
respectively) of the entire field of view in Fig.~\ref{fig:photom_vdBH85} is
just a dispersed star distribution ending in a compact accumulation at
$(B-V)=1$ and below $G=17$ mag approximately. Another clear feature is a
structure at $G=16$ mag in the two CMDs and for $1.2<(B-V)<1.7$ mag, resembling
a red clump.\\

Figure \ref{fig:struct_vdBH85} represents the spatial analysis carried
out by \texttt{ASteCA}. This is: results from the search of a stellar
overdensity, the mean value for the stellar field density, the respective King
profile attempting to fit the radial density profile, and the assumed radius.
\texttt{ASteCA} detected here an overdensity not easily seen in
Fig.~\ref{fig:Vim}, standing out from the stellar background contained
in a radius of 2.2 arcmins. It is characterized by a smooth RDP 
with nearly six times the background density at its peak, 
as shown in Fig.~\ref{fig:struct_vdBH85}.\\

In the following step the removal of interlopers by comparison with the
background field properties yields the field decontaminated CCD
$(U-B)$ vs $(B-V)$ and CMDs, $G$ vs $(B-V)$ and $G$ vs $(V-I)$.
This removal is performed comparing the star density on the cluster
region's photometric diagram (whose stars already have membership
probabilities provided by \texttt{ASteCA}) with that of the surrounding field
regions. These diagrams are shown in Fig. \ref{fig:fundpars_vdBH85}.
We insert in the mid panel of this figure the results from the
best synthetic cluster fitting to the field decontaminated diagrams. In these
three panels we show as well the isochrone curves from which the best
synthetic cluster fit was generated. These isochrones are generated
using the maximum likelihood values found for the metallicity and age, through
averaging of theoretical isochrones taken from the employed grid.
Again, this is just to guide the eye since \texttt{ASteCA} does not fit
isochrones.

Once the membership probabilities are established and the removal of field
interlopers is done, the two CMDs of all stars show a short but evident main
sequence below $G=17$ mag.
Three magnitudes above the cluster turn-off, at $G=14$ mag a handful of stars
appear, possibly part of the bright end of the giant branch. The comparison
with the best fitting of a synthetic cluster throws the following
characteristics for vdBH 85:

\begin{itemize}
\item [a)] the cluster is seen projected against a stellar field with moderate
    to low color excess. The best value corresponds to $E(B-V)=0.3$ in
    correspondence with the maximum value of 0.46 mag stated by S\&F2011.
\item [b)] The free absorption distance modulus of vdBH 85 is
    $13.32\pm0.12$ mag which implies a distance of $4.61\pm0.26$ kpc
    from the Sun. This fact explains by itself the extreme weakness of the
    cluster members.
\end{itemize}

Figure \ref{fig:plx_bys_vdBH85}, finally, includes three panels. The
left one shows the $G$ mag vs Gaia parallax values (uncertainties indicated by
horizontal bars) of cluster members, colored according to the estimated
membership probabilities (colorbar to the right).
The Bayesian distance ($d_{Bayes}$) found by the code is shown here by
a vertical blue dashed line, the equivalent \texttt{ASteCA} distance
($d_{ASteCA}$) with a green dotted line, the weighted average with a red dashed
line (where the weights are the inverse of the parallax
errors), and the naive estimate of obtaining the median of stars with
parallax values greater than zero with black dashed line.
The mid panel is the kernel density estimate of stars in the surrounding
field region and the cluster region, in black and red lines respectively. For
the Anderson-Darling test we used all the stars within the cluster region with
Gaia data. In the right panel we summarize the distances in parsecs and errors,
($d_{Bayes}$) and $d_{ASteCA}$, followed by the corresponding parallax
value, $Plx$, and corrected distance modulus, $\mu_0$. Both fittings are
indicated by the vertical blue and green dashed lines. The final
four text lines in the right panel are the AD values for $Plx$, $PM(\alpha)$,
$PM(\delta)$ from the Anderson-Darling test, followed by the corresponding
$p$-values and finally the combined $p$-value.

The distance estimated with parallax data from Gaia is almost 4 kpc
larger than the one obtained through the photometric analysis. This is most
likely a failure of the Bayesian inference method employed, due to the large
uncertainties associated to most of the probable cluster members. Further
discussion is presented in Sect. \ref{sec:gaia_distances}.
The Anderson-Darling test results in Fig. \ref{fig:plx_bys_vdBH85} suggest the
null hypothesis can be safely rejected given the combined $p$-value of
0.0. The $Plx$, $PM(\alpha)$ and $PM(\delta)$ results from the
Anderson-Darling test leave no doubt in the sense that cluster region and the
surrounding comparison field come from quite different star populations.\\

We conclude that this object is a real and very old cluster, the oldest in our
sample, approximately $7.50\pm0.80\times10^9$ yrs old. This age puts
vdBH85 among the top ten oldest clusters cataloged in the
WEBDA\footnote{\url{https://webda.physics.muni.cz/}} and
DAML\footnote{\url{http://cdsarc.u-strasbg.fr/viz-bin/cat/B/ocl}}
\citep{Dias_2002} databases.

\subsection{NGC 4349}

\begin{figure*}[ht]
    \centering
    \includegraphics[width=\hsize]{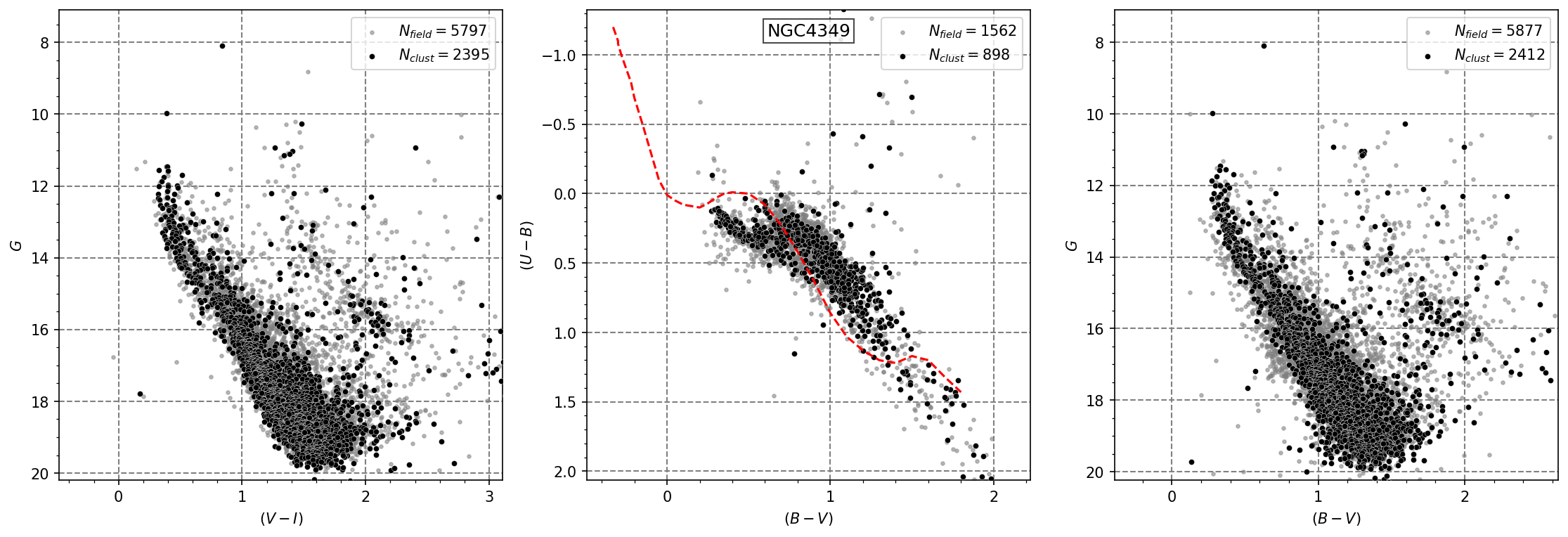}
    \caption{Idem Fig. \ref{fig:photom_vdBH85} for NGC 4349.}
    \label{fig63}
\end{figure*}
\begin{figure*}[ht]
    \centering
    \includegraphics[width=\hsize]{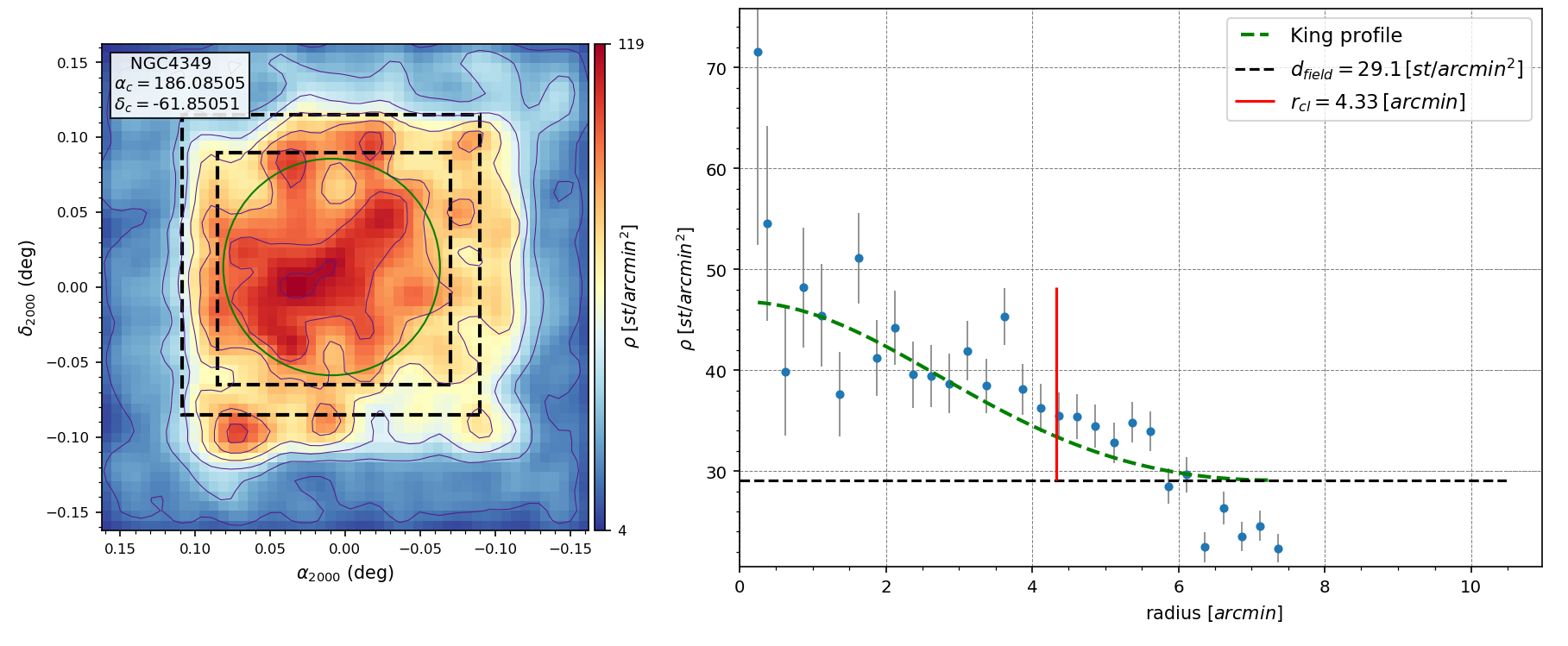}
    \caption{Idem Fig. \ref{fig:struct_vdBH85} for NGC 4349.}
    \label{fig64}
\end{figure*}
\begin{figure*}[ht]
    \centering
    \includegraphics[width=\hsize]{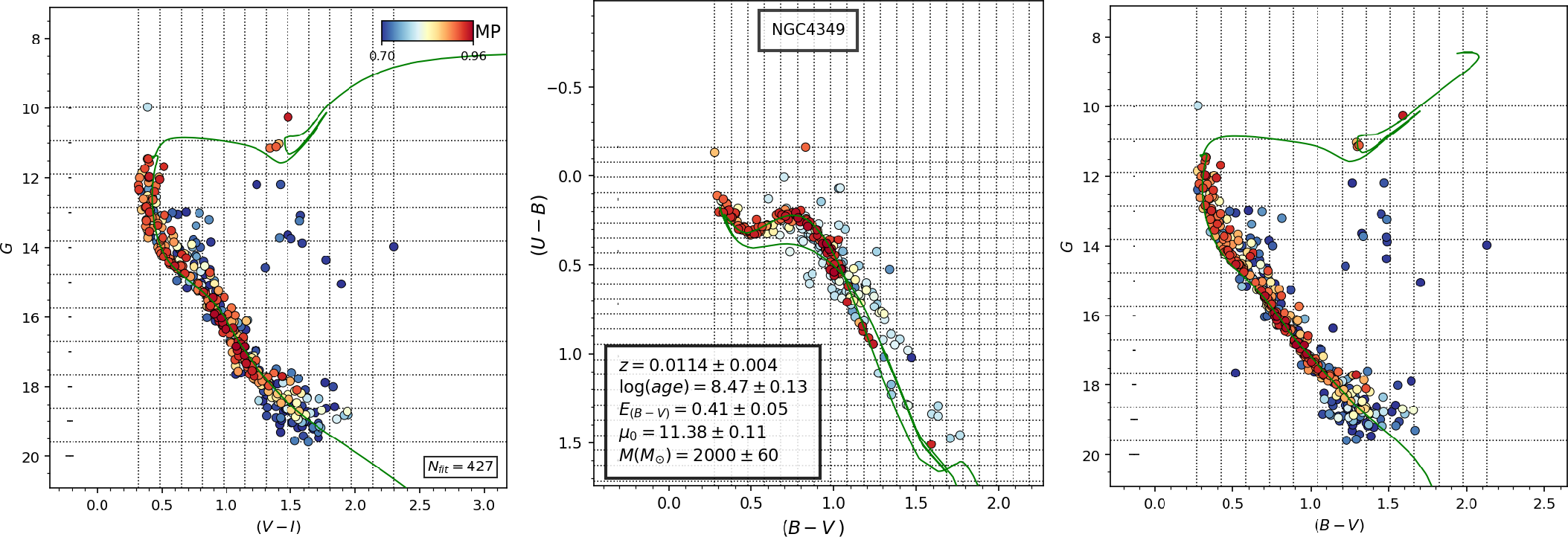}
    \caption{Idem Fig. \ref{fig:fundpars_vdBH85} for NGC 4349.}
    \label{fig65}
\end{figure*}
\begin{figure*}[ht]
    \centering
    \includegraphics[width=\hsize]{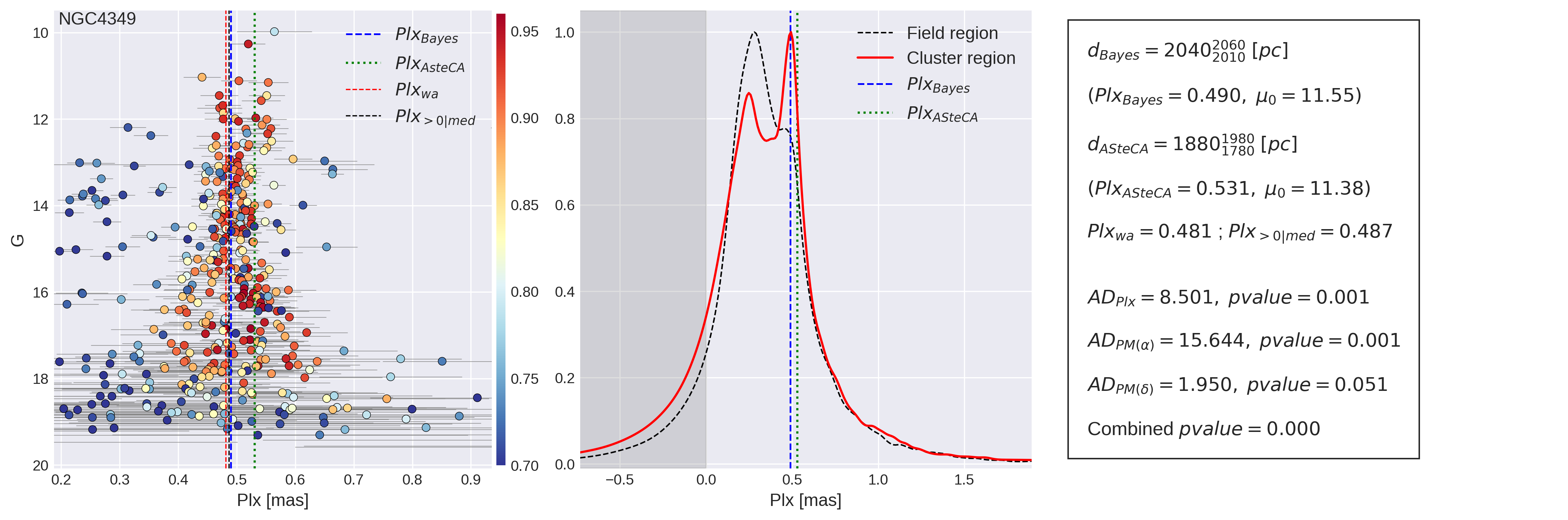}
    \caption{Idem Fig. \ref{fig:plx_bys_vdBH85} for NGC 4349.}
    \label{fig:plx_bys_NGC4349}
\end{figure*}

This is an object in the Crux constellation, placed slightly south of its
geometric center. At first glance, the $V$ image in Fig. \ref{fig:Vim} shows a
distinguishable star accumulation. The overall photometric CCD and CMDs in
Fig.~\ref{fig63} show a prominent star sequence emerging at $G\approx15$ mag
from the usual stellar structure produced by galactic disc stars.
The CCD makes more evident the presence of a reddened but compact sequence of
blue stars placed immediately below the first knee of the intrinsic line. Apart
from this, other bluer stars appear for $(U-B)$ values smaller than 0.0.\\ 

\texttt{ASteCA} analysis revealed an extended overdensity of up to 70 stars per
square arcmin. The observed frame's density map shows two regions with
very distinct mean stellar densities of background. This is just an artifact
generated by combining observations made with two different telescopes, as
detailed in Sect.~\ref{sec:photo_obs}, and is the reason why
the RDP shows such a strange shape, as seen in Fig.~\ref{fig64}. We
settle for a radius of $\sim4$ arcmin, which seems to contain most of the
overdensity, and limit the analysis to the inner frame.
The \texttt{ASteCA} estimation of memberships shows that inside the adopted
cluster radius the probable members of the cluster detach easily from the field
region stars. This is shown in the respective CCD and CMDs of Fig.~\ref{fig65}.
If attention is drawn to the largest probabilities there appears in the three
diagrams a somewhat narrow cluster sequence.
In these cases (i.e., when a cluster sequence can be clearly defined
down to the low mass region) probable members can be identified selecting a
minimum probability value. We used $P>70\%$ which produces a reasonably clean
sequence with an appropriate number of estimated members.

Comparison with synthetic clusters yielded that NGC 4349 is a cluster with the
following properties:

\begin{itemize}
\item [a)] A color excess of $E(B-V)=0.41$ is found for the best
fitting synthetic cluster. Since the maximum color excess provided by
S\&F2011 in this location is 2.83 one concludes that most of the
absorption is produced behind the position of NGC 4349.
\item [b)] The absorption free distance modulus of NGC 4349 is
$11.38\pm0.11$ mag, placing it at a distance of $d=1.88\pm0.05$ kpc from
the Sun.
\end{itemize}

NGC 4349 is the only cluster in our sample with previous photographic photometry
in the $UBV$ system performed by \cite{Lohmann_1961}. Given the usual large
differences between photographic and CCD photometry we performed no comparison
between the Lohmann data-set and ours. According to \cite{Lohmann_1961}, NGC
4349 is located at a distance of $d=1.7$ kpc, almost 200 pc below our estimate.
However, coincidences in terms of reddening, size and background star density
have been found since Lohmann stated a cluster reddening of $E(B-V)=0.38$
and similar cluster size. On the other hand, the Kharchenko
Atlas\footnote{\url{https://webda.physics.muni.cz/cocd.html}}
\citep{Kharchenko_2005} gives a reddening value of $E(B-V)=0.38$ which is
similar to ours with a distance reported of $d=2.1$ kpc, slightly
above our estimate.

The distance found for this cluster using Gaia parallax data with no applied
offset (processed with the Bayesian method described in
Sect.~\ref{ssec:gaia_data}) is $2.04\pm0.03$ kpc, just 160 pc
larger than the photometric distance found by \texttt{ASteCA}.
Notice in Fig.~\ref{fig:plx_bys_NGC4349} that this distance was obtained
respecting the membership selection, thus ensuring that both analysis 
(the photometric analysis and this one) are performed over the exact same set
of stars.

Parallax and proper motion distributions were tested using the Anderson-Darling
statistics. With the exception of the comparison in the case of $PM(\delta)$
(where both samples, cluster and field, seem to come from the same distribution
at a critical value just above 5\%) the remaining two tests report quite
different samples confirming, together with the photometric results, the true
nature of NGC 4349.\\

High probability values for stars inside the overdensity and a clearly traced
cluster sequence confirm the true nature of this object since the over density
and the density profile are followed by a very well defined and extended
photometric counterpart.
Since all these facts are self-consistent, we are confident that NGC 4349 is
an open cluster $0.29\pm0.09\times10^9$ years old.
The Kharchenko Atlas gives quite a similar value for the cluster age, reporting
$\log(t)=8.32$ equivalent to $0.21\times10^9$ yrs.

\subsection{Ruprecht 87}

\begin{figure*}[ht]
    \centering
    \includegraphics[width=\hsize]{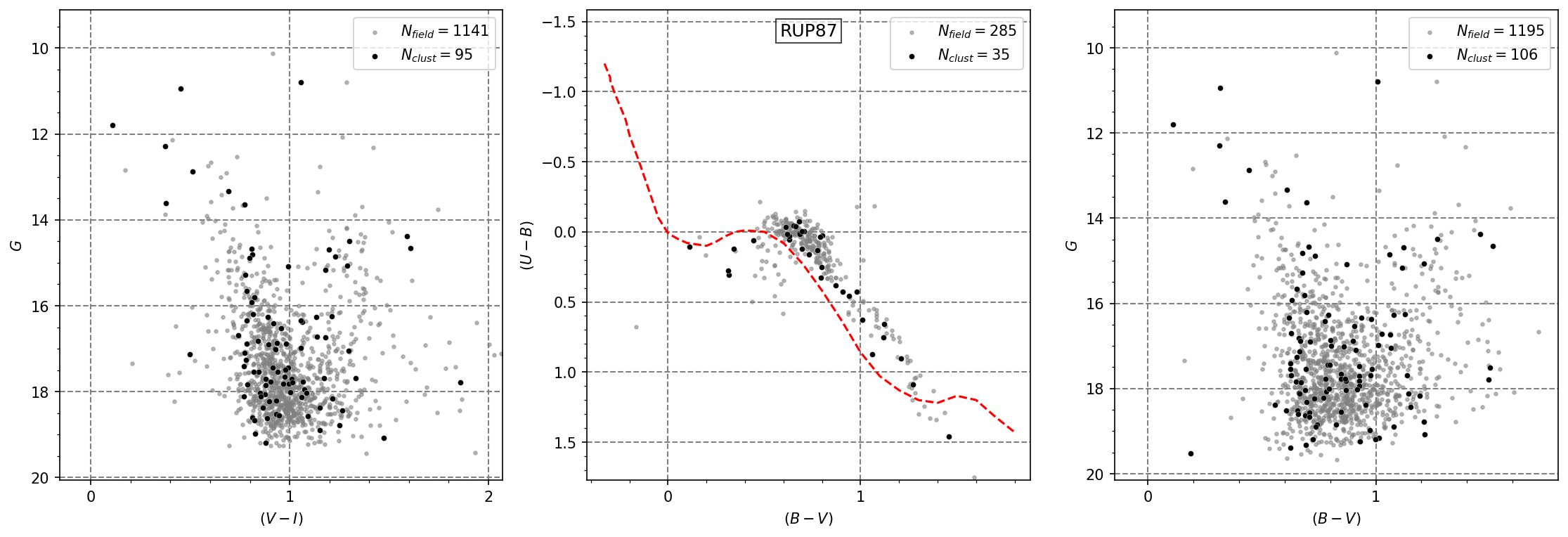}
    \caption{Idem Fig. \ref{fig:photom_vdBH85} for RUP 87.}
    \label{fig:photom_RUP87}
\end{figure*}
\begin{figure*}[ht]
    \centering
    \includegraphics[width=\hsize]{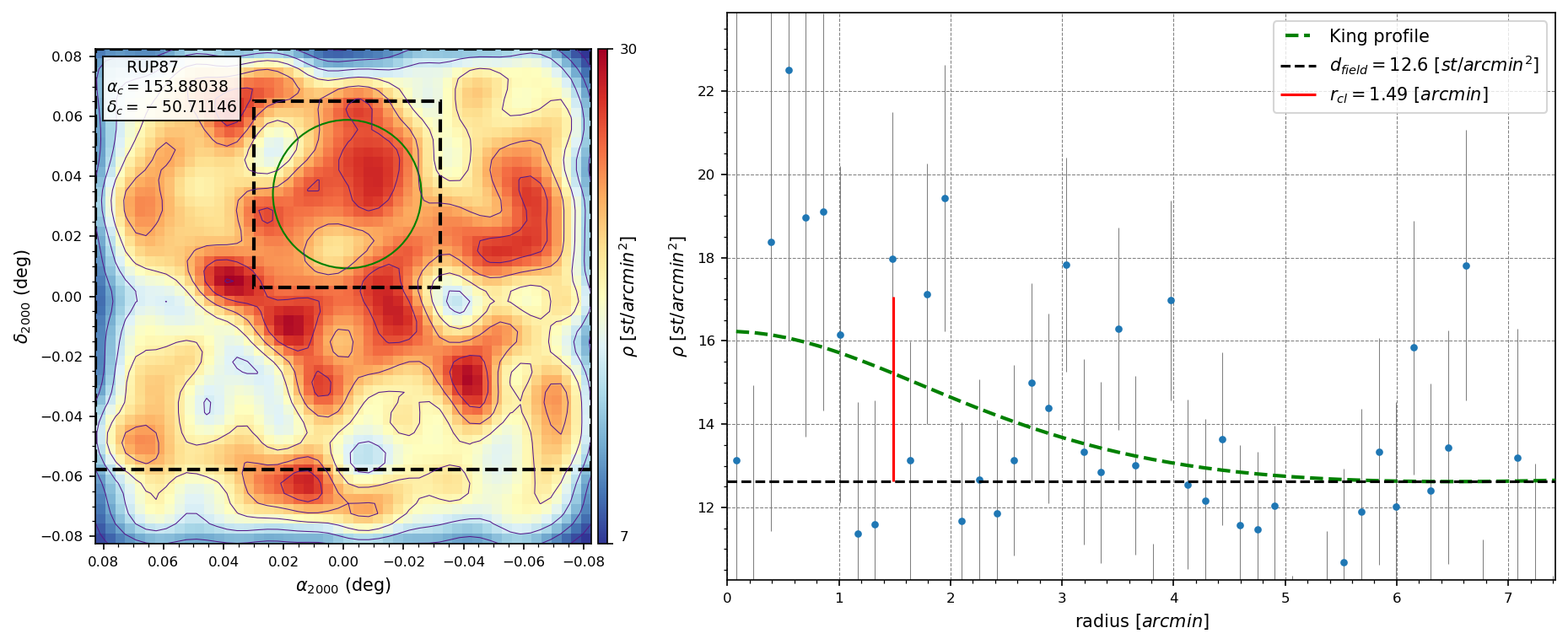}
    \caption{Idem Fig. \ref{fig:struct_vdBH85} for RUP 87.}
    \label{fig:struct_RUP87}
\end{figure*}
\begin{figure*}[ht]
    \centering
    \includegraphics[width=\hsize]{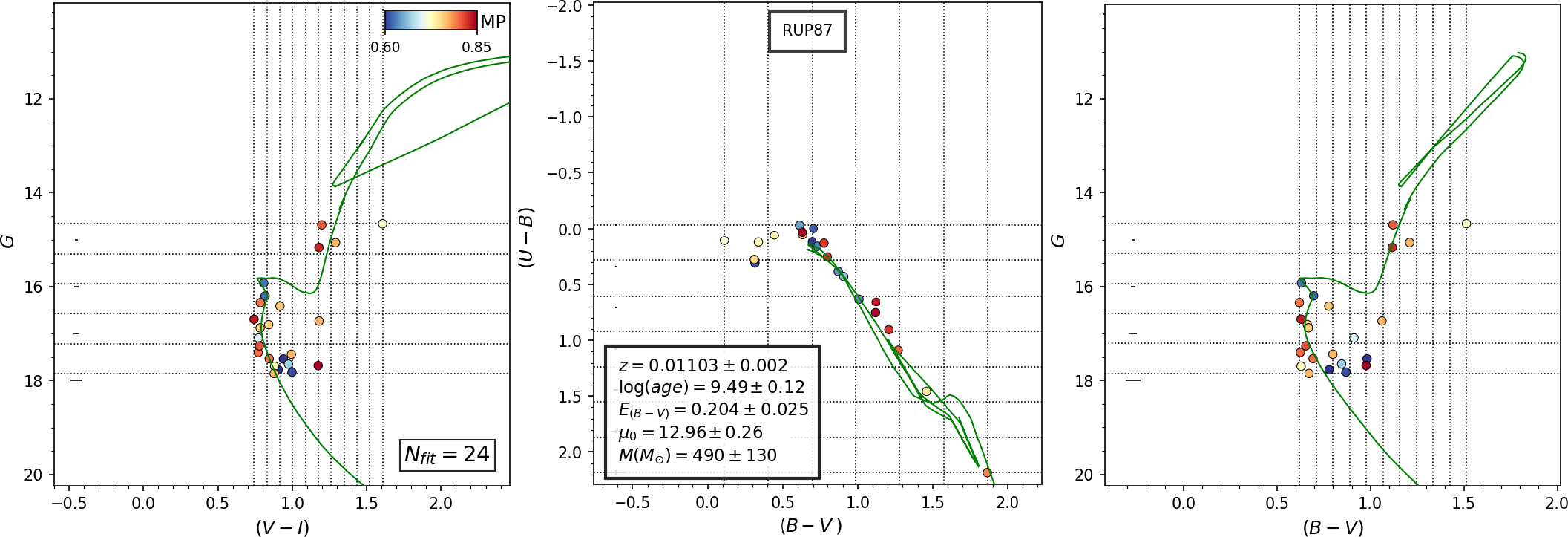}
    \caption{Idem Fig. \ref{fig:fundpars_vdBH85} for RUP 87.}
    \label{fig:fundpars_RUP87}
\end{figure*}
\begin{figure*}[ht]
    \centering
    \includegraphics[width=\hsize]{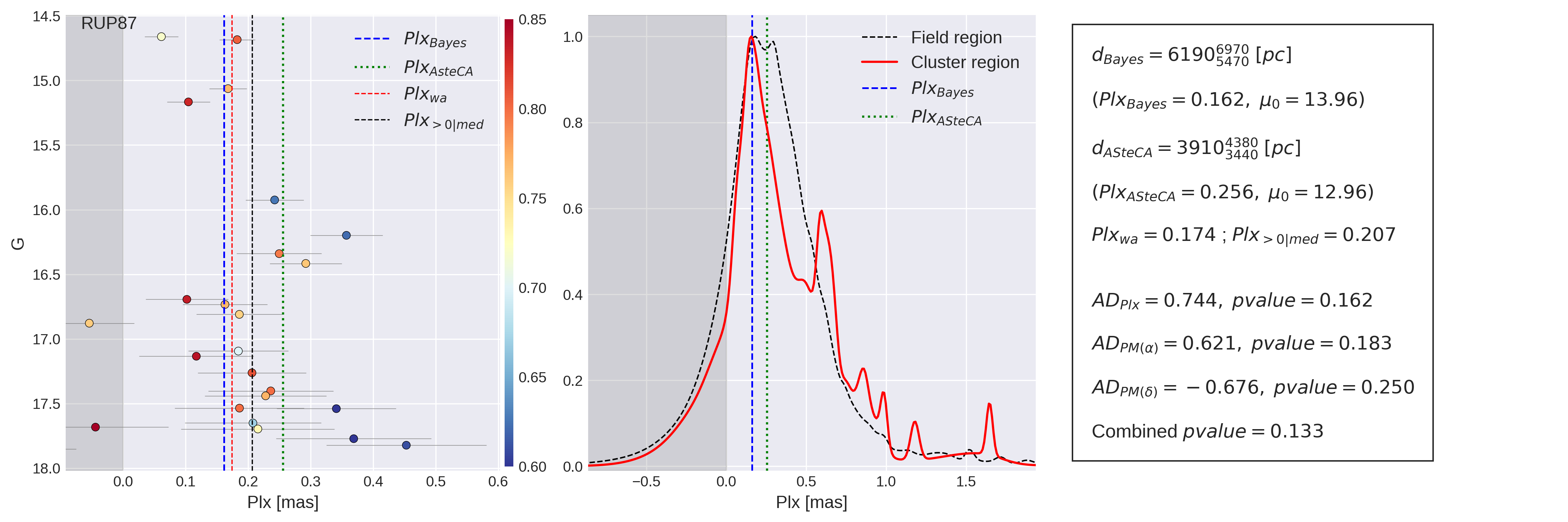}
    \caption{Idem Fig. \ref{fig:plx_bys_vdBH85} for RUP 87.}
    \label{fig:plx_bys_RUP87}
\end{figure*}

RUP 87 is in the east side of the Vela constellation. According to the
respective Fig. \ref{fig:Vim} there is no relevant feature but a rather
poorly populated stellar field with a few bright stars
seemingly grouped  towards the Northern portion of the frame.
The photometric diagrams in Fig.~\ref{fig:photom_RUP87} show no appreciable
stellar structure defining the presence of an open cluster.
The few stars with $(U-B)$ measures plotted in the respective CCD
resemble that of a typical galactic field dominated by a handful of late $F$-
and $G$-type stars followed by a pronounced tail of red stars presumably of
evolved types. Stars in the region $0<(U-B)<0.5$ and $0<(B-V)<0.6$ could be
reddened early $A$- or/and late $B$-types.\\

Accordingly, after many essays \texttt{ASteCA} could not define the presence of
an overdensity as obviously seen in Fig.~\ref{fig:struct_RUP87}.
The inability of our code to identify any overdensity simply means that the
potential locus occupied by the cluster RUP 87 is not unambiguously separated
from the field background stars.
Lacking a clear overdensity we define the
cluster region as that encircled by the green line, i.e., the
sector containing the apparently grouped bright stars. The RDP emerging from
this analysis is quite noisy.

Comparing the density of the defined cluster region with that of the
remaining stellar field, the approximate number of probable members turns out
to be around 20 stars.
When dealing with (purported) clusters with such a low estimated number of
members it is important to be extremely careful with the selection of stars
considered to be ``members''. If we were to simply select a small group
of stars within a similar parallax range and analyzed their photometric
diagram with \texttt{ASteCA}, we will probably obtain a somewhat reasonable
fit. This is because the code will always find the most likely solution,
no matter how dispersed the photometric diagram we give it might be. If we a
priori hand-pick a few stars with a common distance (parallax values), they
will be fitted by a synthetic cluster with a very similar distance modulus as
that defined by the selected parallax values, and some ``best fitted'' values
for the remaining parameters.
Similarly the naive selection of stars with probabilities larger than 0.5 is
not appropriate most of the times (unless a clear sequence can
be defined, as in the case of NGC 4349), since this selection is biased
towards brighter stars.
This is because low mass stars not only have larger associated uncertainties,
they are also located in denser regions of the CMDs. This makes them much more
likely to be assigned lower membership probabilities. A simple cut on 0.5
would generally result in a cluster sequence composed mostly by bright stars,
without respecting the actual photometric density of the purported cluster 
(given by the cluster region vs field region photometric density differences).
Hence, the selected stars within the cluster region should be not only those
with large membership probabilities or sharing a similar physical attribute 
(i.e., parallax). They should also be properly distributed in the photometric
diagrams and as close as possible in number to the estimated number of members.
As stated above this is of particular importance for clusters with few members,
as the process to find their best fit parameters is driven by a handful of
stars which makes the analysis much more delicate.

In the case of RUP 87 we selected stars that had both large membership
values, and were similar in number to the estimated number of members for the
cluster region. The 24 stars that remain in the adopted region along with the
best fit found can be seen in Fig.~\ref{fig:fundpars_RUP87}. The code fits a
somewhat old ($3.1\times10^9$ yrs) synthetic cluster at a distance of
$\sim3900$ pc.\\

As seen in Fig.~\ref{fig:plx_bys_RUP87} the distance estimated through
Gaia parallaxes for the same set of stars is $\sim6200$ pc, which is more
than 2000 pc away from the photometric estimate. This difference is too large to
be consistent with a real cluster, even taking possible offsets into account.
To see if this discrepancy could be solved as we did for vdBH 85 (see
Sect.~\ref{sec:gaia_distances}) we run the same analysis described there using
Bailer-Jones distances. The resulting weighted average for the
distance is $4680_{3090}^{6260}$ pc. This distance is almost 800 pc larger
than the photometric estimate, and 1500 pc smaller than the Gaia parallax
estimate. Such large differences are consistent with the fact that we are not
analyzing an actual cluster.

The Anderson-Darling test values for $Plx$ and proper motions do not confirm
clear differences between the cluster region and the stellar background in
terms of kinematics and distance. The poverty of the photometric diagrams and
the analysis of photometric distances versus parallax distances
are all against the true existence of a cluster in the region RUP 87.
In our interpretation this is not a real entity but a fluctuation of the star
field.

\section{Gaia parallax distances analysis}
\label{sec:gaia_distances}

We shall close our analysis by taking a look at the matter of distances yielded
by \texttt{ASteCA} and those that can be obtained using parallaxes alone.
Specifically, we cross-matched $Plx$ data with our photometry, cluster by
cluster, and processed them within a Bayesian framework (as explained in Sect
\ref{ssec:gaia_data}). The intention is to visualize the change in estimated
distances if no correction is applied to the parallaxes, and when current
values taken from the literature are used.\\

\begin{figure*}[ht]
    \centering
    \includegraphics[width=\hsize]{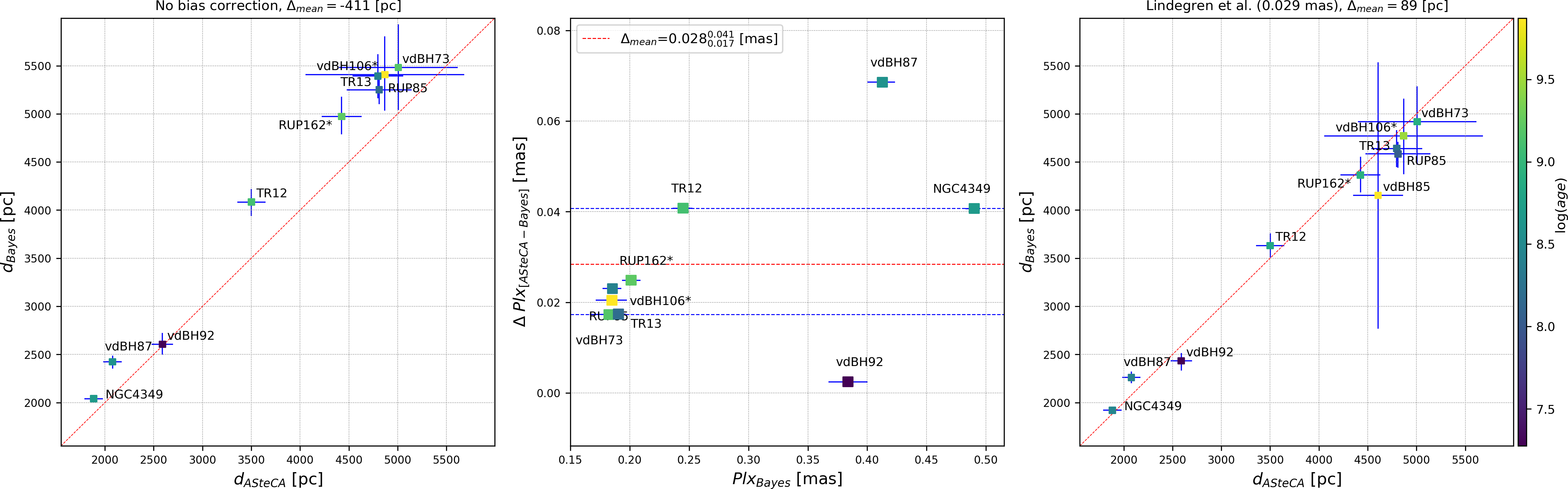}
\caption{Left: \texttt{ASteCA} (photometric) vs Bayesian (parallax)
distances for the clusters listed in Table~\ref{tab:final_tab}, 
confirmed to be real clusters. No bias correction applied to parallax
data. Colorbar to the right indicates $\log(age)$ values.
Center: offset (\texttt{ASteCA} - Bayes) for distances expressed as
parallax in miliarcseconds.
Right: same as left plot, with bias corrections from Lindegren et al.
(+0.029 mas).
The cluster vdBH85 is included here with its distance value estimated from the
list of \cite{BailerJones_2018} individual distances.
}
    \label{fig:prlxbias}
\end{figure*}

In Fig. \ref{fig:prlxbias} we show the \texttt{ASteCA} versus Bayesian 
(parallax) distances with no offset applied (left), and the Bayesian
parallax for each cluster (as the inverse of the distance) versus its
difference with the \texttt{ASteCA} estimate (middle).
It is evident from this figure that \texttt{ASteCA} distances are
systematically smaller than the ones coming from the computation of parallax
alone. The mean of the \texttt{ASteCA} minus parallax differences in
distance is $\sim-411$ pc.
The middle plot with the mean difference suggests that a correction of
+0.028 mas should be applied to the Gaia DR2 parallax values.
The cluster vdBH85 is omitted from Fig.~\ref{fig:prlxbias} (left and
middle plots) because the Bayesian framework applied on its parallax data
yielded results that were clearly wrong. This can be seen in Fig. 
\ref{fig:plx_bys_vdBH85} where the parallax distance estimated is above 8 kpc,
versus the photometric distance obtained by \texttt{ASteCA} of $\sim$4.6 kpc.
Out of the ten clusters in our list of confirmed plus dubious clusters, vdBH85
is the oldest one. This means that its main sequence is quite short and
composed mostly of low mass stars.
More than 60\% of its 146 estimated members have $G>18$ mag, and almost
75\% have Gaia DR2 parallax values with uncertainties larger than 0.1 mas 
(with a mean parallax uncertainty of $\sim$0.16 mas). Because of this, the
Bayesian method fails to estimate a reasonable distance for this cluster,
and we omit it from this analysis.\\

A number of recent articles have found that there is an offset present in
Gaia's parallax data, covering a range of approximately +0.05 mas. We selected
three of these articles that fully cover this range, to compare with our
results obtained with no bias corrections: \cite{Lindegren_2018},
\cite{Schonrich2019}, and \cite{Xu_2019}.
In Lindegren et al. the authors processed the parallax of hundreds of
thousands of quasars deriving a median difference with Gaia data of +0.029 mas.
The work by Sch\"onrich et al. analyzed the radial velocities subset of Gaia
DR2 with their own Bayesian inference tool, and estimated a required +0.054 mas
offset in the parallax data from Gaia DR2. Finally, Xu et al. used $\sim$100
stars with Very Long Baseline Interferometry astrometry, and found an offset of
+0.075 mas with Gaia DR2 parallaxes. 
If we add to the parallax data the offsets given in Lindegren et al., 
Sch\"onrich et al., and Xu et al. (+0.029, +0.054, +0.075 mas,
respectively) the agreement between \texttt{ASteCA} and parallax distances
improves at first and rapidly worsens. The mean differences between photometric
distances and parallax distances are of $\sim0.09$ kpc, $\sim0.39$ kpc, and
$\sim0.62$ kpc, using the Lindegren et al., Sch\"onrich et al., and Xu
et al. corrections, respectively.

In the case of vdBH85, being unable to apply the Bayesian method
described in Sect. \ref{ssec:gaia_data} (as explained above), we turn
to the individual distance values obtained in
\cite{BailerJones_2018}.
In this article the authors use Bayesian inference to estimate distances (in
parsec) to more than 1 billion stars using the Gaia DR2 parallax values,
applying the Lindegren et al. correction.\footnote{This should not be
confused with the Bayesian inference method described in
Sect.~\ref{ssec:gaia_data}. These are two very different processes.}
We cross match our list of members for vdBH85 and approximate the distance to
the cluster as their average distance, weighted by the assigned uncertainties.
Although this is a rather low quality estimate due to the large
uncertainties in the individual distances, as seen by the large error bars
in Fig. \ref{fig:prlxbias} (right plot), it is still close to the
photometric distance estimate.
If we omit vdBH85 entirely, the Lindegren et al. mean difference improves to
$\sim0.05$ kpc.

Our analysis thus points to a required bias correction to Gaia parallaxes of
+0.028 mas, which is very close to the one proposed by Lindegren et al.\\

In Sect \ref{ssec:photom_reduc} we saw that our $(B-V)$ color has a small
offset of $\sim0.0153$ mag when compared to the (transformed) Gaia photometry.
\texttt{ASteCA} employs the extinction law by \citet[][CCC law]{Cardelli_1989}
with the \citet{Odonnell_1994} correction for the near UV, to transform $E
(B-V)$ values into absorptions for any filter.
In our case, we used Gaia's $G$ filter whose absorption $A_G$ is related to $E
(B-V)$ as: $A_G = c_0 \, A_V  = c_0 \, 3.1 \, E(B-V)$, where $c_0\approx0.829$
according to the CCC law. Hence the absorption $A_{G}^{\prime}$, i.e.,
corrected by the offset in $(B-V)$, can be written as: $A_{G}^
{\prime}=0.039+A_G$.
Given the range of distance moduli in this work ($\sim$11 - 14 mag), the impact
of this correction on the distance in parsec goes from $\sim$30 to 100 pc.
If we apply this $(B-V)$ offset to our photometric distances and re-run the
analysis, the +0.028 mas bias in Gaia parallaxes that we found
initially is reduced to +0.023 mas. This is a smaller value, but still very
close to the Lindegren et al. bias.

Certainly, analysis of a more extended sample of clusters is needed for
arriving to conclusive results and establish the detailed relation between
distances from photometry and DR2 parallaxes. The results of the exercise
presented in this section are included in the last 4 columns of Table 
\ref{tab:final_tab}.

\section{Discussion of results and concluding remarks}
\label{sec:results_concl}

We have analyzed the fields of sixteen catalogued open clusters located in a
Galaxy sector covering from $270^\circ$ to $300^\circ$ approximately in
galactic longitude, and mostly close to the formal galactic plane at
$b=0^\circ$.
The cluster parameter estimations presented in this article are based on
precise $UBVI$ photometry analyzed in a automatic way by our code 
\texttt{ASteCA}. The code searches for a meaningful stellar overdensity
assigning membership probabilities by comparison with the surrounding stellar
field.
The next step establishes the physical properties of the best synthetic cluster
that fits the distribution of cluster members in the CMDs and the CCD.
Through this process reddening, distance, age, mass, and metallicity are
given.
The most relevant inconvenience we have found with the present cluster
sample resides in the fact that some of them are extremely faint, which
becomes evident in a visual inspection of their overall CCDs and CMDs.
Things get more difficult because the $(U-B)$ index has been mostly available
only for the bright and blue stars which reduced considerably the data analysis
space. Despite this, we were able to keep the reddening solutions under control
and obtain reliable distances estimations for those objects found true clusters
by our code. This way, we can safely reject RUP 87, vdBH91, RUP 88, Lynga 15,
Loden 565 and NGC 4230 that are most probably random stellar fluctuations. The
results for true and probable open clusters are shown in Table 
\ref{tab:final_tab} in self-explicative format.

\begin{table*}[ht]
\small
\centering
\caption{The symbol ``*'' indicates probable clusters. The $d_{noofset}$ values
are those obtained using the Bayesian method and no bias correction applied on
the Gaia DR2 parallax data. The remaining distances were obtained applying the
indicated offsets to the parallax values.}
\begin{tabular}{lccccccccc}
\hline \hline 
 \emph{Cluster} & z & \emph{Age} & $E(B-V)$ & \emph{Mass} &
 $d_{ASteCA}$ & $d_{noofset}$ & $d_{Lindegren}$ & $d_{Sch\ddot{o}nrich}$ &
 $d_{Xu}$ \\
& & $(10^9 yr)$ & $mag$ & $(10^3\,M_{\odot})$ & $(kpc)$ & $(kpc)$ & $(kpc)$ &
$(kpc)$ & $(kpc)$\\
 \hline
 vdBH 73 & 0.019$\pm0.004$ & 0.78$\pm0.09$ & 1.06$\pm0.04$ & 2.6$\pm0.9$ &
  $5.01\pm0.61$ & $5.48\pm0.44$ & $4.92\pm0.41$ & $4.46\pm0.31$ & $4.05\pm0.33$\\
 RUP 85 & 0.021$\pm0.003$ & 0.18$\pm0.03$ & 1.06$\pm0.03$ & 2.6$\pm0.5$ &
 $4.80\pm0.26$ & $5.39\pm0.23$ & $4.64\pm0.19$ & $4.16\pm0.15$ & $3.83\pm0.14$\\
 vdBH 85 & 0.014$\pm0.002$ & 7.50$\pm0.80$ & 0.30$\pm0.03$ & 2.2$\pm0.5$ &
  $4.61\pm0.26$ & -- & $4.15\pm1.38$ & -- & --\\
 vdBH 87 & 0.025$\pm0.002$ & 0.25$\pm0.08$ & 0.55$\pm0.04$ & 1.4$\pm0.2$ &
 $2.08\pm0.09$ & $2.42\pm0.07$ & $2.26\pm0.06$ & $2.13\pm0.05$ & $2.05\pm0.05$\\
 TR 12 & 0.009$\pm0.002$ & 0.70$\pm0.10$ & 0.31$\pm0.03$ & 0.7$\pm0.1$ &
 $3.50\pm0.15$ & $4.08\pm0.14$ & $3.63\pm0.13$ & $3.31\pm0.10$ & $3.11\pm0.09$\\
 vdBH 92 & 0.009$\pm0.004$ & 0.02$\pm0.01$ & 0.65$\pm0.03$ & 0.4$\pm0.1$ &
 $2.59\pm0.11$ & $2.61\pm0.11$ & $2.43\pm0.09$ & $2.28\pm0.07$ & $2.17\pm0.07$\\
 TR 13 & 0.007$\pm0.004$ & 0.11$\pm0.02$ & 0.56$\pm0.02$ & 0.7$\pm0.2$ &
 $4.81\pm0.33$ & $5.25\pm0.16$ & $4.58\pm0.14$ & $4.10\pm0.11$ & $3.75\pm0.09$\\
 vdBH 106* & 0.012$\pm0.003$ & 3.00$\pm0.80$ & 0.30$\pm0.04$ & 0.5$\pm0.2$ &
 $4.87\pm0.81$ & $5.41\pm0.39$ & $4.77\pm0.39$ & $4.31\pm0.33$ & $4.06\pm0.30$\\
 RUP 162* & 0.009$\pm0.002$ & 0.80$\pm0.20$ & 0.54$\pm0.03$ & 1.2$\pm0.2$ &
 $4.43\pm0.20$ & $4.97\pm0.20$ & $4.37\pm0.18$ & $3.94\pm0.15$ & $3.66\pm0.13$\\
 NGC 4349 & 0.011$\pm0.004$ & 0.29$\pm0.09$ & 0.41$\pm0.05$ & 2.0$\pm0.1$ &
 $1.88\pm0.05$ & $2.04\pm0.03$ & $1.92\pm0.02$ & $1.83\pm0.02$ & $1.76\pm0.01$\\
\hline
\end{tabular}
\label{tab:final_tab}
\end{table*}

If we average the metallicity for each cluster, shown in the second
column of Table \ref{tab:final_tab}, the metal content is $z=0.0136\pm0.006$.
The result is well in agreement with the assumption that the typical Milky Way
open cluster has solar metallicity \citep[$z=0.0152$,][]{Bressan_2012}.\\

Of the remaining ten objects, two are probable clusters with distances in the
4-5 kpc range. Ages of clusters sweep from a few million years
to almost 8 billion years in the case of vdBH 85. The vdBH 106 cluster
is one of the oldest but it is just a probable open cluster, so its age should
be taken with reservation. Two other objects, TR 13 and vdBH 92, are young with
ages close to and under 100 million years respectively, while the rest are all
less than 1 billion years old.

\begin{figure}[ht]
    \centering
    \includegraphics[width=\hsize]{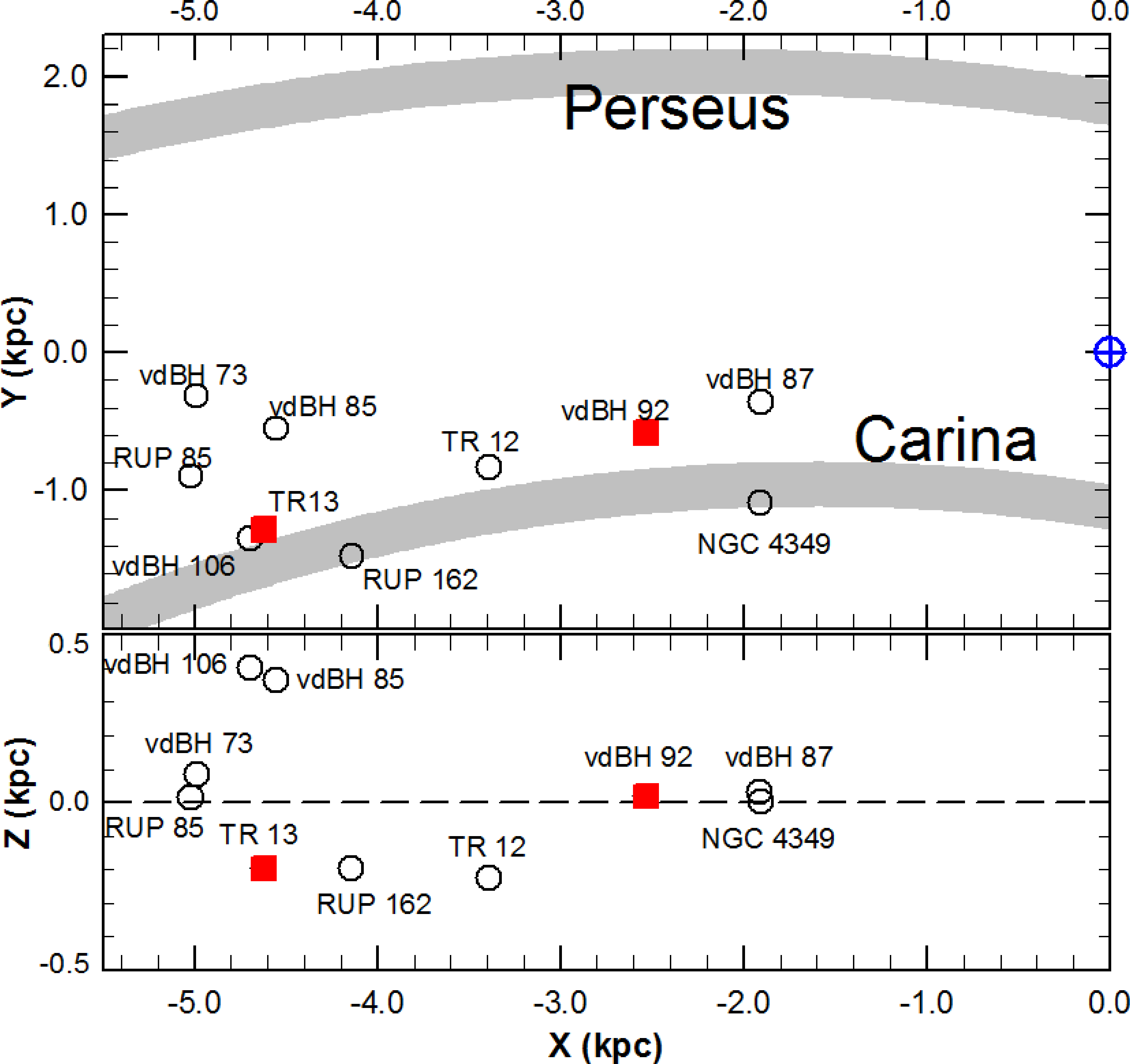}
\caption{The X-Y (upper panel) and X-Z (lower plane) projection of the true and
probable clusters in our sample (open circles). The red squares
enclose the youngest clusters in our list (see Table~\ref{tab:final_tab}).
In the upper panel the thick grey lines shows the trace of the Perseus and
Carina arms according to \cite{valle_2005}.
The position of the Sun is shown by a blue circle with a cross inside.
Dashed line in the lower panel depicts the galactic equator.}
\label{fig68}
\end{figure}

A final remark concerns the spatial distribution of the
eight real clusters plus two probable ones indicated in Table 
\ref{tab:final_tab}. These objects  are plotted in Fig~\ref{fig68} in the X-Y
(upper) and X-Z (lower) planes of the Milky Way -following the usual signs
convention- where the Sun is placed at $(0, 0)$. Superposed in this
figure is the outline of the Carina Arm taken from \cite{valle_2005}. All
these objects are plotted with open circles except the two youngest,
which are shown with red squares. TR 13, one of the youngest (0.1 Gyr)
and farthest (4.8 kpc) objects, is located along the external side of the
Carina arm but appears well below the Galaxy plane at about -0.2 kpc, thus
accompanying the warp of this arm already mentioned by, among others,
\cite{Cersosimo_2009}. 
The other young cluster, vdBH 92 (0.02 Gyr), is relatively far from the Carina
Nebula nucleus in an intermediate zone between that region and the Sun but
still seen close to the northwest side of the Carina Nebula at a distance
that is comprised within the estimated maximum and minimum distance for Carina.
vdBH 106 (3 Gyr) and vdBH 85 (7.5 Gyr) are the oldest objects found in our
search and are, in turn, placed well above the formal galactic equator (0.3-0.4
kpc). TR 12 (0.7 Gyr) is another quite old object placed below the plane (-0.2
kpc) together with RUP 162 (1 Gyr). The rest of the clusters are of middle age
and relatively close to the Galaxy plane.

With respect to photometric versus parallax distances, we can conclude that by
adding $\sim+0.028$ mas to the cluster computed parallaxes from Gaia DR2
the level of agreement with the photometric distances improves considerably.
Taking into account the small offset found for the $(B-V)$ color, this
value drops to +0.023 mas, which is only 0.006 mas smaller than the
Lindegren et al. +0.029 mas correction.
This reinforces the evidence pointing to this offset over larger
values proposed in the literature. Our cluster sample is not large enough
to permit us drawing stronger conclusions on this matter, particularly
regarding the possible dependence of the correction with distance.

\begin{acknowledgements}

G.P., E.E.G., M.S.P. and R.A.V. acknowledge the financial support from CONICET 
(PIP317) and the UNLP.
AM acknowledges the support from the Portuguese Strategic Programme
UID/FIS/00099/2019 for CENTRA.
The authors are very much indebted with the anonymous referee for the helpful
comments and suggestions that contributed to greatly improving the manuscript.
This research was made possible through the use of the AAVSO Photometric
All-Sky Survey (APASS), funded by the Robert Martin Ayers Sciences Fund and NSF
AST-1412587.
This research has made use of the WEBDA database, operated at the Department of
Theoretical Physics and Astrophysics of the Masaryk University.
This research has made use of the VizieR catalogue access tool, operated at CDS,
Strasbourg, France~\citep{Ochsenbein_2000}.
This research has made use of ``Aladin sky atlas'' developed at
CDS, Strasbourg Observatory, France~\citep{Bonnarel2000,Boch2014}.
This research has made use of NASA's Astrophysics Data System.
This research made use of the Python language v3.7.3~\citep{vanRossum_1995}
and the following packages:
NumPy\footnote{\url{http://www.numpy.org/}}~\citep{vanDerWalt_2011};
SciPy\footnote{\url{http://www.scipy.org/}}~\citep{Jones_2001};
Astropy\footnote{\url{http://www.astropy.org/}}, a community-developed core
Python package for Astronomy \citep{Astropy_2013};
matplotlib\footnote{\url{http://matplotlib.org/}}~\citep{hunter_2007};
emcee\footnote{\url{http://emcee.readthedocs.io}}~\citep{emcee};
corner.py\footnote{\url{https://corner.readthedocs.io}}~\citep{corner}.
\end{acknowledgements}

\bibliographystyle{aa}
\bibliography{biblio}

\begin{thebibliography}{55}
\expandafter\ifx\csname natexlab\endcsname\relax\def\natexlab#1{#1}\fi

\bibitem[{{Aller} {et~al.}(1982){Aller}, {Appenzeller}, {Baschek}, {Duerbeck},
  {Herczeg}, {Lamla}, {Meyer-Hofmeister}, {Schmidt-Kaler}, {Scholz},
  {Seggewiss}, {Seitter}, \& {Weidemann}}]{Aller1982}
{Aller}, L.~H., {Appenzeller}, I., {Baschek}, B., {et~al.}, eds. 1982,
  {Landolt-B{\"o}rnstein: Numerical Data and Functional Relationships in
  Science and Technology - New Series '' Gruppe/Group 6 Astronomy and
  Astrophysics '' Volume 2 Schaifers/Voigt: Astronomy and Astrophysics /
  Astronomie und Astrophysik '' Stars and Star Clusters / Sterne und
  Sternhaufen}

\bibitem[{{Astropy Collaboration} {et~al.}(2013){Astropy Collaboration},
  {Robitaille}, {Tollerud}, {Greenfield}, {Droettboom}, {Bray}, {Aldcroft},
  {Davis}, {Ginsburg}, {Price-Whelan}, {Kerzendorf}, {Conley}, {Crighton},
  {Barbary}, {Muna}, {Ferguson}, {Grollier}, {Parikh}, {Nair}, {Unther},
  {Deil}, {Woillez}, {Conseil}, {Kramer}, {Turner}, {Singer}, {Fox}, {Weaver},
  {Zabalza}, {Edwards}, {Azalee Bostroem}, {Burke}, {Casey}, {Crawford},
  {Dencheva}, {Ely}, {Jenness}, {Labrie}, {Lim}, {Pierfederici}, {Pontzen},
  {Ptak}, {Refsdal}, {Servillat}, \& {Streicher}}]{Astropy_2013}
{Astropy Collaboration}, {Robitaille}, T.~P., {Tollerud}, E.~J., {et~al.} 2013,
  \aap, 558, A33

\bibitem[{{Avedisova}(2002)}]{Avedisova_2002}
{Avedisova}, V.~S. 2002, Astronomy Reports, 46, 193

\bibitem[{{Bailer-Jones} {et~al.}(2018){Bailer-Jones}, {Rybizki}, {Fouesneau},
  {Mantelet}, \& {Andrae}}]{BailerJones_2018}
{Bailer-Jones}, C.~A.~L., {Rybizki}, J., {Fouesneau}, M., {Mantelet}, G., \&
  {Andrae}, R. 2018, \aj, 156, 58

\bibitem[{{Boch} \& {Fernique}(2014)}]{Boch2014}
{Boch}, T. \& {Fernique}, P. 2014, in Astronomical Society of the Pacific
  Conference Series, Vol. 485, Astronomical Data Analysis Software and Systems
  XXIII, ed. N.~{Manset} \& P.~{Forshay}, 277

\bibitem[{{Bonnarel} {et~al.}(2000){Bonnarel}, {Fernique}, {Bienaym{\'e}},
  {Egret}, {Genova}, {Louys}, {Ochsenbein}, {Wenger}, \&
  {Bartlett}}]{Bonnarel2000}
{Bonnarel}, F., {Fernique}, P., {Bienaym{\'e}}, O., {et~al.} 2000, AAPS, 143,
  33

\bibitem[{{Bossini} {et~al.}(2019){Bossini}, {Vallenari}, {Bragaglia},
  {Cantat-Gaudin}, {Sordo}, {Balaguer-N{\'u}{\~n}ez}, {Jordi}, {Moitinho},
  {Soubiran}, {Casamiquela}, {Carrera}, \& {Heiter}}]{Bossini_2019}
{Bossini}, D., {Vallenari}, A., {Bragaglia}, A., {et~al.} 2019, \aap, 623, A108

\bibitem[{{Bressan} {et~al.}(2012){Bressan}, {Marigo}, {Girardi}, {Salasnich},
  {Dal Cero}, {Rubele}, \& {Nanni}}]{Bressan_2012}
{Bressan}, A., {Marigo}, P., {Girardi}, L., {et~al.} 2012, \mnras, 427, 127

\bibitem[{{Cantat-Gaudin} {et~al.}(2018){Cantat-Gaudin}, {Jordi}, {Vallenari},
  {Bragaglia}, {Balaguer-N{\'u}{\~n}ez}, {Soubiran}, {Bossini}, {Moitinho},
  {Castro-Ginard}, {Krone-Martins}, {Casamiquela}, {Sordo}, \&
  {Carrera}}]{2018A&A...618A..93C}
{Cantat-Gaudin}, T., {Jordi}, C., {Vallenari}, A., {et~al.} 2018, A\&A, 618,
  A93

\bibitem[{{Cardelli} {et~al.}(1989){Cardelli}, {Clayton}, \&
  {Mathis}}]{Cardelli_1989}
{Cardelli}, J.~A., {Clayton}, G.~C., \& {Mathis}, J.~S. 1989, \apj, 345, 245

\bibitem[{{Carraro} {et~al.}(2010){Carraro}, {Costa}, \&
  {Ahumada}}]{Carraro_2010}
{Carraro}, G., {Costa}, E., \& {Ahumada}, J.~A. 2010, \aj, 140, 954

\bibitem[{{Carraro} {et~al.}(2005){Carraro}, {V{\'a}zquez}, {Moitinho}, \&
  {Baume}}]{Carraro_2005}
{Carraro}, G., {V{\'a}zquez}, R.~A., {Moitinho}, A., \& {Baume}, G. 2005,
  \apjl, 630, L153

\bibitem[{{Cersosimo} {et~al.}(2009){Cersosimo}, {Mader}, {Figueroa},
  {V{\'e}lez}, {Soto}, \& {Azc{\'a}rate}}]{Cersosimo_2009}
{Cersosimo}, J.~C., {Mader}, S., {Figueroa}, N.~S., {et~al.} 2009, \apj, 699,
  469

\bibitem[{Charbonneau(1995)}]{Charbonneau_1995}
Charbonneau, P. 1995, The Astrophysical Journal Supplement Series, 101, 309

\bibitem[{{Dias} {et~al.}(2002){Dias}, {Alessi}, {Moitinho}, \&
  {L{\'e}pine}}]{Dias_2002}
{Dias}, W.~S., {Alessi}, B.~S., {Moitinho}, A., \& {L{\'e}pine}, J.~R.~D. 2002,
  \aap, 389, 871

\bibitem[{Dolphin(2002)}]{Dolphin_2002}
Dolphin, A.~E. 2002, Monthly Notices of the Royal Astronomical Society, 332, 91

\bibitem[{Efron \& Tibshirani(1986)}]{efron1986}
Efron, B. \& Tibshirani, R. 1986, Statist. Sci., 1, 54

\bibitem[{Foreman-Mackey(2016)}]{corner}
Foreman-Mackey, D. 2016, The Journal of Open Source Software, 1, 24

\bibitem[{{Foreman-Mackey} {et~al.}(2013){Foreman-Mackey}, {Hogg}, {Lang}, \&
  {Goodman}}]{emcee}
{Foreman-Mackey}, D., {Hogg}, D.~W., {Lang}, D., \& {Goodman}, J. 2013, PASP,
  125, 306

\bibitem[{{Gaia Collaboration} {et~al.}(2018){Gaia Collaboration}, {Brown},
  {Vallenari}, {Prusti}, {de Bruijne}, {Babusiaux}, \&
  {Bailer-Jones}}]{GaiaDR2_2018}
{Gaia Collaboration}, {Brown}, A.~G.~A., {Vallenari}, A., {et~al.} 2018, ArXiv
  e-prints [\eprint[arXiv]{1804.09365}]

\bibitem[{Hunter {et~al.}(2007)}]{hunter_2007}
Hunter, J.~D. {et~al.} 2007, Computing in science and engineering, 9, 90

\bibitem[{Janes \& Adler(1982)}]{Janes_1982}
Janes, K. \& Adler, D. 1982, The Astrophysical Journal Supplement Series, 49,
  425

\bibitem[{Jones {et~al.}(2001)Jones, Oliphant, Peterson, {et~al.}}]{Jones_2001}
Jones, E., Oliphant, T., Peterson, P., {et~al.} 2001, {SciPy}: Open source
  scientific tools for {Python}, [Online; accessed 2016-06-21]

\bibitem[{{Kharchenko} {et~al.}(2005){Kharchenko}, {Piskunov}, {R{\"o}ser},
  {Schilbach}, \& {Scholz}}]{Kharchenko_2005}
{Kharchenko}, N.~V., {Piskunov}, A.~E., {R{\"o}ser}, S., {Schilbach}, E., \&
  {Scholz}, R.-D. 2005, A\&A, 438, 1163

\bibitem[{{King}(1962)}]{King_1962}
{King}, I. 1962, \aj, 67, 471

\bibitem[{{Kroupa}(2002)}]{Kroupa_2002}
{Kroupa}, P. 2002, Science, 295, 82

\bibitem[{{Landolt}(1992)}]{1992AJ....104..340L}
{Landolt}, A.~U. 1992, AJ, 104, 340

\bibitem[{{Lindegren} {et~al.}(2018){Lindegren}, {Hern{\'a}ndez}, {Bombrun},
  {Klioner}, {Bastian}, {Ramos-Lerate}, {de Torres}, {Steidelm{\"u}ller},
  {Stephenson}, {Hobbs}, {Lammers}, {Biermann}, {Geyer}, {Hilger}, {Michalik},
  {Stampa}, {McMillan}, {Casta{\ ~n}eda}, {Clotet}, {Comoretto}, {Davidson},
  {Fabricius}, {Gracia}, {Hambly}, {Hutton}, {Mora}, {Portell}, {van Leeuwen},
  {Abbas}, {Abreu}, {Altmann}, {Andrei}, {Anglada}, {Balaguer-N{\'u}{\ ~n}ez},
  {Barache}, {Becciani}, {Bertone}, {Bianchi}, {Bouquillon}, {Bourda},
  {Br{\"u}semeister}, {Bucciarelli}, {Busonero}, {Buzzi}, {Cancelliere},
  {Carlucci}, {Charlot}, {Cheek}, {Crosta}, {Crowley}, {de Bruijne}, {de
  Felice}, {Drimmel}, {Esquej}, {Fienga}, {Fraile}, {Gai}, 49~{Garralda},
  {Gonz{\'a}lez-Vidal}, {Guerra}, {Hauser}, {Hofmann}, {Holl}, {Jordan},
  {Lattanzi}, {Lenhardt}, {Liao}, {Licata}, {Lister}, {L{\"o}ffler},
  {Marchant}, {Martin- Fleitas}, {Messineo}, {Mignard}, {Morbidelli}, {Poggio},
  {Riva}, {Rowell}, {Salguero}, {Sarasso}, {Sciacca}, {Siddiqui}, {Smart},
  {Spagna}, {Steele}, {Taris}, {Torra}, {van Elteren}, {van Reeven}, \&
  {Vecchiato}}]{Lindegren_2018}
{Lindegren}, L., {Hern{\'a}ndez}, J., {Bombrun}, A., {et~al.} 2018, A\&A, 616,
  A2

\bibitem[{{Lohmann}(1961)}]{Lohmann_1961}
{Lohmann}, W. 1961, Astronomische Nachrichten, 286, 105

\bibitem[{{Luri} {et~al.}(2018){Luri}, {Brown}, {Sarro}, {Arenou},
  {Bailer-Jones}, {Castro-Ginard}, {de Bruijne}, {Prusti}, {Babusiaux}, \&
  {Delgado}}]{Luri_2018}
{Luri}, X., {Brown}, A.~G.~A., {Sarro}, L.~M., {et~al.} 2018, ArXiv e-prints
  [\eprint[arXiv]{1804.09376}]

\bibitem[{{Magrini} {et~al.}(2009){Magrini}, {Stanghellini}, {Corbelli},
  {Galli}, \& {Villaver}}]{2009yCat..35120063M}
{Magrini}, L., {Stanghellini}, L., {Corbelli}, E., {Galli}, D., \& {Villaver},
  E. 2009, VizieR Online Data Catalog, J/A+A/512/A63

\bibitem[{{Moitinho}(2010)}]{Moitinho_2010}
{Moitinho}, A. 2010, in IAU Symposium, Vol. 266, Star Clusters: Basic Galactic
  Building Blocks Throughout Time and Space, ed. R.~{de Grijs} \& J.~R.~D.
  {L{\'e}pine}, 106--116

\bibitem[{{Moitinho} {et~al.}(2006){Moitinho}, {V{\'a}zquez}, {Carraro},
  {Baume}, {Giorgi}, \& {Lyra}}]{Moitinho_2006}
{Moitinho}, A., {V{\'a}zquez}, R.~A., {Carraro}, G., {et~al.} 2006, \mnras,
  368, L77

\bibitem[{{Monteiro} \& {Dias}(2019)}]{Monteiro_2019}
{Monteiro}, H. \& {Dias}, W.~S. 2019, \mnras, 487, 2385

\bibitem[{{Ochsenbein} {et~al.}(2000){Ochsenbein}, {Bauer}, \&
  {Marcout}}]{Ochsenbein_2000}
{Ochsenbein}, F., {Bauer}, P., \& {Marcout}, J. 2000, \aaps, 143, 23

\bibitem[{{O'Donnell}(1994)}]{Odonnell_1994}
{O'Donnell}, J.~E. 1994, \apj, 422, 158

\bibitem[{{Perren} {et~al.}(2017){Perren}, {Piatti}, \&
  {V{\'a}zquez}}]{Perren_2017}
{Perren}, G.~I., {Piatti}, A.~E., \& {V{\'a}zquez}, R.~A. 2017, A\&A, 602, A89

\bibitem[{Perren {et~al.}(2015)Perren, V{\'{a}}zquez, \& Piatti}]{Perren_2015}
Perren, G.~I., V{\'{a}}zquez, R.~A., \& Piatti, A.~E. 2015, A\&A, 576, A6

\bibitem[{{Ruprecht} {et~al.}(1996){Ruprecht}, {Balazs}, \&
  {White}}]{Ruprecht_1996}
{Ruprecht}, J., {Balazs}, B., \& {White}, R.~E. 1996, VizieR Online Data
  Catalog, VII/101A

\bibitem[{{Schlafly} \& {Finkbeiner}(2011)}]{Schlafly_2011}
{Schlafly}, E.~F. \& {Finkbeiner}, D.~P. 2011, ApJ, 737, 103

\bibitem[{{Sch{\"o}nrich} {et~al.}(2019){Sch{\"o}nrich}, {McMillan}, \&
  {Eyer}}]{Schonrich2019}
{Sch{\"o}nrich}, R., {McMillan}, P., \& {Eyer}, L. 2019, \mnras, 487, 3568

\bibitem[{{Sollima} {et~al.}(2010){Sollima}, {Carballo-Bello}, {Beccari},
  {Ferraro}, {Pecci}, \& {Lanzoni}}]{Sollima_2010}
{Sollima}, A., {Carballo-Bello}, J.~A., {Beccari}, G., {et~al.} 2010, MNRAS,
  401, 577

\bibitem[{{Soubiran} {et~al.}(2018){Soubiran}, {Cantat-Gaudin},
  {Romero-G{\'o}mez}, {Casamiquela}, {Jordi}, {Vallenari}, {Antoja},
  {Balaguer-N{\'u}{\~n}ez}, {Bossini}, {Bragaglia}, {Carrera}, {Castro-Ginard},
  {Figueras}, {Heiter}, {Katz}, {Krone-Martins}, {Le Campion}, {Moitinho}, \&
  {Sordo}}]{Soubiran_2018}
{Soubiran}, C., {Cantat-Gaudin}, T., {Romero-G{\'o}mez}, M., {et~al.} 2018,
  \aap, 619, A155

\bibitem[{{Sp\"ath}(2004)}]{Spath2004}
{Sp\"ath}, H. 2004, Mathematical Communications, 9, 27

\bibitem[{{Stetson}(1987)}]{Stetson_1987}
{Stetson}, P.~B. 1987, \pasp, 99, 191

\bibitem[{{Stetson} {et~al.}(1990){Stetson}, {Davis}, \&
  {Crabtree}}]{Stetson_1990}
{Stetson}, P.~B., {Davis}, L.~E., \& {Crabtree}, D.~R. 1990, in Astronomical
  Society of the Pacific Conference Series, Vol.~8, CCDs in astronomy, ed.
  G.~H. {Jacoby}, 289--304

\bibitem[{{Tadross}(2011)}]{Tadross_2011}
{Tadross}, A.~L. 2011, Journal of Korean Astronomical Society, 44, 1

\bibitem[{{Trumpler}(1930)}]{Trumpler_1930}
{Trumpler}, R.~J. 1930, Lick Observatory Bulletin, 420, 154

\bibitem[{{Vall{\'e}e}(2005)}]{valle_2005}
{Vall{\'e}e}, J.~P. 2005, \aj, 130, 569

\bibitem[{{van den Bergh} \& {Hagen}(1975)}]{vdBH1975}
{van den Bergh}, S. \& {Hagen}, G.~L. 1975, \aj, 80, 11

\bibitem[{Van Der~Walt {et~al.}(2011)Van Der~Walt, Colbert, \&
  Varoquaux}]{vanDerWalt_2011}
Van Der~Walt, S., Colbert, S.~C., \& Varoquaux, G. 2011, Computing in Science
  \& Engineering, 13, 22

\bibitem[{{van Rossum}(1995)}]{vanRossum_1995}
{van Rossum}, G. 1995, {Python} tutorial, Report CS-R9526, pub-CWI, pub-CWI:adr

\bibitem[{{V{\'a}zquez} {et~al.}(2008){V{\'a}zquez}, {May}, {Carraro},
  {Bronfman}, {Moitinho}, \& {Baume}}]{Vazquez2008}
{V{\'a}zquez}, R.~A., {May}, J., {Carraro}, G., {et~al.} 2008, ApJ, 672, 930

\bibitem[{{Xu} {et~al.}(2019){Xu}, {Zhang}, {Reid}, {Zheng}, \&
  {Wang}}]{Xu_2019}
{Xu}, S., {Zhang}, B., {Reid}, M.~J., {Zheng}, X., \& {Wang}, G. 2019, arXiv
  e-prints, arXiv:1903.04105

\bibitem[{{Yen} {et~al.}(2018){Yen}, {Reffert}, {Schilbach}, {R{\"o}ser},
  {Kharchenko}, \& {Piskunov}}]{Yen_2018}
{Yen}, S.~X., {Reffert}, S., {Schilbach}, E., {et~al.} 2018, ArXiv e-prints
  [\eprint[arXiv]{1802.04234}]

\end{thebibliography}

\appendix

The thirteen clusters in this Appendix are ordered according to their
longitude, as shown in Table \ref{tab:clust_list}. The remaining three
analyzed clusters were presented in Sect. \ref{sec:cluster_discuss}.

\section{van den Bergh-Hagen 73}

\begin{figure*}[ht]
    \centering
    \includegraphics[width=\hsize]{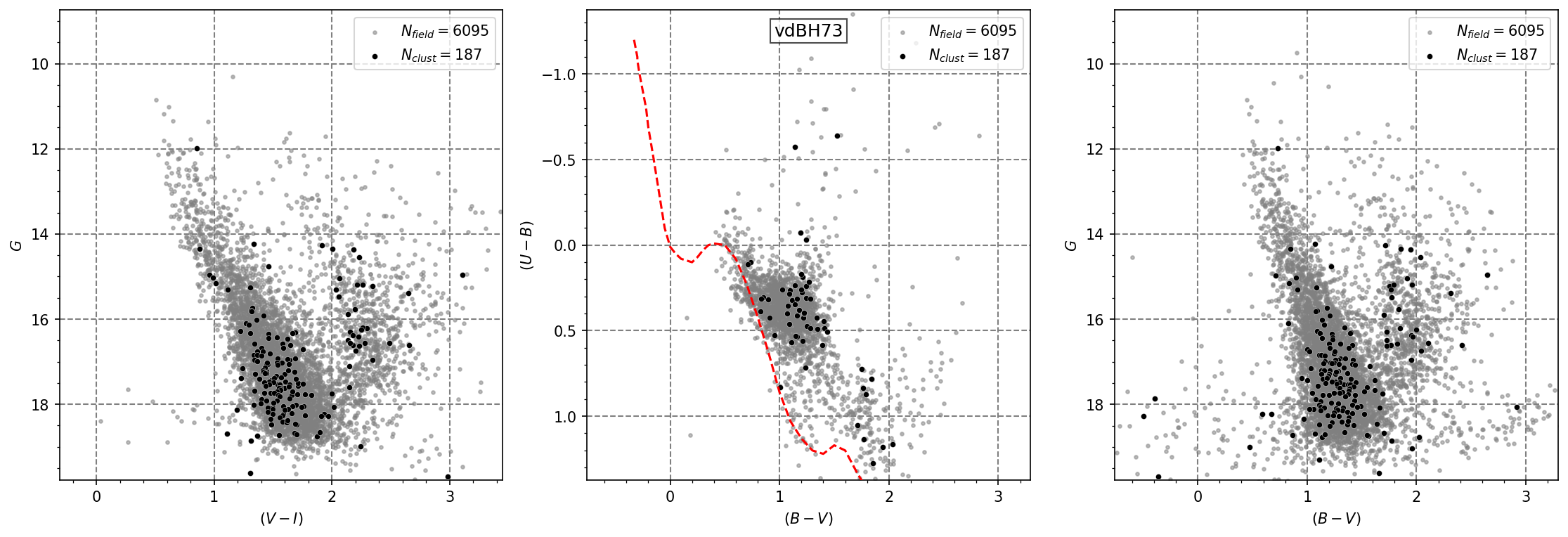}
\caption{Idem Fig. \ref{fig:photom_vdBH85} for vdBH 73.}
    \label{fig:photom_vdBH73}
\end{figure*}

The cluster vdBH 73 is placed in almost the center of the Vela
constellation well at the northeast border of the Carina Constellation. The
visual chart of the region in Fig. \ref{fig:Vim} shows a small and compact
grouping of stars at the very center of the frame surrounded by a dense stellar
field.
The inspection of the CCD and CMDs for all the stars observed
in the targeted region in Fig. \ref{fig:photom_vdBH73} gives no clear
hints about the presence of a cluster there, likely due to the
effect of field star contamination.
There are in the CMDs of Fig.~\ref{fig:photom_vdBH73} few stars above $G=15$
mag and at larger magnitudes the CMDs strongly widen.
The reddening in the CCD, right panel in Fig. \ref{fig:photom_vdBH73}, is quite
strong and displaces the bulk of stars entirely toward the red side. A
few blue stars with negative $(U-B)$ values appear strongly affected by
variable reddening.\\

\begin{figure*}[ht]
    \centering
    \includegraphics[width=\hsize]{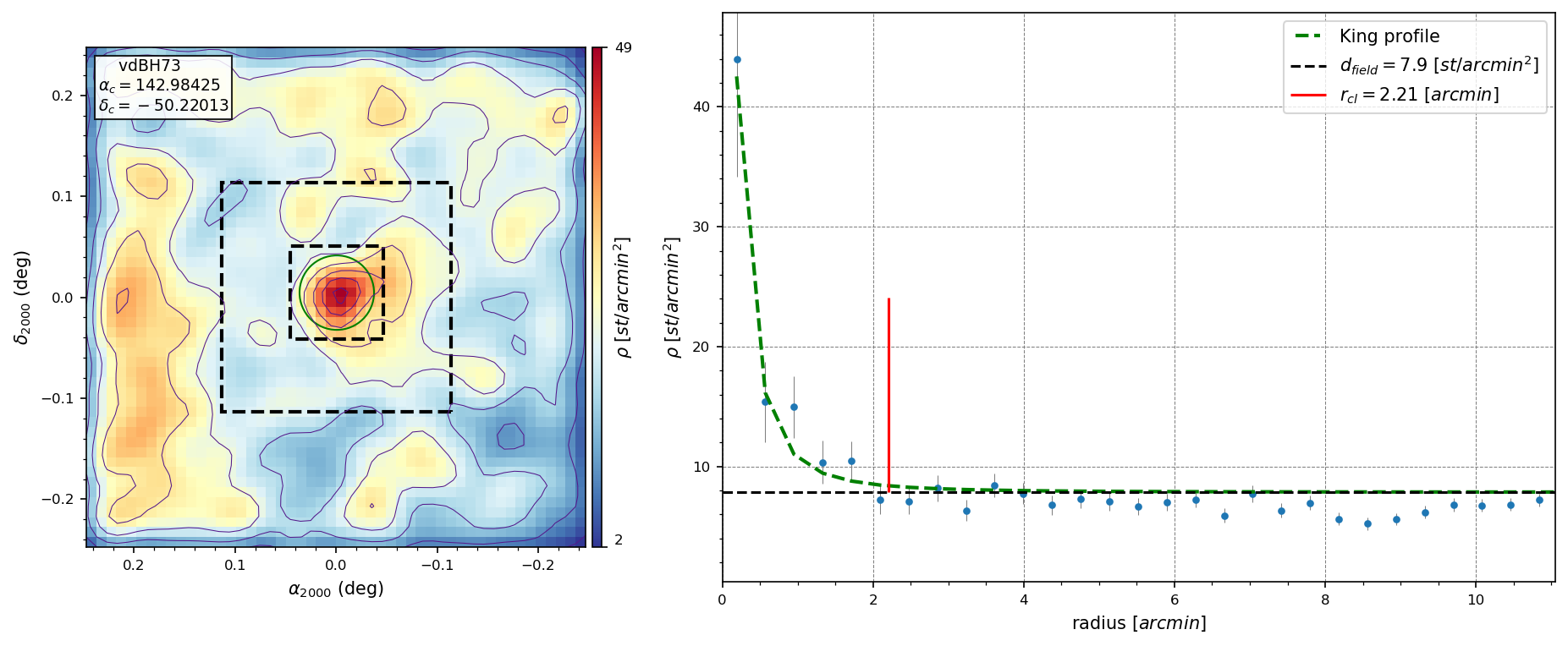}
\caption{Idem Fig. \ref{fig:struct_vdBH85} for vdBH 73.}
    \label{fig:struct_vdBH73}
\end{figure*}

The left panel in Fig.~\ref{fig:struct_vdBH73} shows a pronounced star
overdensity of 2.2 arcmin radius, coincident with the location expected for
vdBH 73. This overdensity appears immersed in a region of large field
star contamination. As seen in the RDP to the right, the density peak is about
four times above the mean for the field.\\

\begin{figure*}[ht]
    \centering
    \includegraphics[width=\hsize]{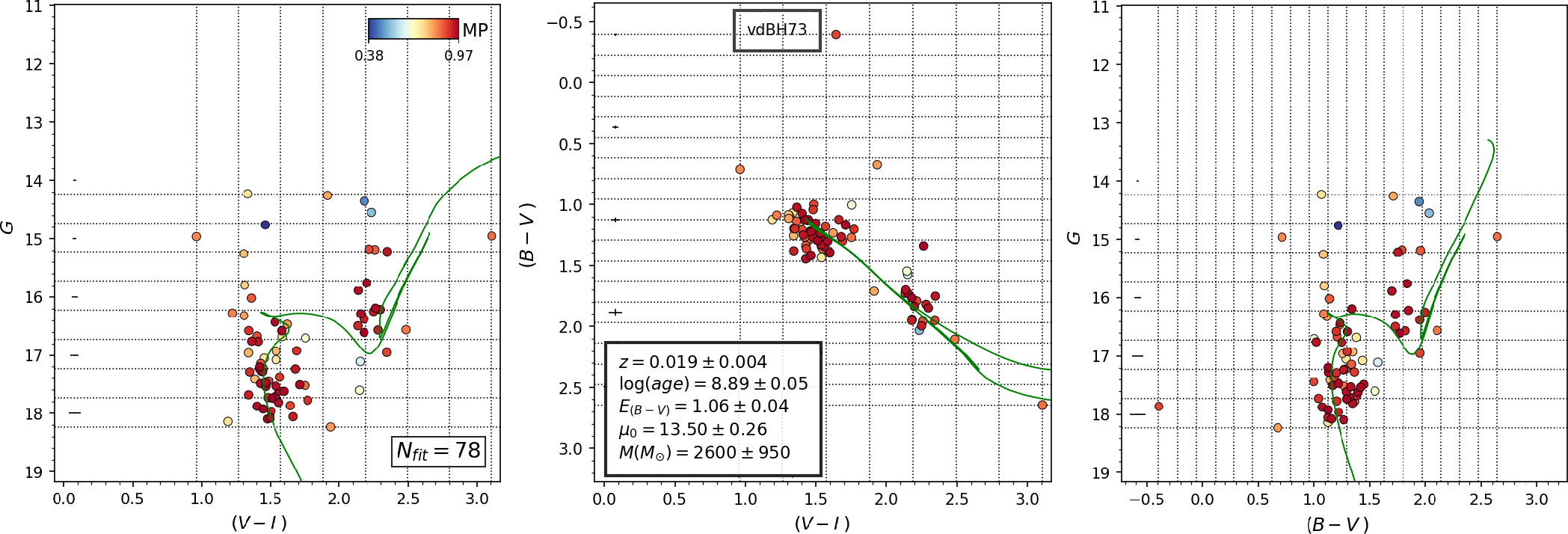}
\caption{Idem Fig. \ref{fig:fundpars_vdBH85} for vdBH 73 with the $(B-V)$ vs
$(V-I)$ diagram instead of the $(B-V)$ vs $(U-B)$ diagram.}
    \label{fig:fundpars_vdBH73}
\end{figure*}

The CMDs in Fig.~\ref{fig:fundpars_vdBH73}, left and right panels, put in
evidence a cluster main sequence subtending 1.5 magnitudes and a faint giant
branch with stars up to $G=15$ mag.
The $(B-V)$ vs $(V-I)$ CCD is shown in the middle panel instead of the
$(B-V)$ vs $(U-B)$ diagram because the latter did not contain enough stars to
be of use in the extinction estimation process.
Although the CMDs after the removal of interlopers look somewhat noisy,
those stars with membership probabilities above $\sim0.7$ clearly
trace the sequence of an evolved cluster.

The best fitting of a synthetic cluster yields the following results:

\begin{itemize}
\item [a)] The cluster is immersed in a region of moderate absorption
since the mean of reddening comes to be $E(B-V)=1.06$, a value compatible
with the ones provided by \cite{Schlafly_2011} (hereafter S\&F2011) who
found a maximum $E(B-V)$ of about 1.2 mag towards vdBH 73.
\item [b)] The absorption-free distance modulus turns out to be
$13.50\pm0.26$ mag placing this object at $5.01\pm0.61$ kpc from the Sun.
\end{itemize}

From the photometric point of view the existence of a well outlined cluster main
sequence and the high probability memberships of the stars seen in it confirm
the real entity of vdBH 73.

The usage of parallax data from Gaia shows a good agreement in distance
reaching up $5.48\pm0.44$ kpc in the sense that Gaia parallaxes place the
cluster farther than photometry does. This difference improves when offset is
applied to the parallax data, as shown in Sect. \ref{sec:results_concl}.
The Anderson-Darling test applied to parallax and proper motion data
demonstrates that the null hypothesis can indeed be rejected with a
combined $p$-value of 0.0, pointing to a real cluster present in this
region.\\

We conclude from our analysis that van den Bergh-Hagen 73 is an intermediate
aged cluster around $0.78\pm0.09\times10^9$ years old.

\begin{figure*}[ht]
    \centering
    \includegraphics[width=\hsize]{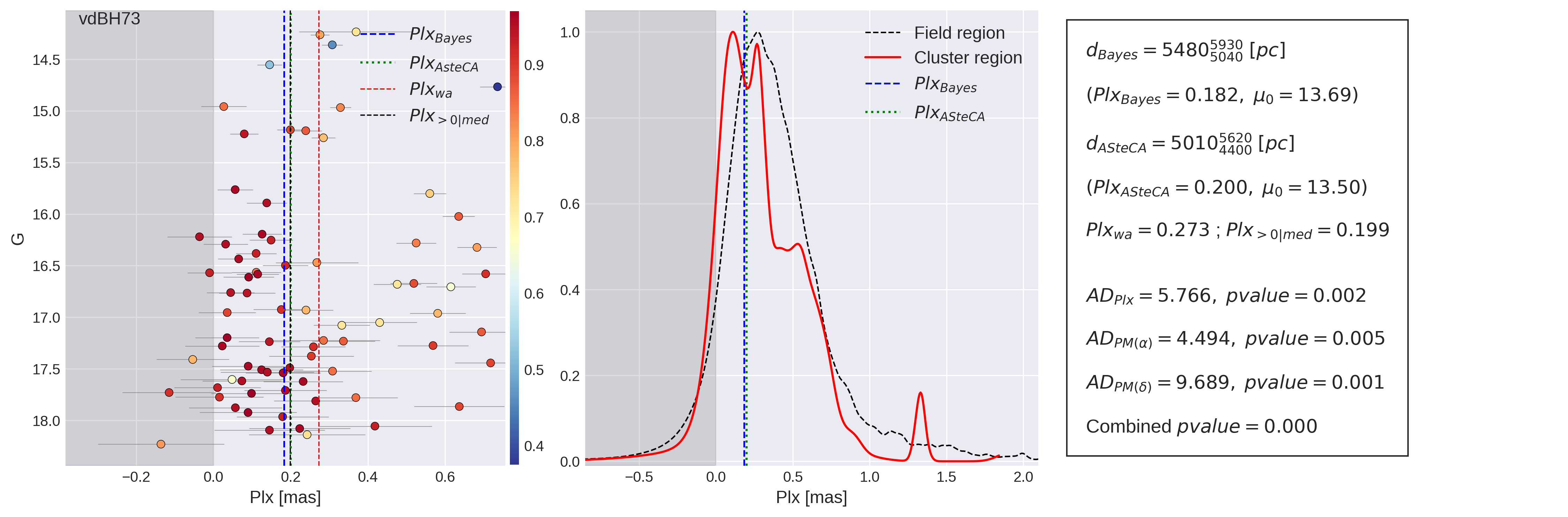}
\caption{Idem Fig. \ref{fig:plx_bys_vdBH85} for vdBH 73.}
\label{fig:plx_bys_vdBH73}
\end{figure*}

\section{Ruprecht 85}

\begin{figure*}[ht]
    \centering
    \includegraphics[width=\hsize]{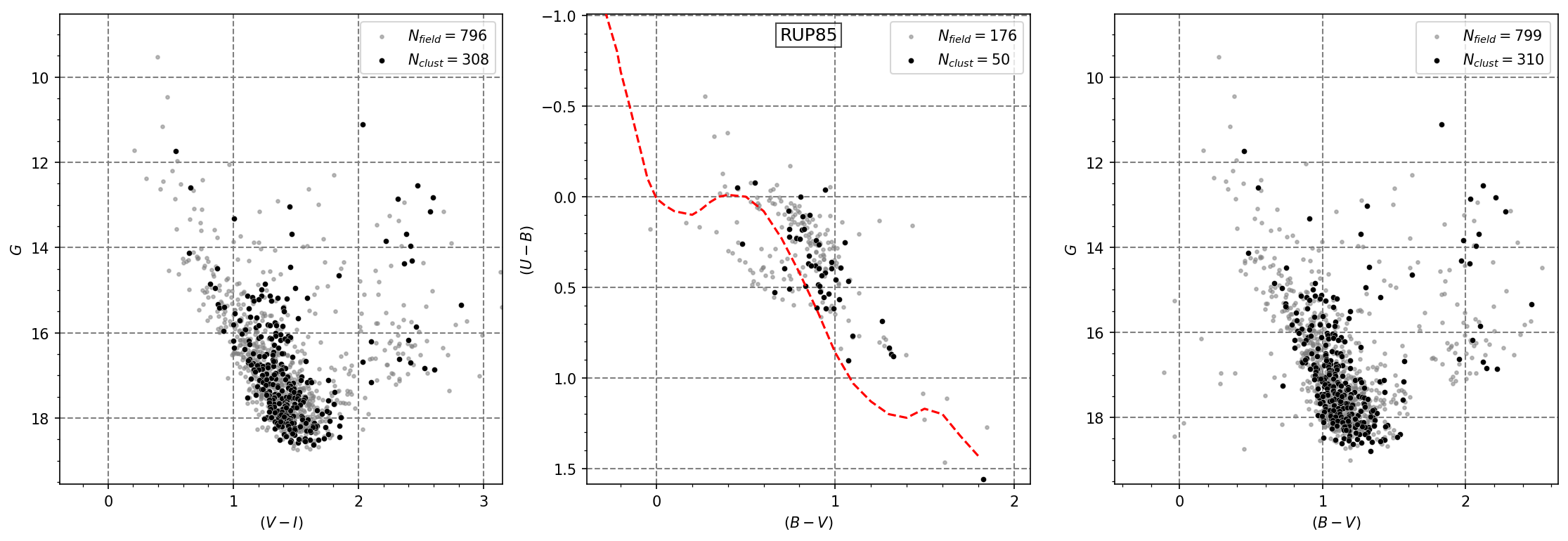}
    \caption{Idem Fig. \ref{fig:photom_vdBH85} for RUP 85.}
    \label{fig7}
\end{figure*}

Ruprecht 85 belongs to the south side of the Vela Constellation close to the
border of the Carina region. This cluster appears in Fig. \ref{fig:Vim} as a
slight increment in the stellar field towards the north part in the respective
frame.
The overall stars photometric diagrams as shown in Fig. \ref{fig7} do not show
the presence of any cluster sequence but a vertical strip of stars emerging from
a poorly populated stellar field above $G=14$ mag defined by disk stars.\\

\begin{figure*}[ht]
    \centering
    \includegraphics[width=\hsize]{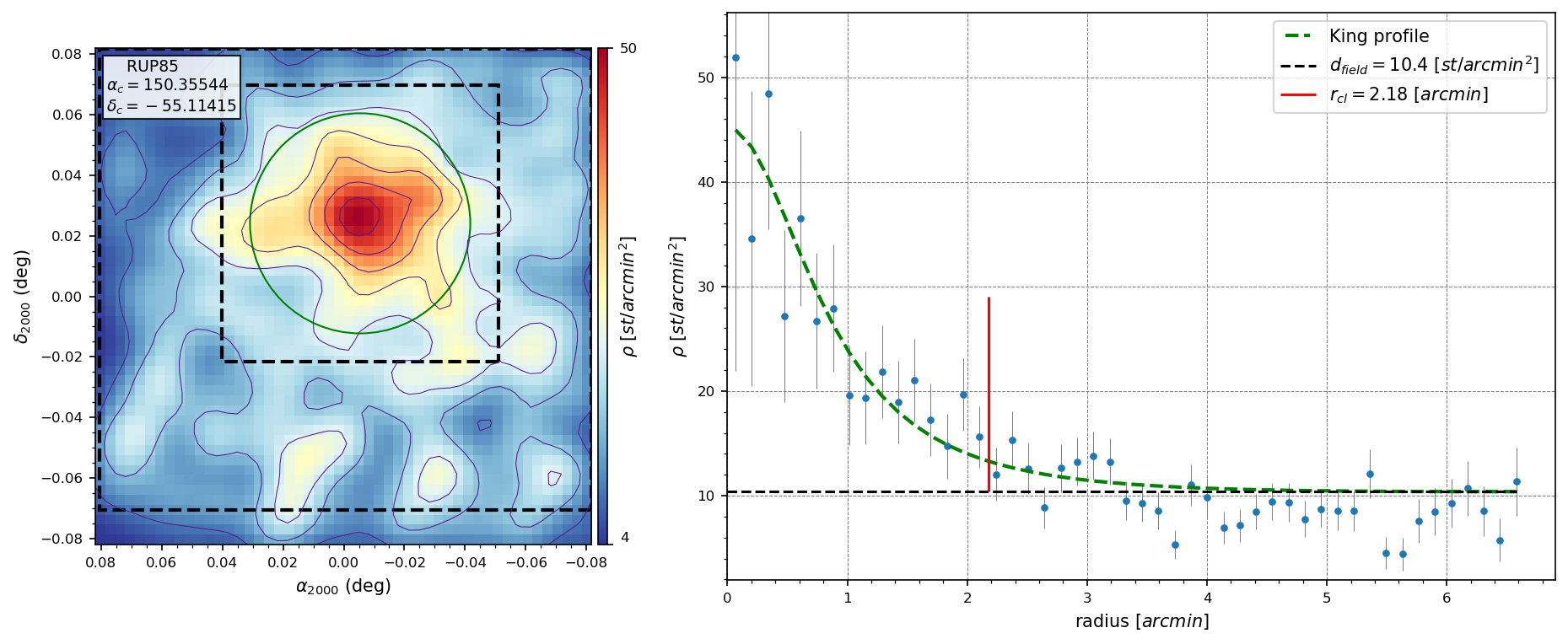}
    \caption{Idem Fig. \ref{fig:struct_vdBH85} for RUP 85.}
    \label{fig8}
\end{figure*}

\begin{figure*}[ht]
    \centering
    \includegraphics[width=\hsize]{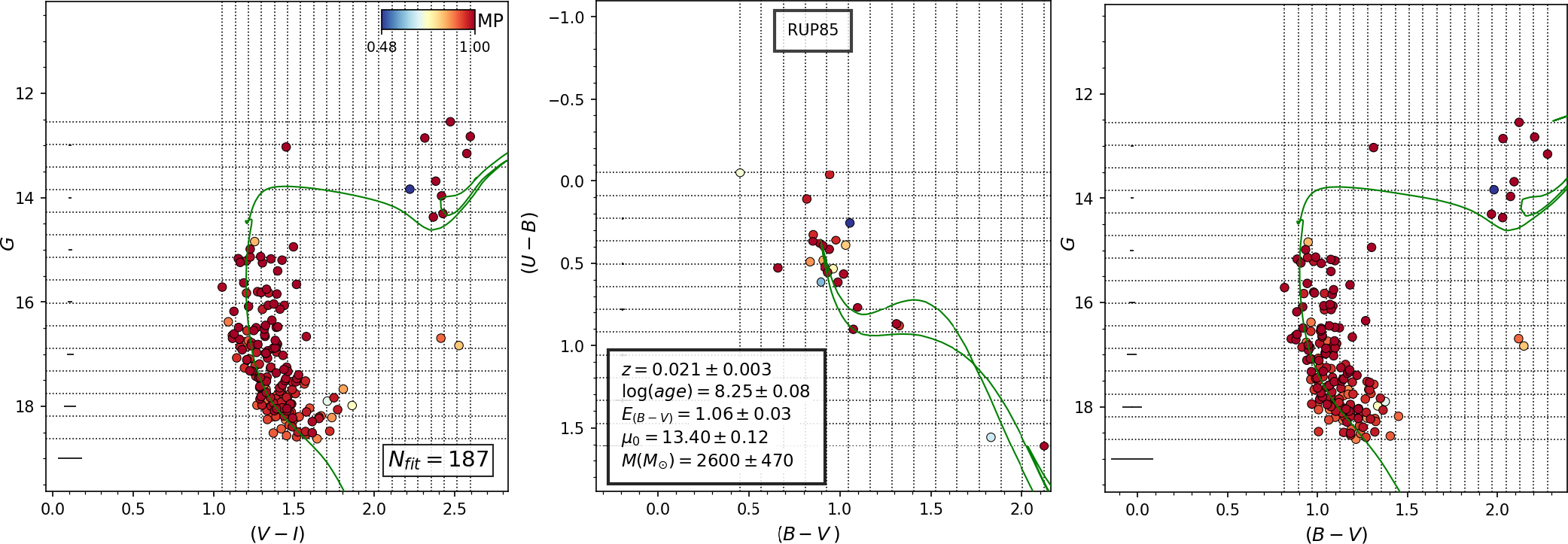}
    \caption{Idem Fig. \ref{fig:fundpars_vdBH85} for RUP 85.}
    \label{fig9}
\end{figure*}

The structural analysis performed by \texttt{ASteCA} yields a clean overdensity
at the location of this object that appears subtending an almost circular area
with a radius between 2-3 arcmins, Fig.~\ref{fig8} left panel.
As shown in Fig.~\ref{fig8} right panel, the RDP is well developed
and with a star density five times above the background level. The
photometric diagrams, CCD and CMDs of stars with membership probabilities above
0.48 and up to 1.0 shown in Fig. \ref{fig9} depict a rather noisy main
sequence sweeping 3.5 magnitudes.
Combining structural evidences with evidences coming from the photometric
diagrams we conclude that RUP 85 is a real entity. As for the cluster
parameters of the best synthetic cluster fitting the observations it is found
that:

\begin{itemize}
\item [a)] As was the case with vdBH 73, RUP 85 is also placed
    in a region of moderate color excess. The cluster has $E(B-V)=1.06$
    also entirely in line with a maximum $E(B-V)$ of 2 mag according to
    S\&F2011.
\item [b)] The free absorption distance modulus is $13.40\pm0.12$ mag
corresponding to a distance $d=4.80\pm0.26$ kpc.
\end{itemize}

\begin{figure*}[ht]
    \centering
    \includegraphics[width=\hsize]{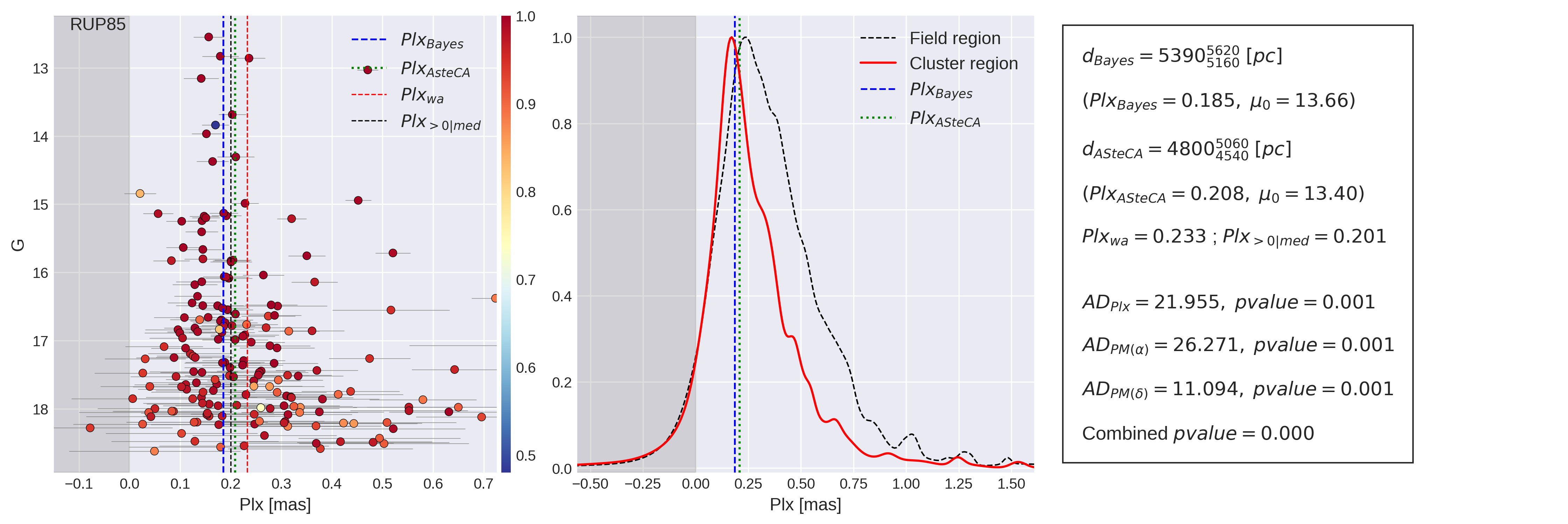}
    \caption{Idem Fig. \ref{fig:plx_bys_vdBH85} for RUP 85.}
    \label{fig10}
\end{figure*}

The results from the Anderson-Darling test in Fig. \ref{fig10} applied to $Plx$,
$PM(\alpha)$ and $PM(\delta)$ indicate clearly that the cluster region and the
surrounding background population come from quite different star populations.
Therefore, the null hypothesis can be rejected.\\

We conclude that RUP 85 is a real open cluster around $0.18\pm0.03\times10^9$
years old.

\section{van den Bergh-Hagen 87}

\begin{figure*}[ht]
    \centering
    \includegraphics[width=\hsize]{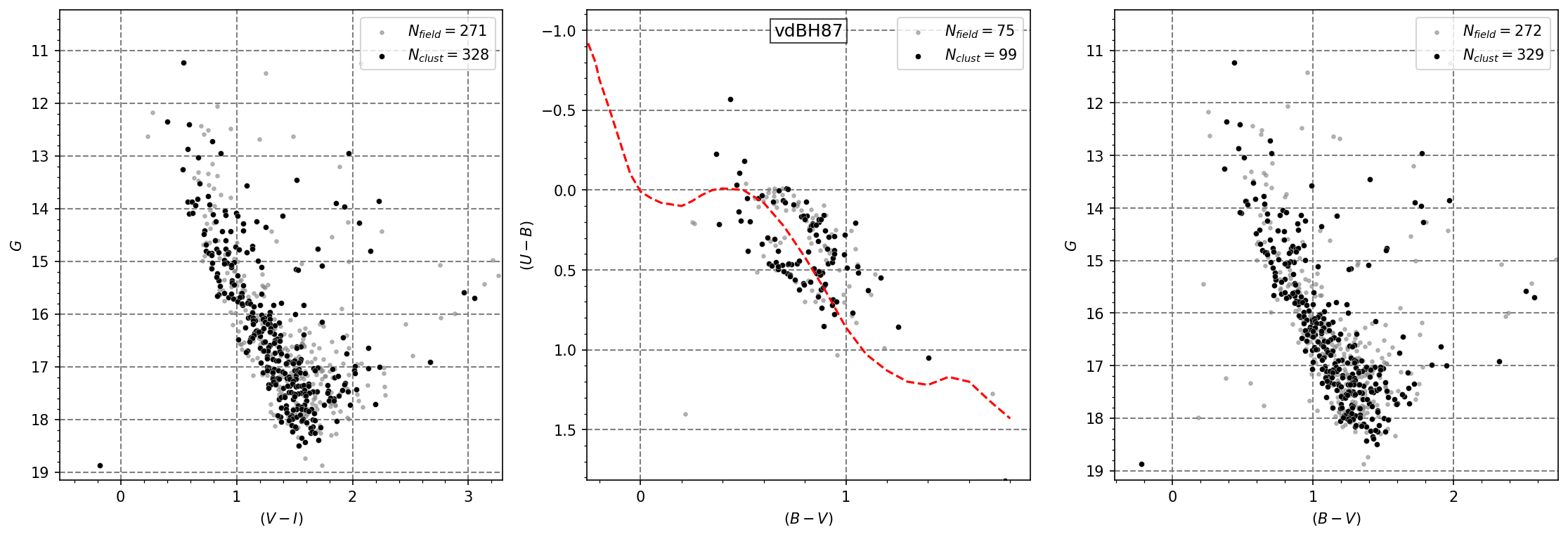}
    \caption{Idem Fig. \ref{fig:photom_vdBH85} for vdBH 87.}
    \label{fig15}
\end{figure*}

Like RUP 85, vdBH 87 is seen toward the south of Vela Constellation close to
the border with Carina. A weak grouping of stars placed towards the north of
the frame is seen in Fig. \ref{fig:Vim}. In turn, the CMDs in Fig. \ref{fig15}
seem to reflect a typical star disk sequence up to $G=15$ mag approximately
with an amorphous distribution at the bright end. The CCD is, on the other
side, rather poor.\\

\begin{figure*}[ht]
    \centering
    \includegraphics[width=\hsize]{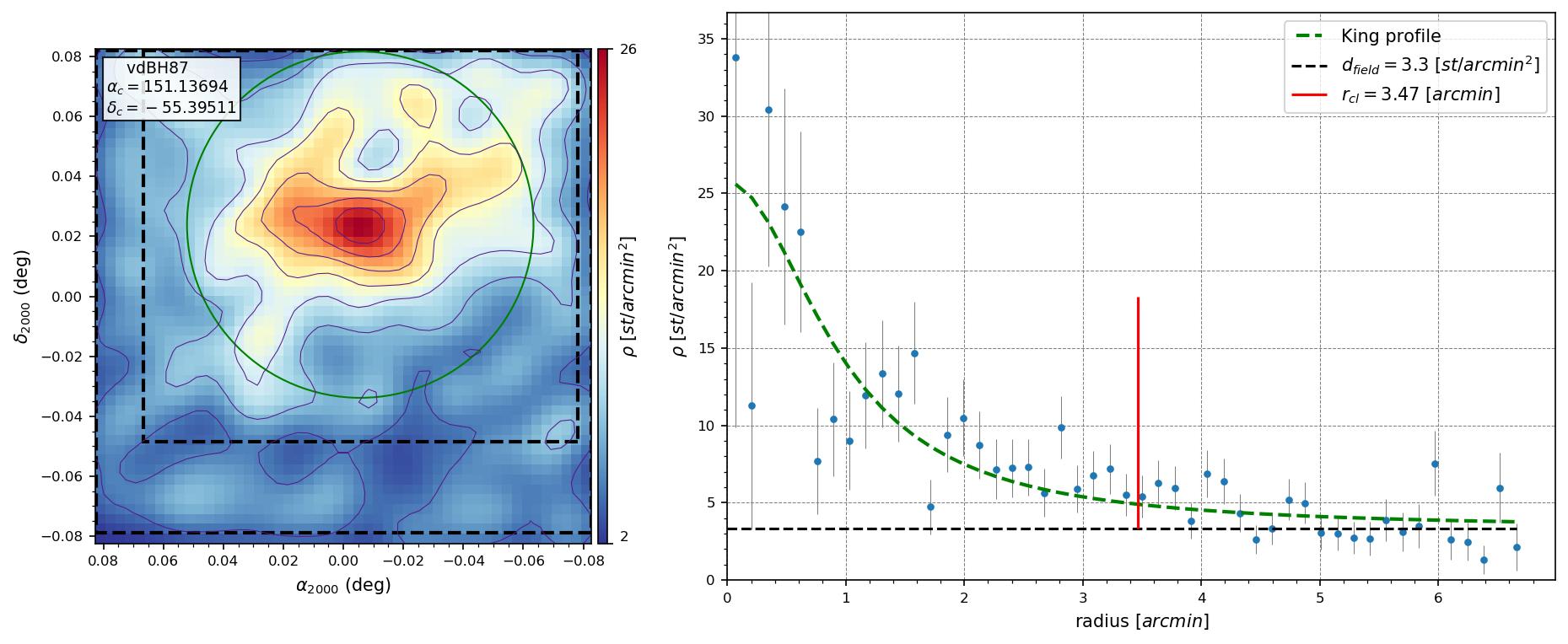}
    \caption{Idem Fig. \ref{fig:struct_vdBH85} for vdBH 87.}
    \label{fig16}
\end{figure*}

\begin{figure*}[ht]
    \centering
    \includegraphics[width=\hsize]{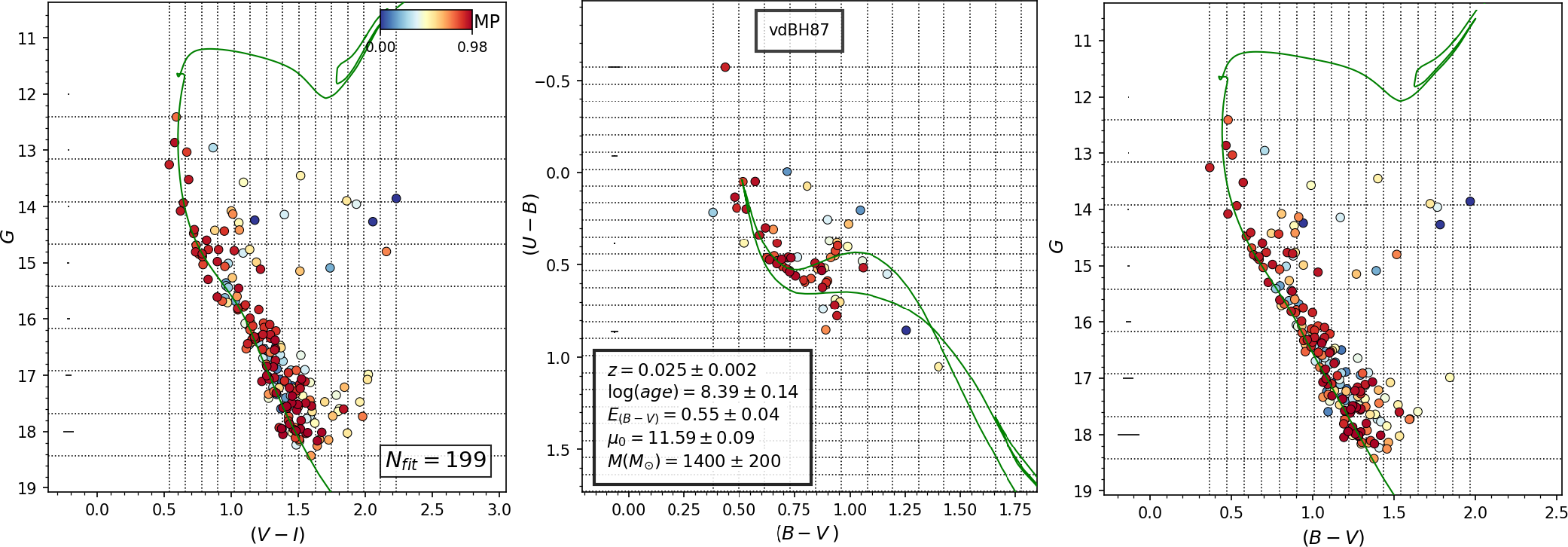}
    \caption{Idem Fig. \ref{fig:fundpars_vdBH85} for vdBH 87.}
    \label{fig17}
\end{figure*}

A stellar overdensity reaching $\sim7$ times the field star density is
seen in Fig. \ref{fig16}. The spatial structure of this overdensity suggests an
elongation in right ascension and a RDP characterized by a very narrow density
peak followed by a star coronal distribution at about 1.5 arcmins from the
center.
The clean CMDs in Fig. \ref{fig17} leave no doubt as for the nature of vdBH 87
since inside this overdensity a robust and narrow cluster main
sequence is evident. Its sequence extends for more than 5 mag in the
CMDs, including stars with very low membership probabilities well detached
from the sequence, in the range from 0.0 to 0.98. The parameters of the
synthetic cluster that best fits the real star distributions are:

\begin{itemize}
\item [a)] The color excess is $E(B-V)=0.56$ indicating thus a moderate
    absorption in the cluster direction. In turn, this color excess value is
    below the maximum reddening $E(B-V)=2.9$ computed in the region for 
    S\&F2011.
\item [b)] The corrected distance modulus is $11.59\pm0.09$ mag implying a
distance of $d=2.08\pm0.09$ kpc. The cluster is not far from the Sun and
this closeness explains the moderate color excess found.
\end{itemize}

\begin{figure*}[ht]
    \centering
    \includegraphics[width=\hsize]{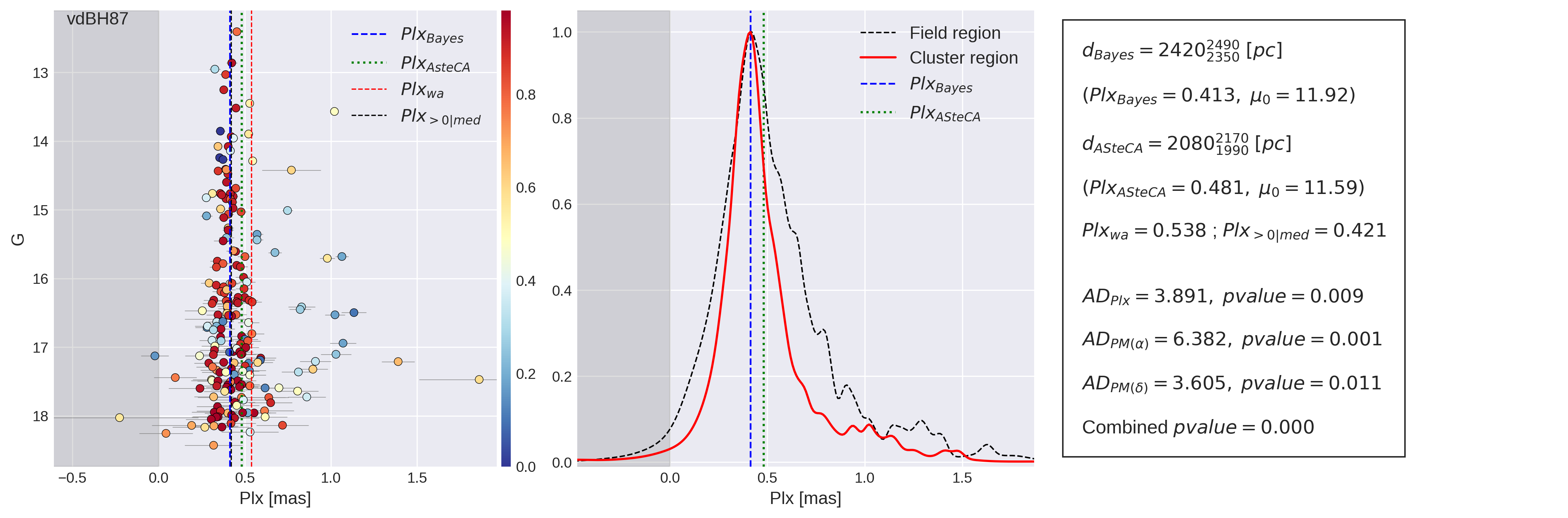}
    \caption{Idem Fig. \ref{fig:plx_bys_vdBH85} for vdBH 87.}
    \label{fig18}
\end{figure*}

The results of the application of the Anderson-Darling test in Fig. \ref{fig18}
are coincident with what \texttt{ASteCA} have found. This is that cluster and
field regions are quite different not only from the photometric perspective but
also from a kinematic view.

In conclusion vdBH 87 is a real open cluster $0.25\pm0.08\times10^9$
years old.


\section{van den Bergh-Hagen 92}

\begin{figure*}[ht]
    \centering
    \includegraphics[width=\hsize]{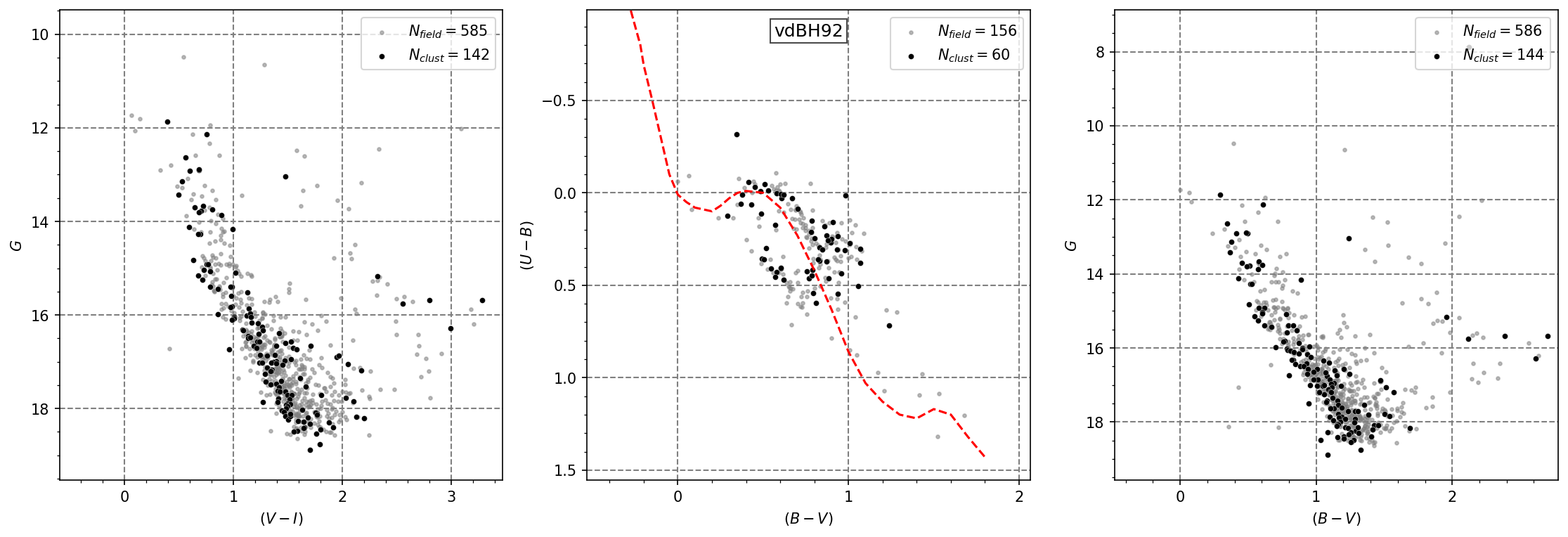}
    \caption{Idem Fig. \ref{fig:photom_vdBH85} for vdBH 92.}
    \label{fig35}
\end{figure*}

Placed south of Vela, near the eastern border with Carina, vdBH
92 is a relevant handful of bright stars as seen in the $V$ image of Fig.
\ref{fig:Vim}. The CMDs and CCD for all stars in the region, as shown in Fig.
\ref{fig35}, depict a narrow star sequence with some scatter at their respective
bright ends. Particularly the CCD shows, not far from the intrinsic line, a
group of $F$- and $G$-type stars and another group of stars below the intrinsic
line that could be $B$- and $A$- type stars displaced by the reddening effect.\\

\begin{figure*}[ht]
    \centering
    \includegraphics[width=\hsize]{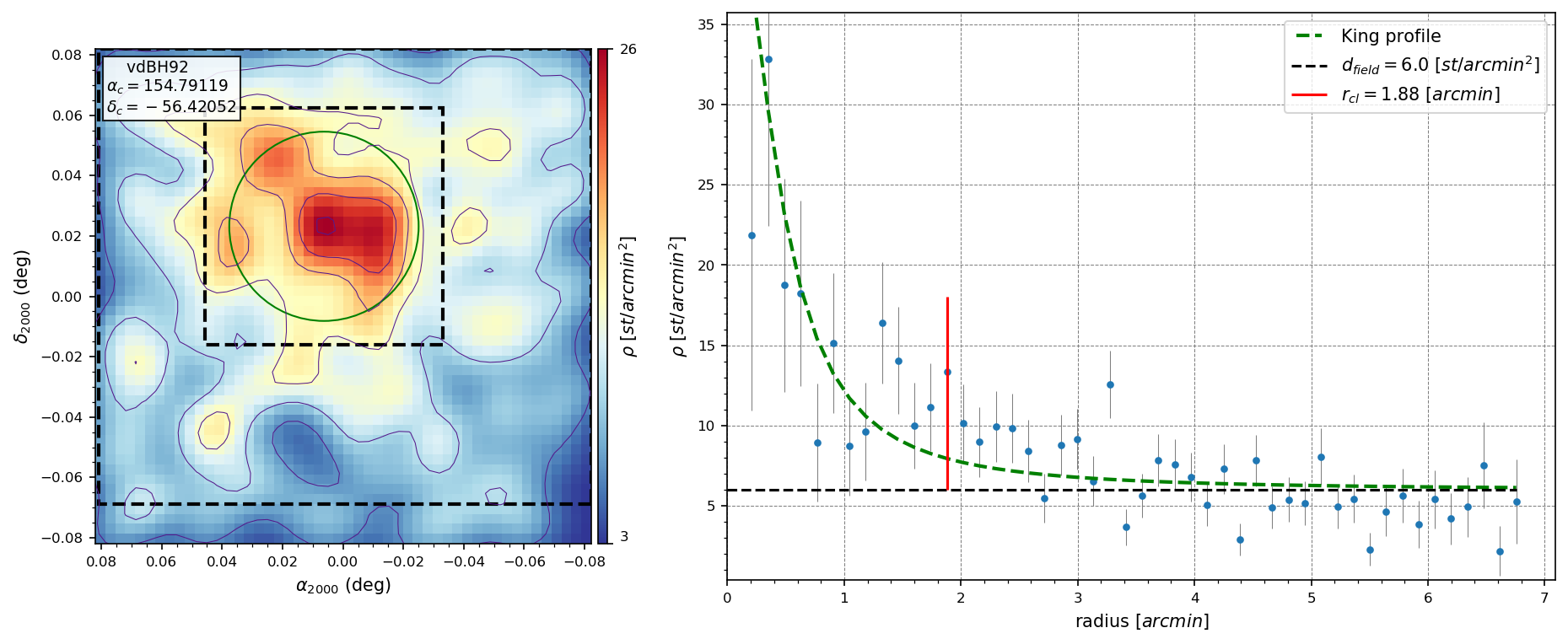}
    \caption{Idem Fig. \ref{fig:struct_vdBH85} for vdBH 92.}
    \label{fig36}
\end{figure*}

\begin{figure*}[ht]
    \centering
    \includegraphics[width=\hsize]{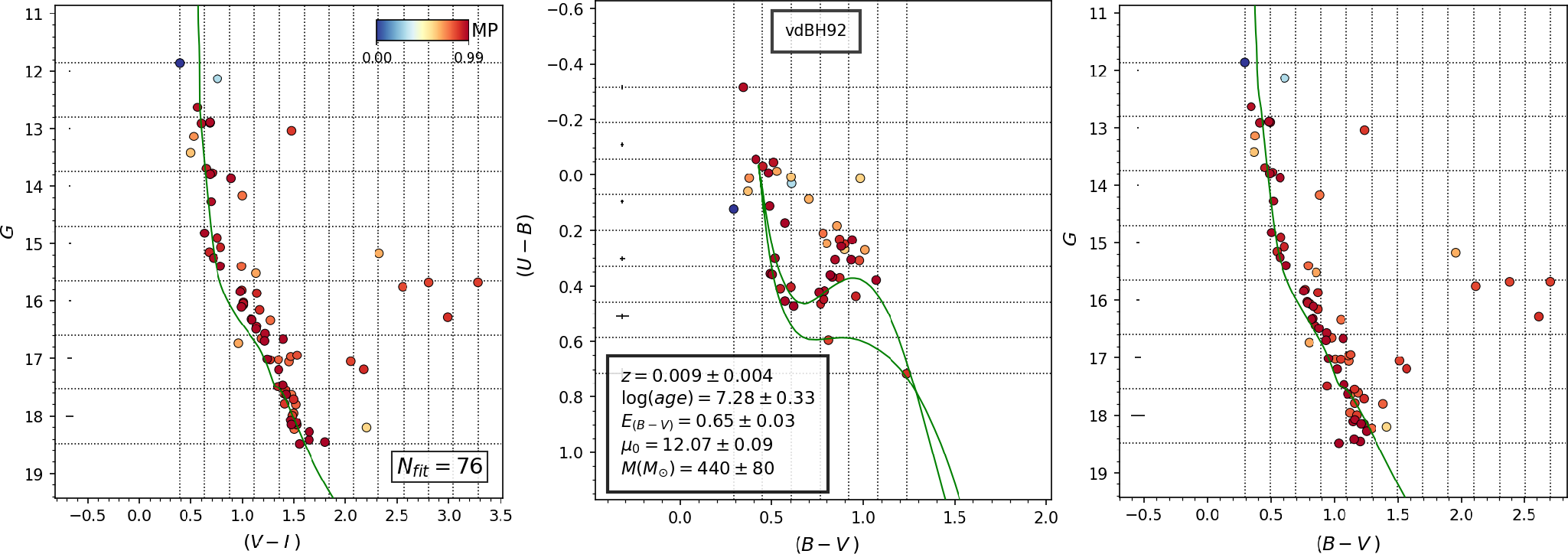}
    \caption{Idem Fig. \ref{fig:fundpars_vdBH85} for vdBH 92.}
    \label{fig37}
\end{figure*}

The \texttt{ASteCA} analysis in Fig. \ref{fig36} revealed the presence of a
well isolated star overdensity rising above the field stars density of about 6
stars per square arcmin. We identify this overdensity with vdBH
92. Notwithstanding the noisy RDP the limits of the overdensity can be well
established. As indicated in Fig. \ref{fig37} few stars have been found inside
the cluster limits with mostly large membership values. Despite the
low number of members, a 7 magnitude extended cluster main sequence can be
seen. The comparison with synthetic clusters made by \texttt{ASteCA} yields:

\begin{itemize}
\item [a)] the best fitting of a synthetic cluster to the clean data in Fig. 
    \ref{fig37} indicates a color excess of $E(B-V)=0.65$. Since the
    maximum color excess provided by S\&F2011 is 2.34 for this zone we
    conclude that most of the absorption is produced behind the position of
    vdBH 92. This object is therefore placed in front of a strong
    absorption region.
\item [b)] The absorption free distance modulus becomes
$12.07\pm0.09$ mag, which places vdBH 92 at a distance of $d=2.59\pm0.11$ kpc.
\end{itemize}

\begin{figure*}[ht]
    \centering
    \includegraphics[width=\hsize]{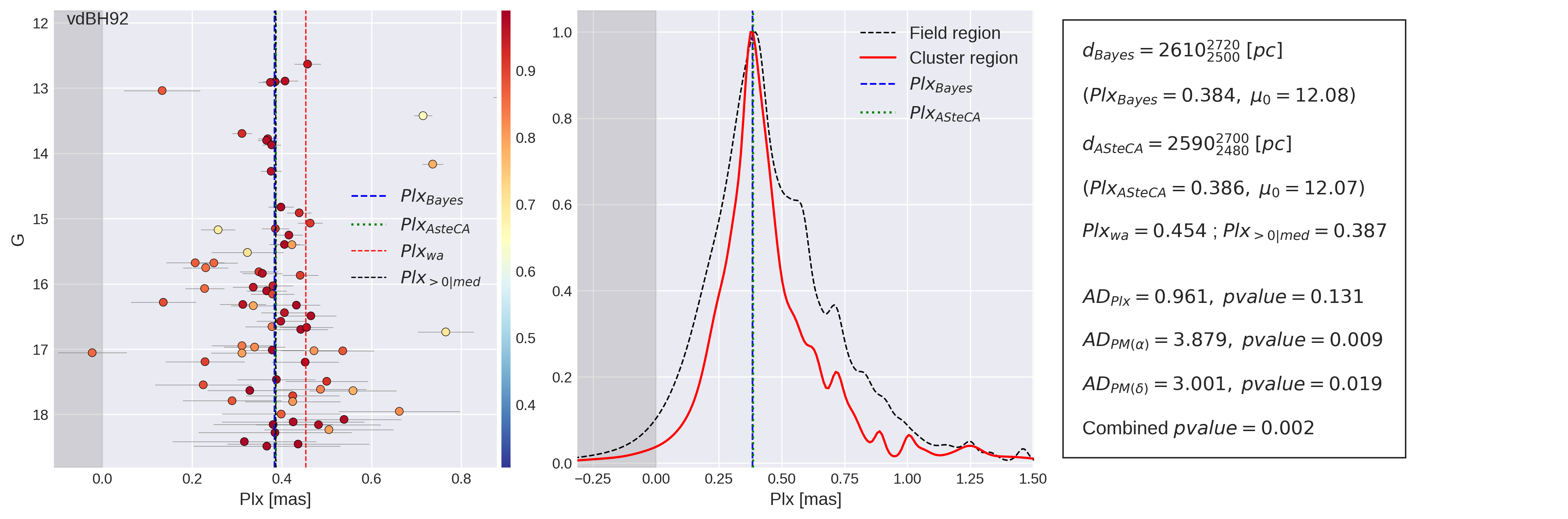}
    \caption{Idem Fig. \ref{fig:plx_bys_vdBH85} for vdBH 92.}
    \label{fig38}
\end{figure*}

By applying the Anderson-Darling test it is noticed that the parallax
distributions for stars inside and outside the cluster boundaries are not
sufficiently different from each other to reach the 5\% critical value, as
indicated in the right panel of Fig. \ref{fig38}. However, proper motions are
quite different in both regions. We combine this last finding with the
presence of a well defined overdensity that, in turn, shows a reasonable and
extended cluster main sequence to conclude that both samples come from
different populations.\\

These results together confirm the true nature of vdBH 92. This is a
young cluster $0.02\pm0.01\times10^9$ years old, the youngest true cluster
in our sample.


\section{Trumpler 12}

\begin{figure*}[ht]
    \centering
    \includegraphics[width=\hsize]{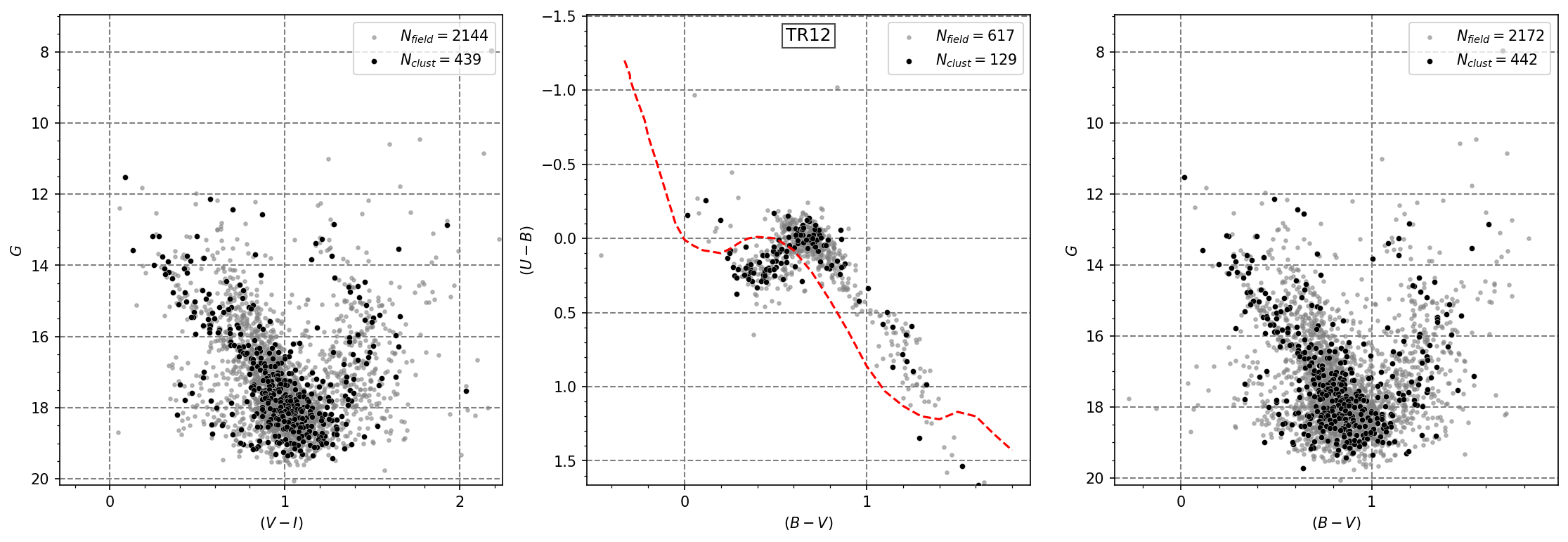}
    \caption{Idem Fig. \ref{fig:photom_vdBH85} for TR 12.}
    \label{fig19}
\end{figure*}

This object is placed in the west side of the Carina HII region where it appears
as a sparse handful of bright stars in Fig. \ref{fig:Vim}. The CMDs in Fig.
\ref{fig19}, including all stars in the region, show the following patterns:
there is a wide grouping of stars below $G=18$ mag but to the right side of
it, and also at this magnitude value, a narrow structure of stars up to $G=14$
mag slightly displaced to the blue side emerges. From $G=18$ mag a typical
vertical galactic disk population rises too.\\

\begin{figure*}[ht]
    \centering
    \includegraphics[width=\hsize]{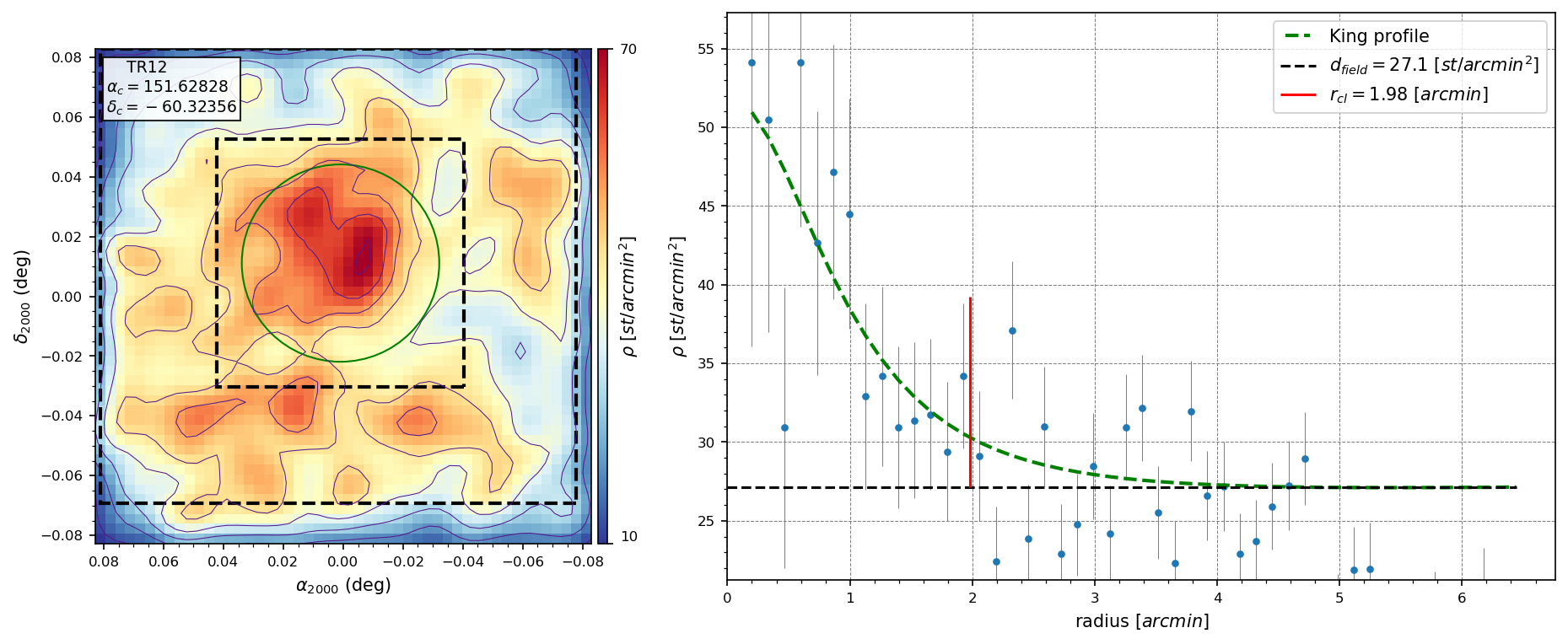}
    \caption{Idem Fig. \ref{fig:struct_vdBH85} for TR 12.}
    \label{fig20}
\end{figure*}

\begin{figure*}[ht]
    \centering
    \includegraphics[width=\hsize]{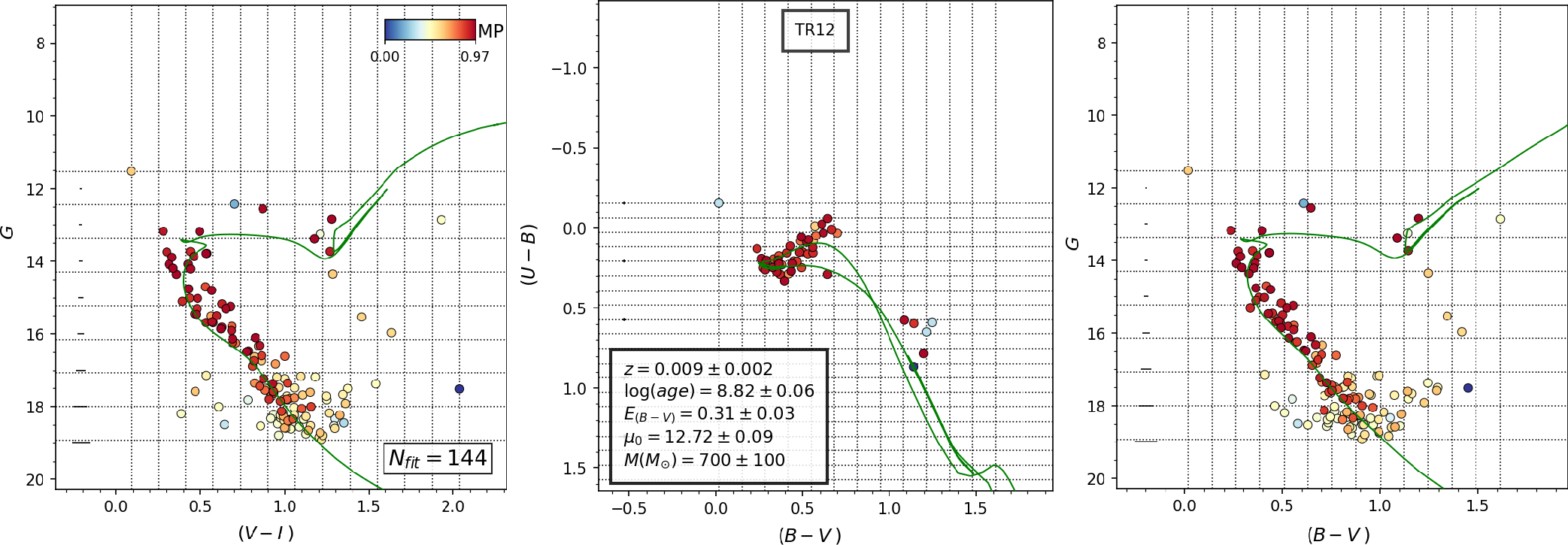}
    \caption{Idem Fig. \ref{fig:fundpars_vdBH85} for TR 12.}
    \label{fig21}
\end{figure*}

\texttt{ASteCA} detected a main overdensity in a region of high
stellar contamination, as shown in Fig.~\ref{fig20}.
This overdensity is characterized by a quite noisy RDP, a fact
explained in part because at the peak of the RDP there is less than twice the
density of the background.
Under this condition it is not an easy task to fix an appropriate
radius for the overdensity. We tentatively adopt $\sim2$ arcmin radius as a
reasonable compromise.
The membership probabilities in the zone of the overdensity are mostly
above 0.5 as indicated in Fig. \ref{fig21}. Again, as in vdBH 87, the
handful of low membership stars are very well detached from the main
cluster sequence.
A clear cluster main sequence can be seen in Fig. \ref{fig21} spanning roughly
4-5 mag. These stars belong to the tiny blue and narrow sequence detected
easily in the diagrams of Fig.~\ref{fig19} between $G=12$ and $G=16$
mag. Comparison with synthetic clusters yields the following values:

\begin{itemize}
\item [a)] A color excess of $E(B-V)=0.31$ is found for the best fitting.
    Since the maximum color excess provided by S\&F2011 is 0.50 we find that
    TR12 is placed in a zone of low absorption.
\item [b)] The absorption free distance modulus is $12.7\pm0.09$ mag,
representing a distance of $3.50\pm0.15$ kpc. At such a distance and with
low absorption it is reasonable to find a high background stellar density
as seen in Fig.~\ref{fig20}.
\end{itemize}

From the Anderson-Darling statistics shown in Fig.~\ref{fig21} we see that
proper motions for the cluster and for the field population belong to different
samples. On the other hand, the parallaxes can not be safely separated into
distinct stellar regions.\\

The clear cluster sequence and the low $p$-value (0.003) obtained with the AD
test, leads us to conclude that TR 12 is a real cluster about
$0.70\pm0.10\times10^9$ years old.

\begin{figure*}[ht]
    \centering
    \includegraphics[width=\hsize]{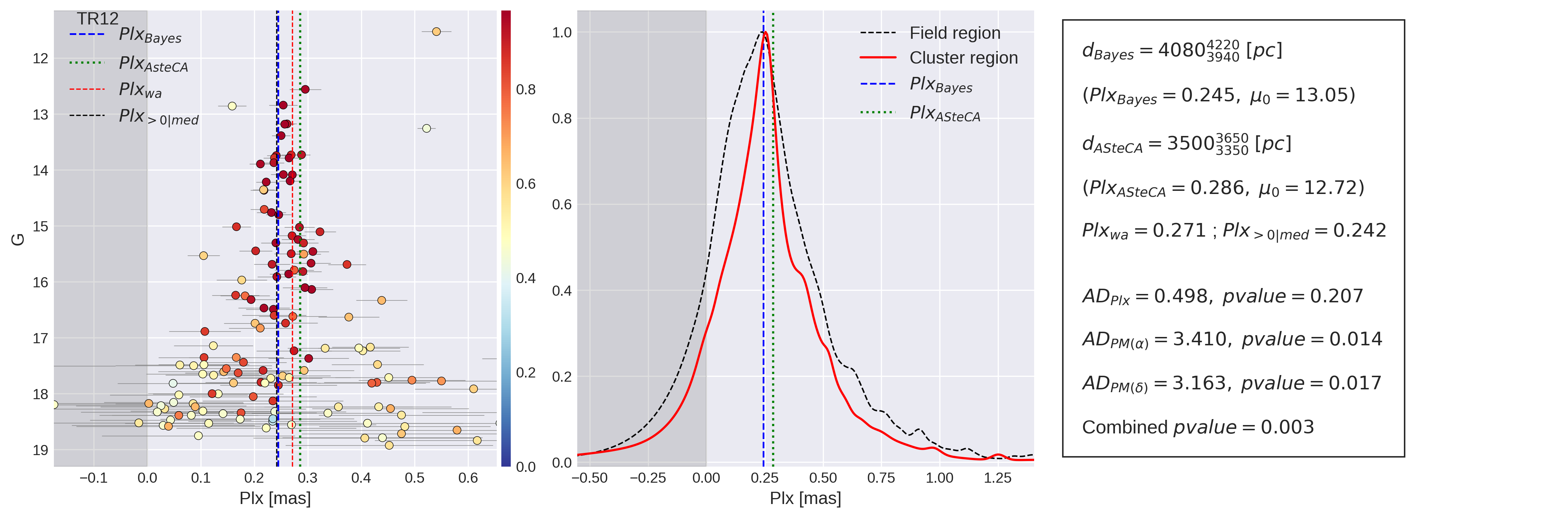}
    \caption{Idem Fig. \ref{fig:plx_bys_vdBH85} for TR 12.}
    \label{fig22}
\end{figure*}

\begin{figure*}[ht]
    \centering
    \includegraphics[width=\hsize]{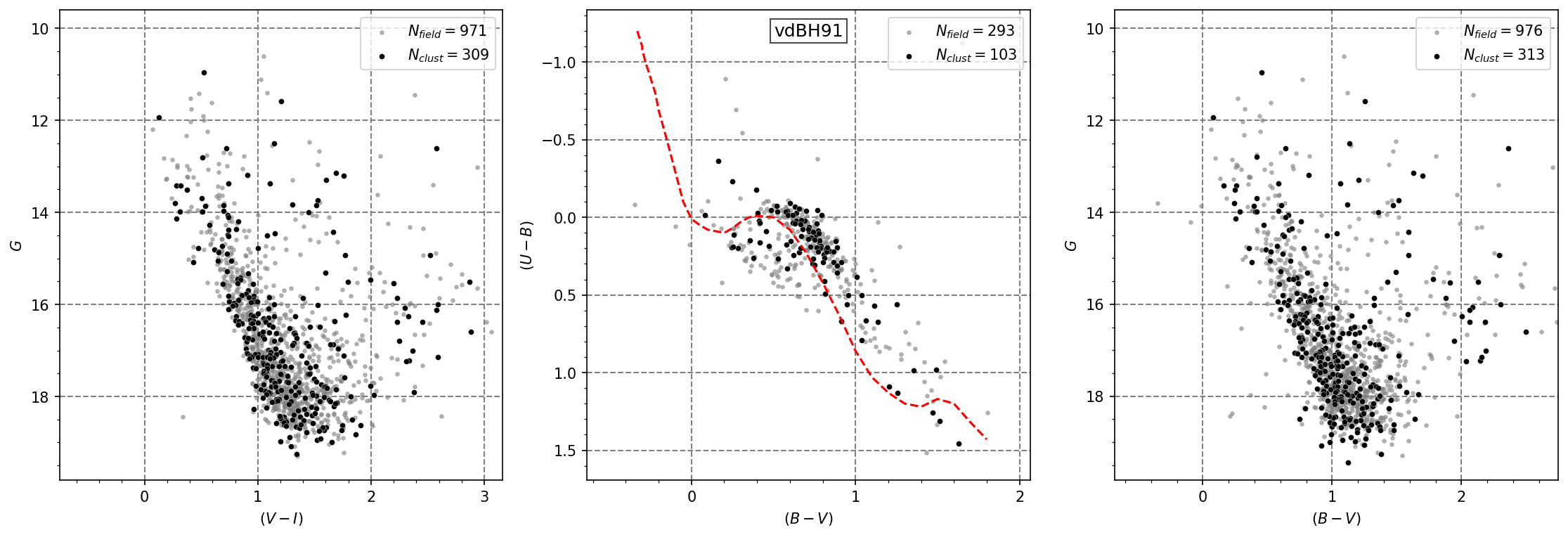}
    \caption{Idem Fig. \ref{fig:photom_vdBH85} for vdBH 91.}
    \label{fig27}
\end{figure*}

\section{van den Bergh-Hagen 91}

vdBH 91 is a potential cluster at the west side of Carina HII
region, specifically near the northern border of this constellation with Vela.
No relevant stellar structure appears in the $V$ image of Fig. \ref{fig:Vim}
but a common pattern of a galactic field star near the galactic plane.
The overall CMDs in Fig.~\ref{fig27} show a stellar sequence that, at first
sight, resemble the usual diagrams for open clusters. In turn, the CCD
is dominated by a tail of $F$- and $G$-type stars prolonged by red stars. It is
noticed as well the presence of some reddened early type stars for negative
$(U-B)$ indices.\\

\begin{figure*}[ht]
    \centering
    \includegraphics[width=\hsize]{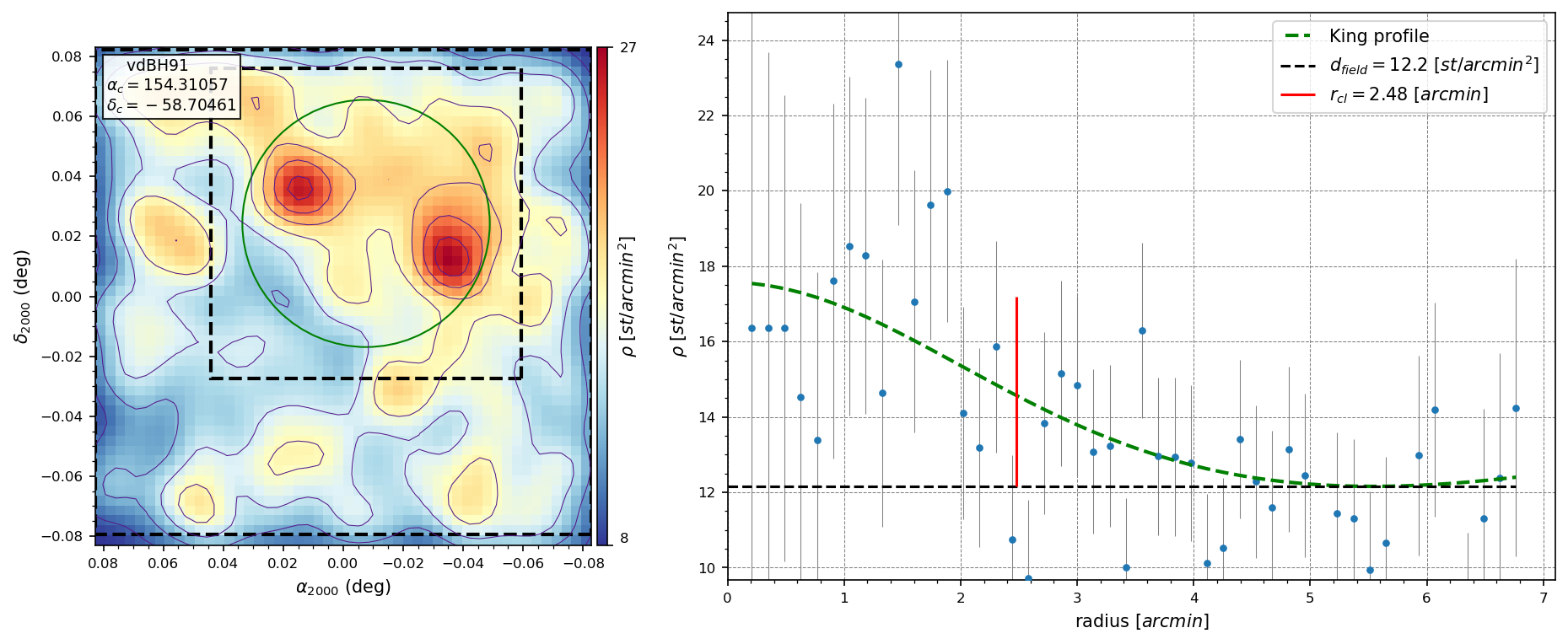}
    \caption{Idem Fig. \ref{fig:struct_vdBH85} for vdBH 91.}
    \label{fig28}
\end{figure*}

\begin{figure*}[ht]
    \centering
    \includegraphics[width=\hsize]{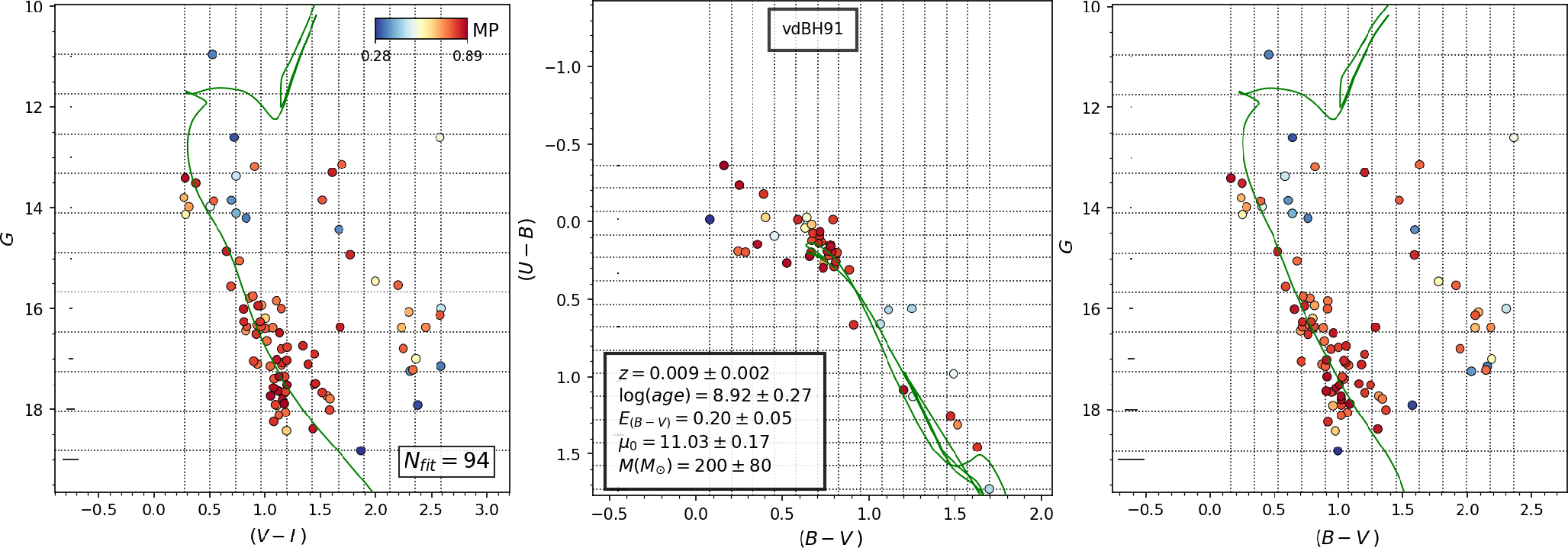}
    \caption{Idem Fig. \ref{fig:fundpars_vdBH85} for vdBH 91.}
    \label{fig29}
\end{figure*}

\texttt{ASteCA} found two well separated stellar overdensity peaks in
Fig.~\ref{fig28} whose relevance in terms of structure is not
important given the overall low stellar density of the field.
The noisy RDP proves by itself the poverty of the entire field surveyed in term
of star number.
After some attempts looking for a cluster sequence we ask \texttt{ASteCA} to
estimate the probabilities for stars inside an adopted radius
of $\sim2.5$ arcmin, shown in Fig.~\ref{fig28} (right).
As seen in Fig.~\ref{fig29} almost one hundred stars inside the circle
associated to vdBH 91 were found in the CMDs.
No clear cluster sequence is traced by stars with large probabilities, which
are scattered across the entire CMDs. The absence of a cluster sequence
combined with the poor and noisy overdensity are all against the reliability of
this cluster.

\begin{figure*}[ht]
    \centering
    \includegraphics[width=\hsize]{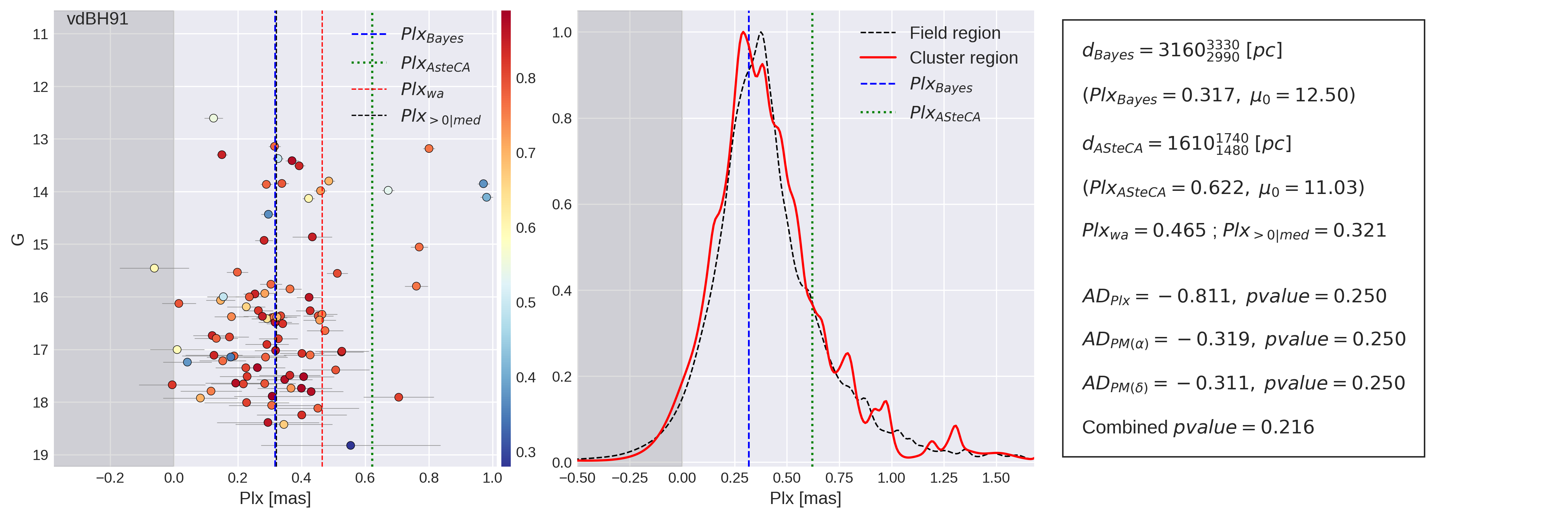}
    \caption{Idem Fig. \ref{fig:plx_bys_vdBH85} for vdBH 91.}
    \label{fig30}
\end{figure*}

The Anderson-Darling test in Fig. \ref{fig30}, right panel, is clear regarding
the true nature of vdBH 91 since the high combined $p$-value
indicates that the null hypothesis (cluster and field areas come from the same
originating distribution) con not be reasonably rejected. This result is against
the \cite{Kharchenko_2005} study where the authors found that vdBH 91
is a cluster at 0.75 kpc, approximately $0.16\times10^9$ yr old and affected by
a mean color excess $E(B-V)=0.08$.\\

We conclude that vdBH 91 is a random fluctuation of the stellar
foreground/background, and not a real entity.


\section{Trumpler 13}

\begin{figure*}[ht]
    \centering
    \includegraphics[width=\hsize]{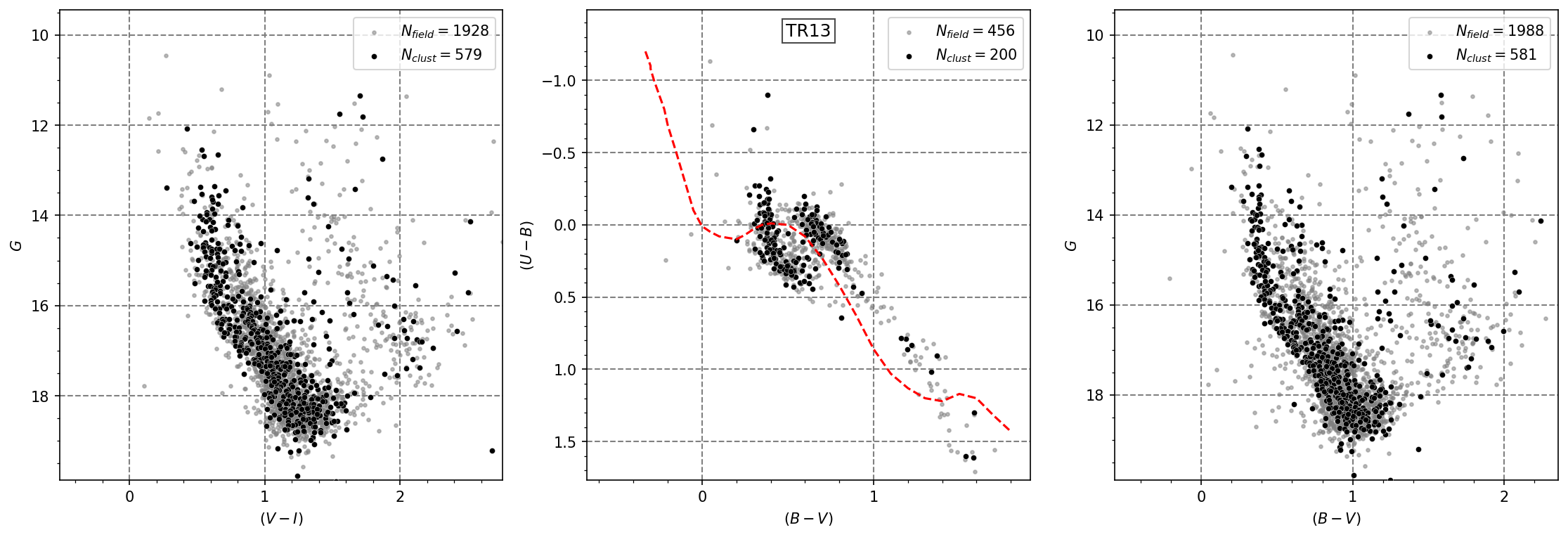}
    \caption{Idem Fig. \ref{fig:photom_vdBH85} for TR 13.}
    \label{fig39}
\end{figure*}

TR 13 is a weak object also at the south west of the Carina HII
region, seen as a diffuse but extended star accumulation near the center of the
$V$ image in Fig.~\ref{fig:Vim}. The two CMDs in Fig. \ref{fig39} show an
uncommon pattern: we see that above $G=17.5$ mag the star sequence splits
into two branches with one of them extending to the bluest side while other
branch follows the common representation of galaxy disc stars.
In the CCD the situation is the same: a wide and reddened band of potential
$B$-type stars is placed for $(B-V)<0.45$ and for $-0.25<(U-B)<0.5$ with a few
more stars at negative $(U-B)$ index while another strip of stars goes from the
characteristic place for $F$-type stars extending to the red tail including
probable giant stars.\\

\begin{figure*}[ht]
    \centering
    \includegraphics[width=\hsize]{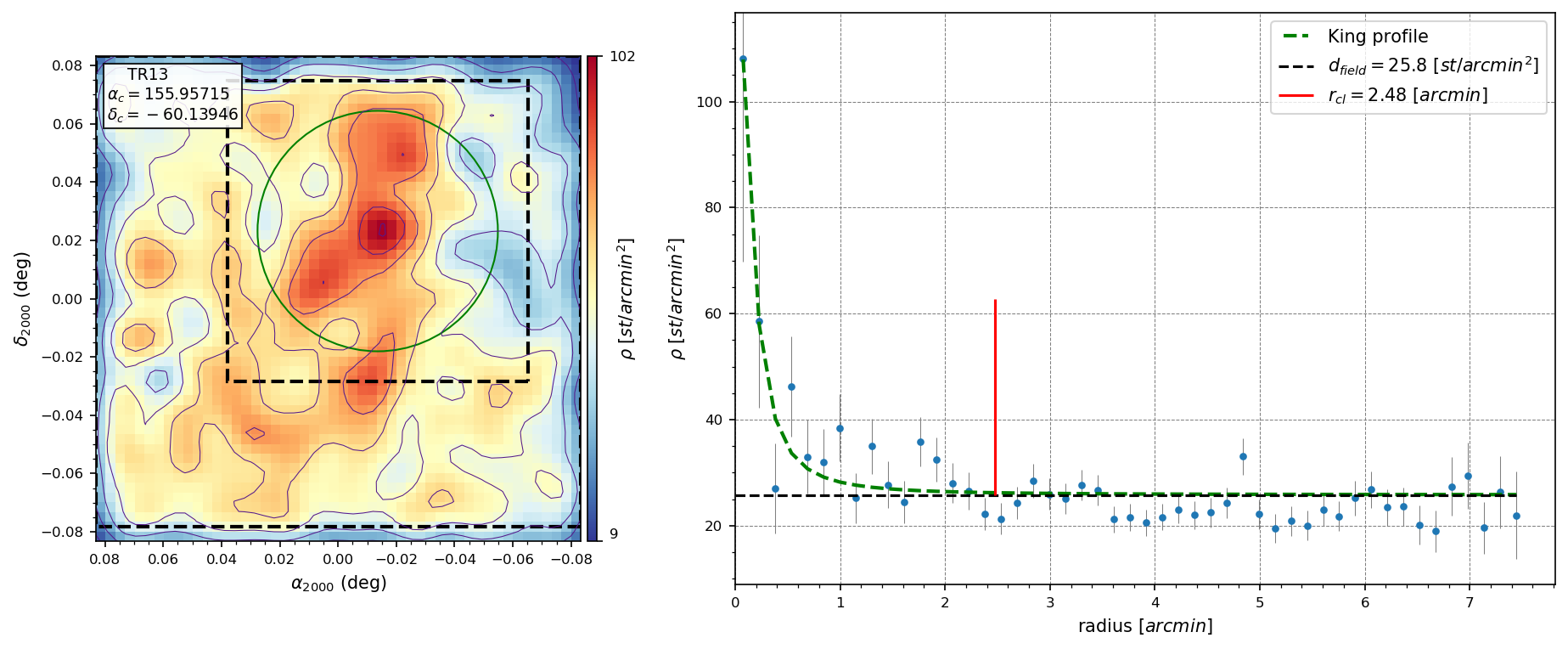}
    \caption{Idem Fig. \ref{fig:struct_vdBH85} for TR 13.}
    \label{fig40}
\end{figure*}

\begin{figure*}[ht]
    \centering
    \includegraphics[width=\hsize]{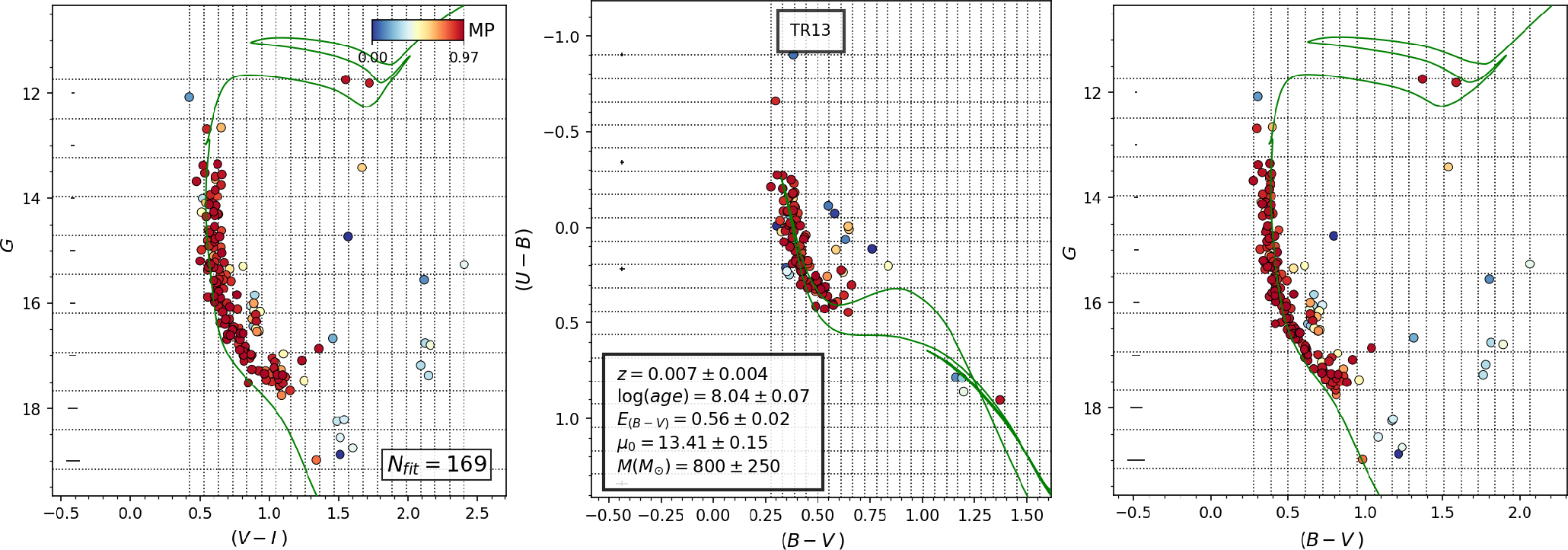}
    \caption{Idem Fig. \ref{fig:fundpars_vdBH85} for TR 13.}
    \label{fig41}
\end{figure*}

Fig. \ref{fig40} indicates that \texttt{ASteCA} found a spatially extended
overdensity mostly elongated north-south, which, at its peak is nearly 4 times
above a mean field stellar density of $\sim26$ stars per square arcmin. Given
the shape and extension of the overdensity we adopted a formal radius of
$\sim2.5$ arcmin and asked \texttt{ASteCA} to compute the membership
probabilities for those stars inside the area. 
ig. \ref{fig41} shows that after the removal of field interlopers, almost 170
stars are left composing a narrow cluster main sequence extending for more than
5 magnitudes. Consequently, when comparing with synthetic clusters the results
yield:

\begin{itemize}
\item [a)] A color excess of $E(B-V)=0.56$ is found for the best fitting
of a synthetic cluster. Since the maximum color excess provided by S\&F2011
is 1.94 it is reasonable to conclude that most of the absorption is produced
behind the position of TR 13.
\item [b)] The absorption free distance modulus of TR 13 is estimated to be
$13.41\pm0.15$ mag, placing it at a distance of $4.81\pm0.33$ kpc from
the Sun.
\end{itemize}

\begin{figure*}[ht]
    \centering
    \includegraphics[width=\hsize]{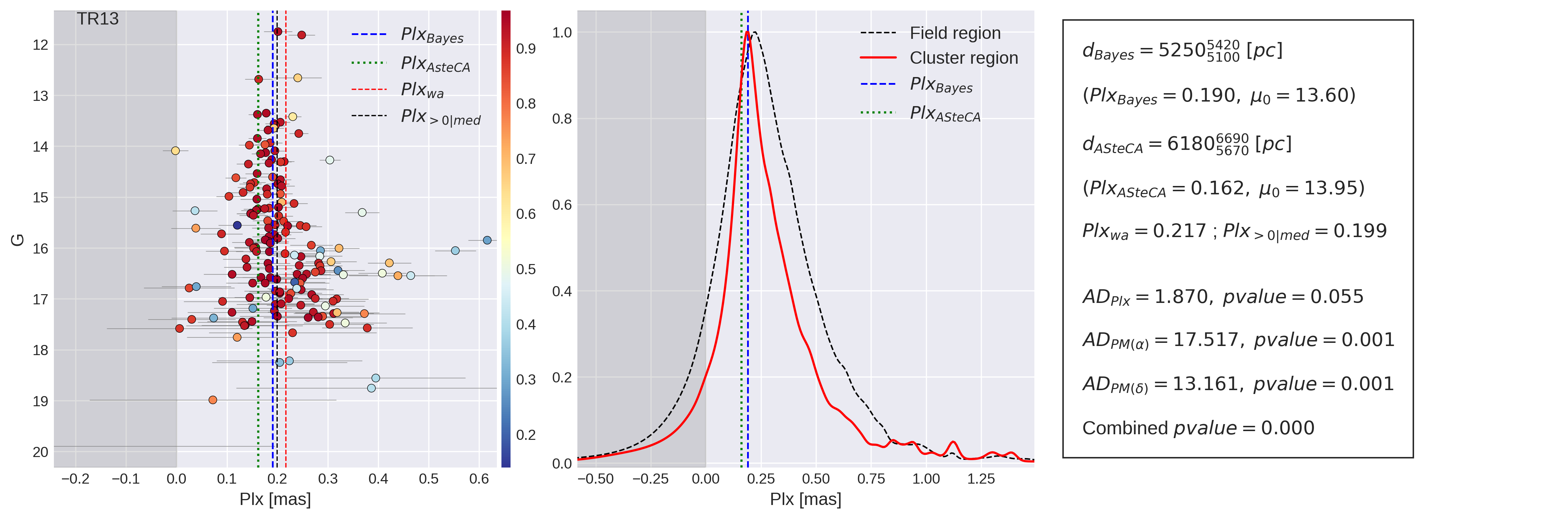}
    \caption{Idem Fig. \ref{fig:plx_bys_vdBH85} for TR 13.}
    \label{fig42}
\end{figure*}

The Anderson-Darling statistics in Fig.~\ref{fig42}, right panel, confirm the
photometric results: cluster area and the surrounding field region
possess quite different properties.\\

The selected probable members inside the overdensity confirm the
true nature of this object since the over density and the density profile are
followed by a very well defined and extended photometric counterpart. All these
facts combined with the results from the Anderson-Darling test are
self-consistent, so that we are confident that TR 13 is a young cluster of
$0.11\pm0.02\times10^9$ years old.


\section{van den Bergh-Hagen 106}

\begin{figure*}[ht]
    \centering
    \includegraphics[width=\hsize]{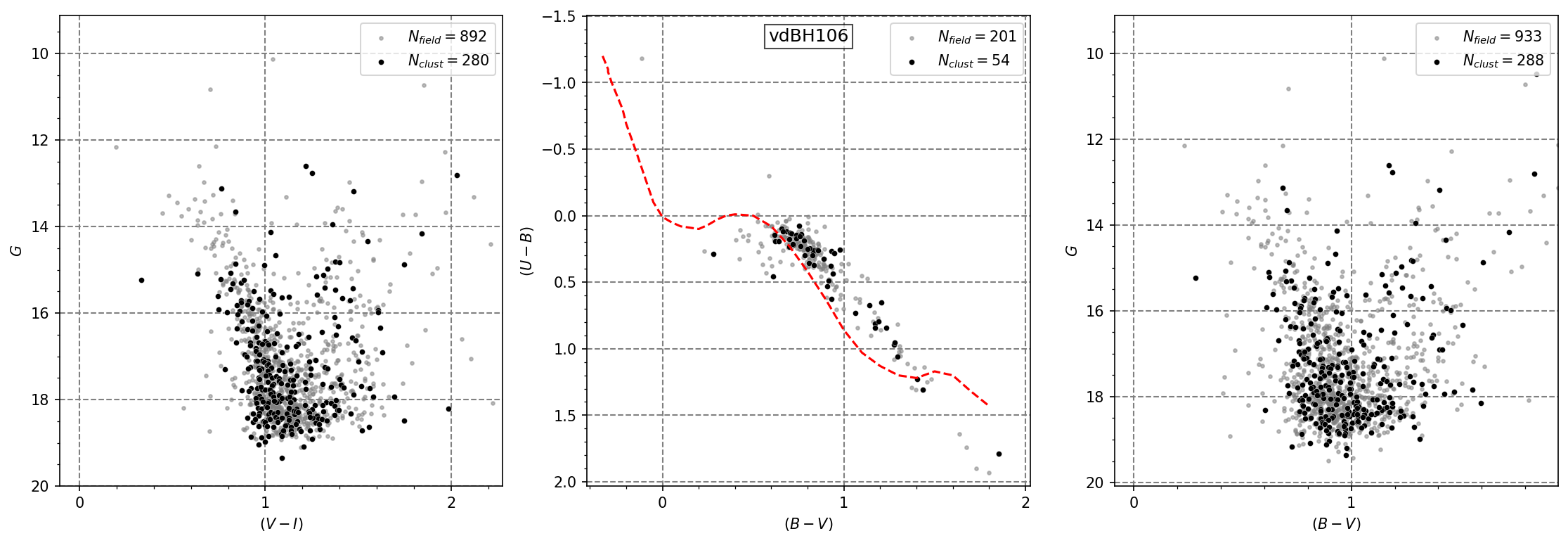}
    \caption{Idem Fig. \ref{fig:photom_vdBH85} for vdBH 106.}
    \label{fig43}
\end{figure*}

\begin{figure*}[ht]
    \centering
    \includegraphics[width=\hsize]{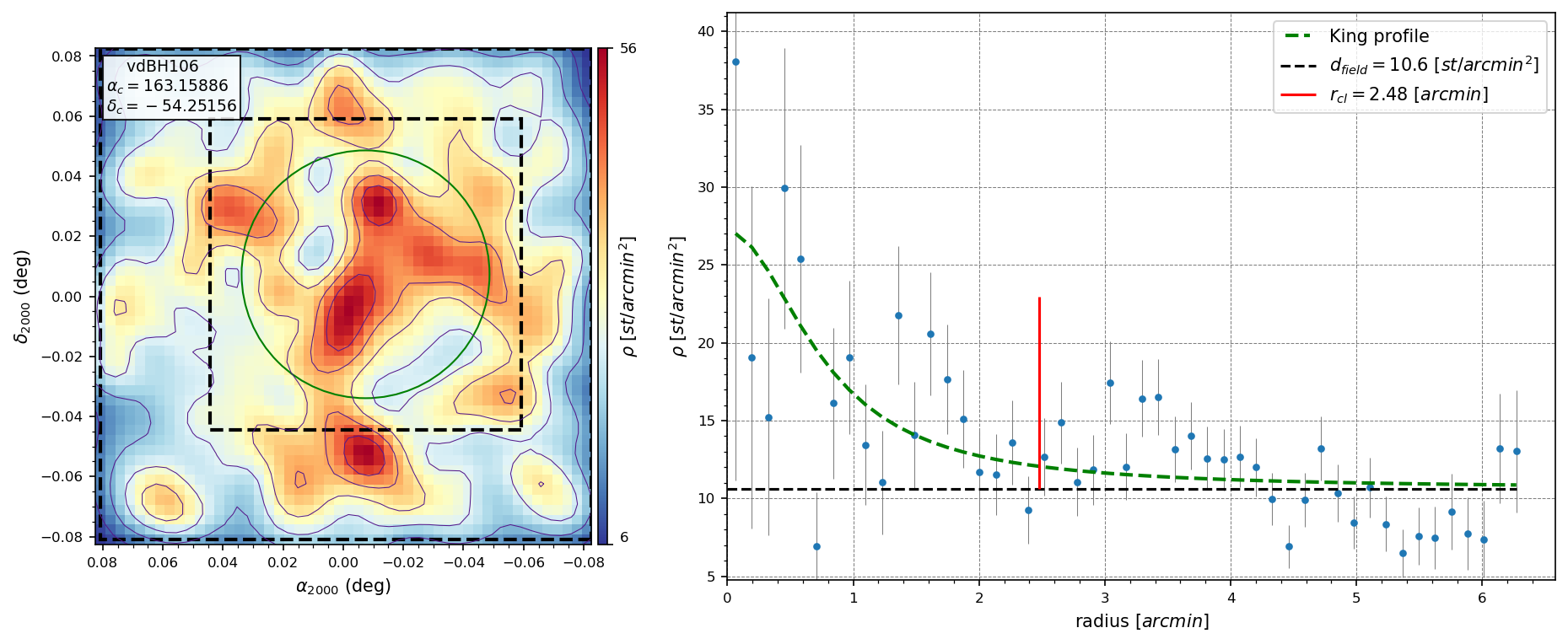}
    \caption{Idem Fig. \ref{fig:struct_vdBH85} for vdBH 106.}
    \label{fig44}
\end{figure*}

This cluster is placed at the south-east of the Vela constellation. The not so
dense stellar field where it is placed has no relevant features except a few
moderately bright stars as shown in Fig. \ref{fig:Vim}.
The CMDs shown in Fig.~\ref{fig43} represent typical photometric features
structures of galactic fields with no cluster inside. As for the CCD in the
same figure it shows a reduced number of stars below the intrinsic line 
(probably reddened late $B$- and $A$-types) and a tail of stars from of late
$F$-types to $M$-type stars –some of them probably giant- at the red end.
\texttt{ASteCA} spatial analysis found some star clumps as seen in
Fig.~\ref{fig44}, left panel. We focus the attention on the main one at the very
center of the frame, since here we see the highest overdensity peak with
$\sim3$ times more stars than at the mean stellar background density of
$\sim11$ stars per square arcmin.
We assume that most of stars in vdBH 106 must be included there so the cluster
parameters should be well established. The RDP to the right appears not well
defined since it reflects the irregular and poor star density even inside the
zone selected to investigate the cluster parameters.
Only 82 stars have been selected as probable members inside this area.
Those stars having probabilities near the maximum values in this region would
seem to outline a (rather noisy) cluster sequence that can be fitted with a
synthetic cluster which yields the following parameters:

\begin{itemize}
\item [a)] A color excess of $E(B-V)=0.30$ has been found to affect the
cluster. This value is well in line with the maximum color excess provided
by S\&F2011, $E(B-V)=0.57$ in this direction.
\item [b)] The absorption free distance modulus of vdBH 106 was found to be
$13.44\pm0.36$ mag, putting the cluster at a distance of
$d=4.87\pm0.81$ kpc from the Sun.
\end{itemize}

In this region we found from the application of the Anderson-Darling test that
the parallax and proper motion distributions seem to belong to the same
originating distribution, as seen in Fig. \ref{fig46}. Indeed, the
large combined $p$-value makes the rejection of the null hypothesis difficult,
if not impossible\\

\begin{figure*}[ht]
    \centering
    \includegraphics[width=\hsize]{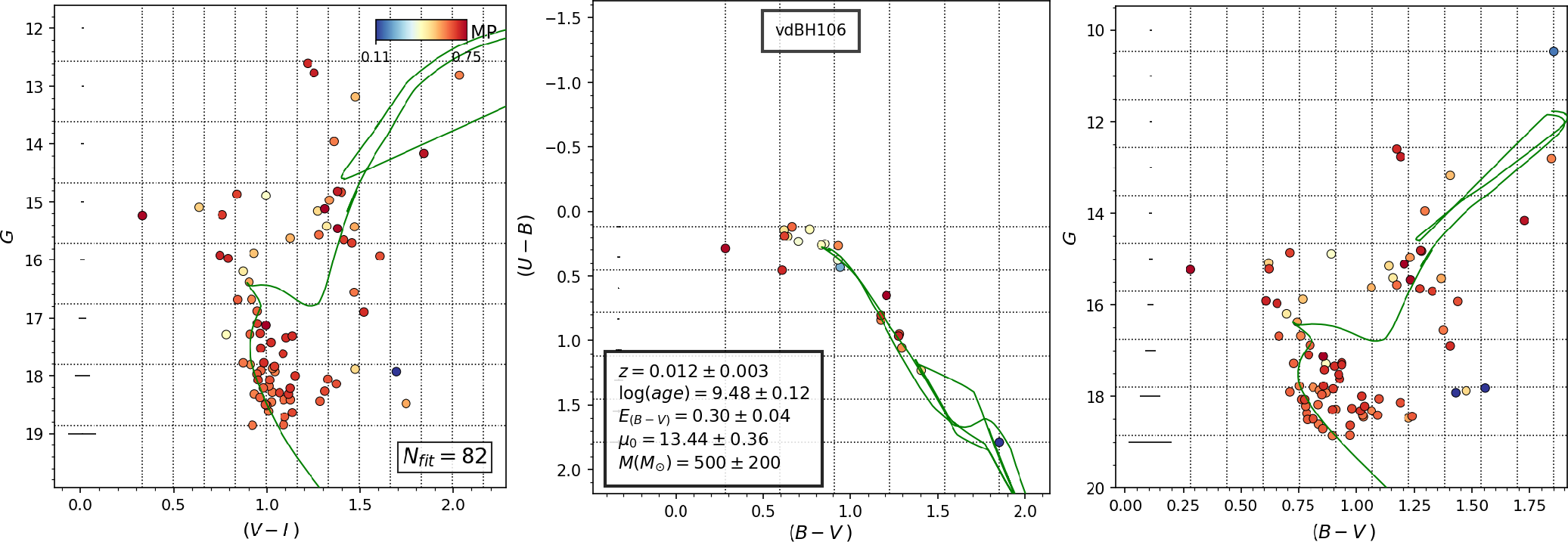}
    \caption{Idem Fig. \ref{fig:fundpars_vdBH85} for vdBH 106.}
    \label{fig45}
\end{figure*}

\begin{figure*}[ht]
    \centering
    \includegraphics[width=\hsize]{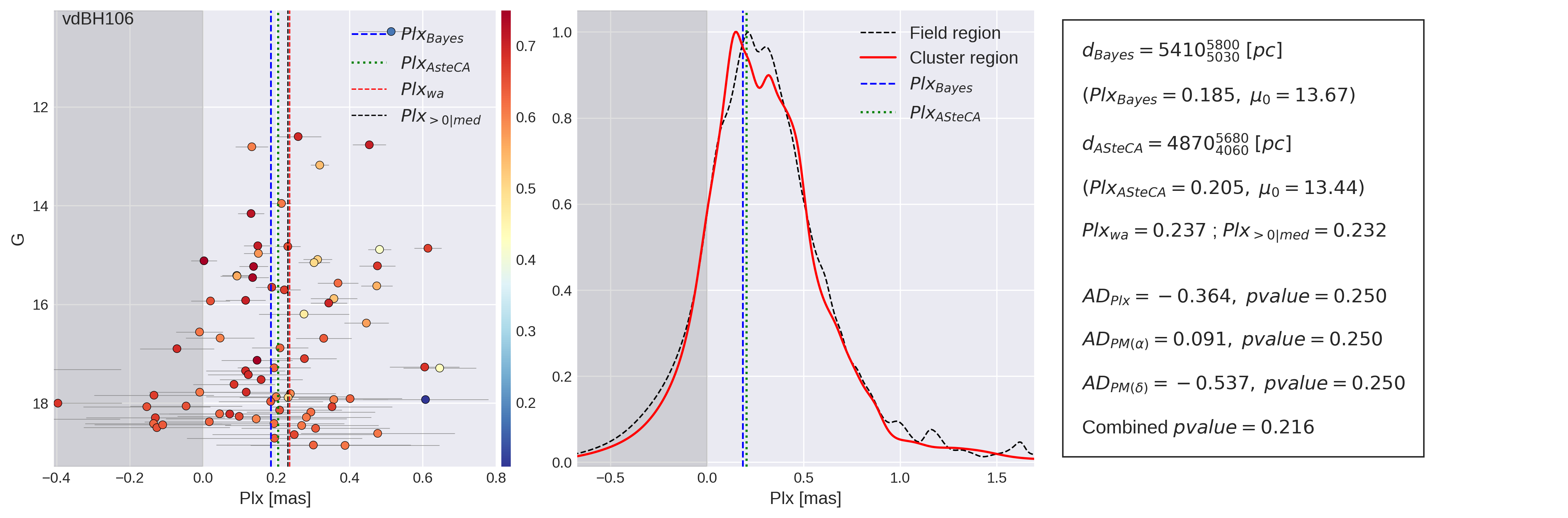}
    \caption{Idem Fig. \ref{fig:plx_bys_vdBH85} for vdBH 106.}
    \label{fig46}
\end{figure*}

Despite a trace of a sequence belonging to a typical old cluster is
noticeable in Fig.~\ref{fig45}, we are cautious as to confirm its nature.
Clearly deeper photometric observations (particularly in the $U$ filter) are
needed. Meanwhile and under the assumption that we are facing a true object
vdBH 106 could be an old open cluster around $3.00\pm0.80\times10^9$ years
old.


\section{Ruprecht 88}

\begin{figure*}[ht]
    \centering
    \includegraphics[width=\hsize]{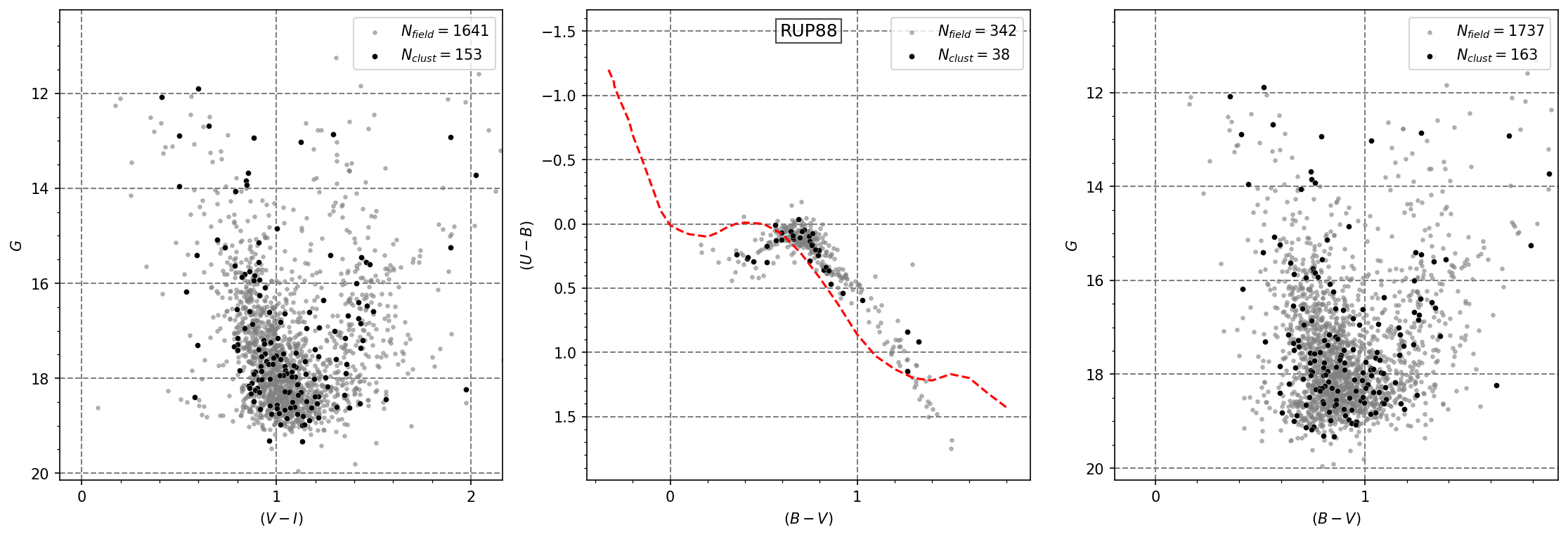}
    \caption{Idem Fig. \ref{fig:photom_vdBH85} for RUP 88.}
    \label{fig31}
\end{figure*}

RUP 88 is another potential cluster south of the Carina HII region. Like
other objects in this paper no obvious stellar grouping is perceived in the $V$
image of Fig. \ref{fig:Vim}. The overall star CMDs in Fig.~\ref{fig31} show a
scattered star distribution above $G=16$ mag. From this magnitude down the
common pattern of galactic disc stars takes place in the CMDs. The CCD in Fig.
\ref{fig31} suggests that no blue and therefore young star is present in the
region of RUP 88. In the range $0.2<(B-V)<0.8$ we see a handful of stars that
could be reddened late of late $B$- types or $A$-$F$-type stars. The remaining of
this diagram is a trace composed by $A$- to $M$-type stars.\\

\begin{figure*}[ht]
    \centering
    \includegraphics[width=\hsize]{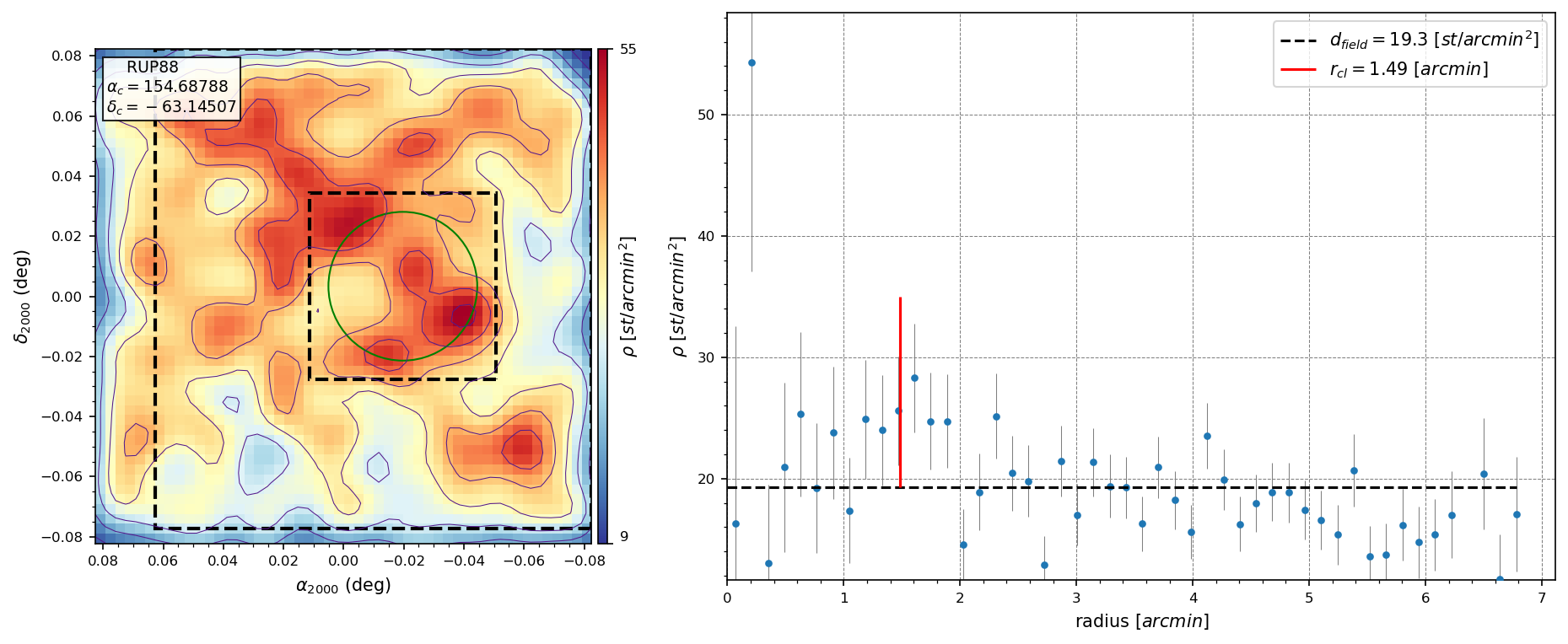}
    \caption{Idem Fig. \ref{fig:struct_vdBH85} for RUP 88.}
    \label{fig32}
\end{figure*}

\begin{figure*}[ht]
    \centering
    \includegraphics[width=\hsize]{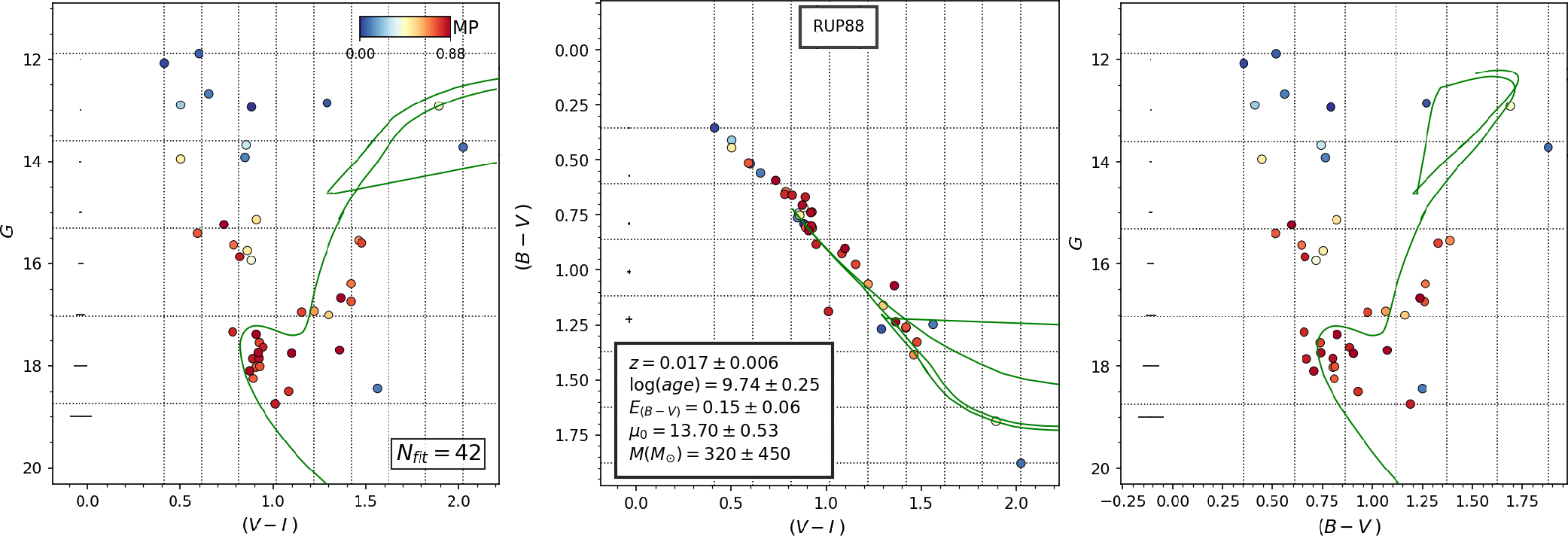}
    \caption{Idem Fig. \ref{fig:fundpars_vdBH85} for RUP 88
with the $(B-V)$ vs $(V-I)$ diagram instead of the $(B-V)$ vs $(U-B)$
diagram.}
    \label{fig33}
\end{figure*}

As with other clusters in the present sample, when the spatial distribution
of stars in the frame is analyzed no clear star overdensity
appears in the location where RUP 88 is supposed to exist. In fact, the
contour plot in Fig. \ref{fig32}, left panel, shows a poor star number
enhancement from south west to northeast of the frame extending north west.
Given the difficulties to state the position of the cluster center (if it
exists) we ask \texttt{ASteCA} to inspect the region encircled in green in Fig.
\ref{fig32}, where a reasonable density profile could be found. The
RDP is still noisy because of a rather poor star number contained between the
assumed cluster limits.
If we look at the CMDs in Fig.~\ref{fig33} only 42 stars with a wide
range of probabilities remain inside the adopted cluster region 
after interlopers are removed, with no trace of a cluster sequence
found.
The three photometric diagrams in Fig. \ref{fig33} confirm this point as only
an amorphous distribution of stars scarcely resembling a cluster main sequence
can be seen.\\

\begin{figure*}[ht]
    \centering
    \includegraphics[width=\hsize]{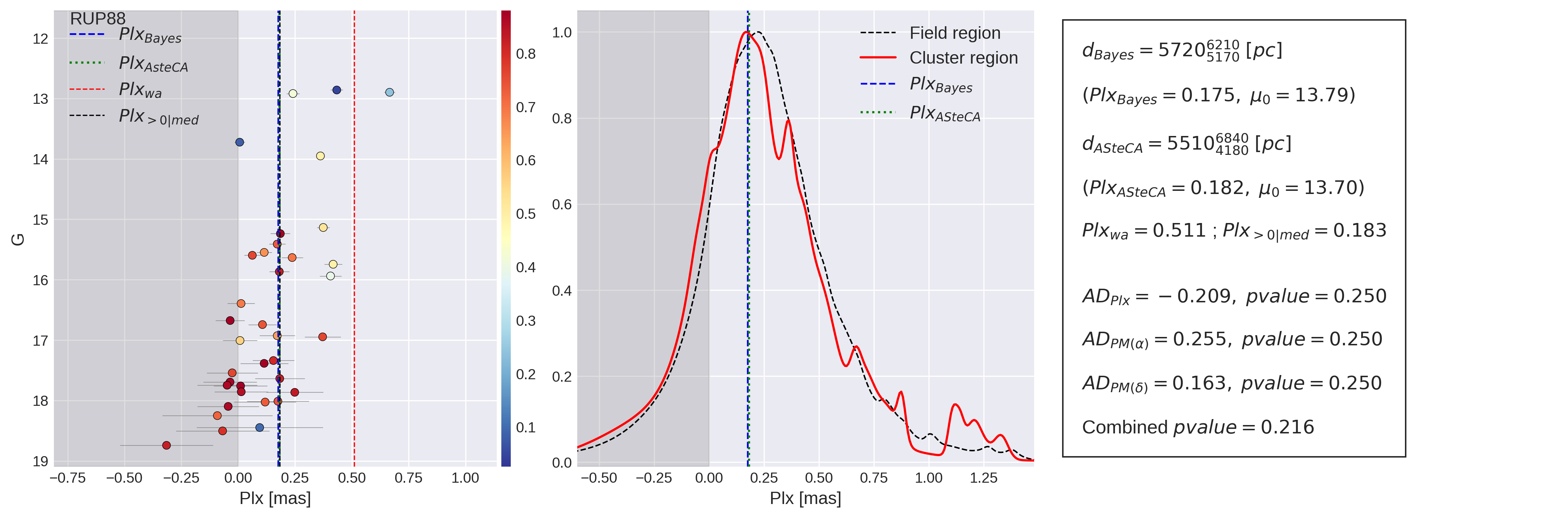}
    \caption{Idem Fig. \ref{fig:plx_bys_vdBH85} for RUP 88.}
    \label{fig34}
\end{figure*}

The Anderson-Darling test in Fig. \ref{fig34}, right panel is unable to
separate the cluster population from the one from the field region, for the
three explored dimensions. The combined $p$-value for proper motions and
parallaxes is large, suggesting that both samples come from the same
population. The necessary requirement that there is a reasonable main
sequence is absent and, combined with this result, precludes
concluding that RUP 88 is a true cluster.


\section{Ruprecht 162}
\label{app:rup162}

\begin{figure*}[ht]
    \centering
    \includegraphics[width=\hsize]{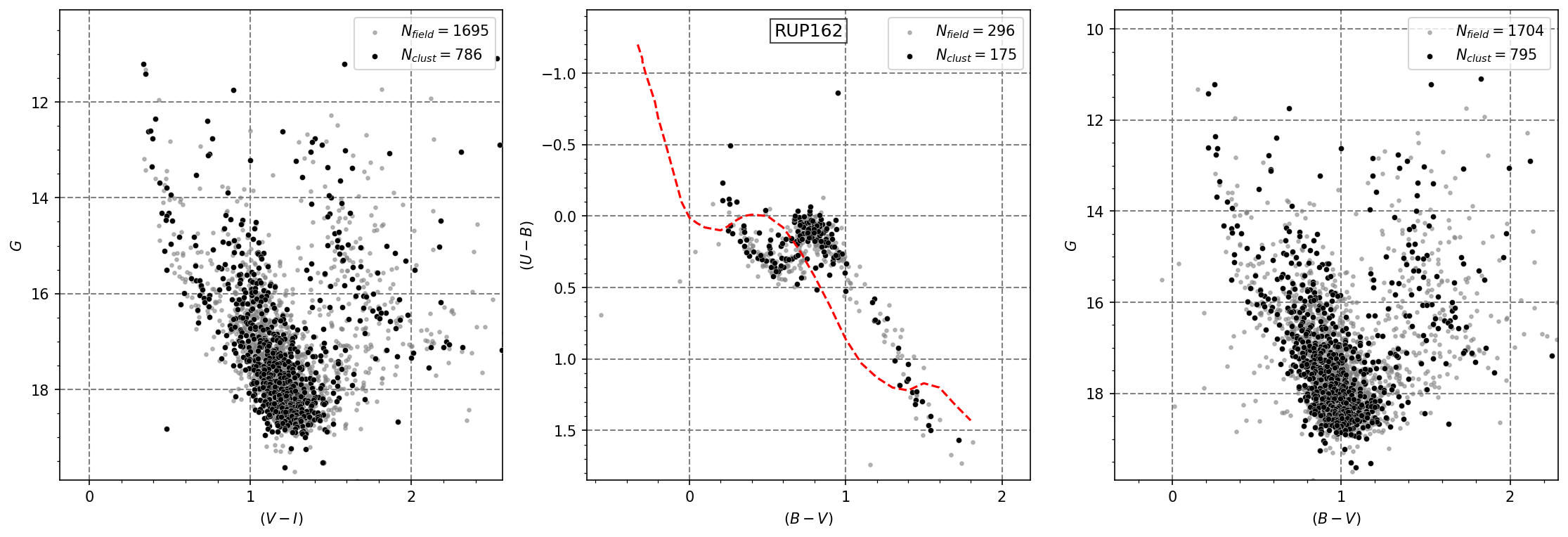}
    \caption{Idem Fig. \ref{fig:photom_vdBH85} for RUP 162.}
    \label{fig47}
\end{figure*}

Placed to the south east of the Carina HII region, the $V$ image of the region
in Fig. \ref{fig:Vim} where the cluster is supposed to exist shows a moderate
number of stars resembling a star grouping placed at the north-west in the
frame. At first glance the CMDs in Fig.~\ref{fig47} for the overall stars look
as if a cluster main sequence is emerging from the trace of the disk star
distribution.
In the same figure, middle panel, the CCD splits into two star groups: one of
them is mostly placed below the intrinsic line for $0.0<(B-V)< 0.8$ and
resembles a strip of reddened blue stars (including early and late $B$-types
and, perhaps, some $A$-type stars); the other group shows a distribution of $F$-
to $M$-type stars strongly affected by reddening in appearance.\\

\begin{figure*}[ht]
    \centering
    \includegraphics[width=\hsize]{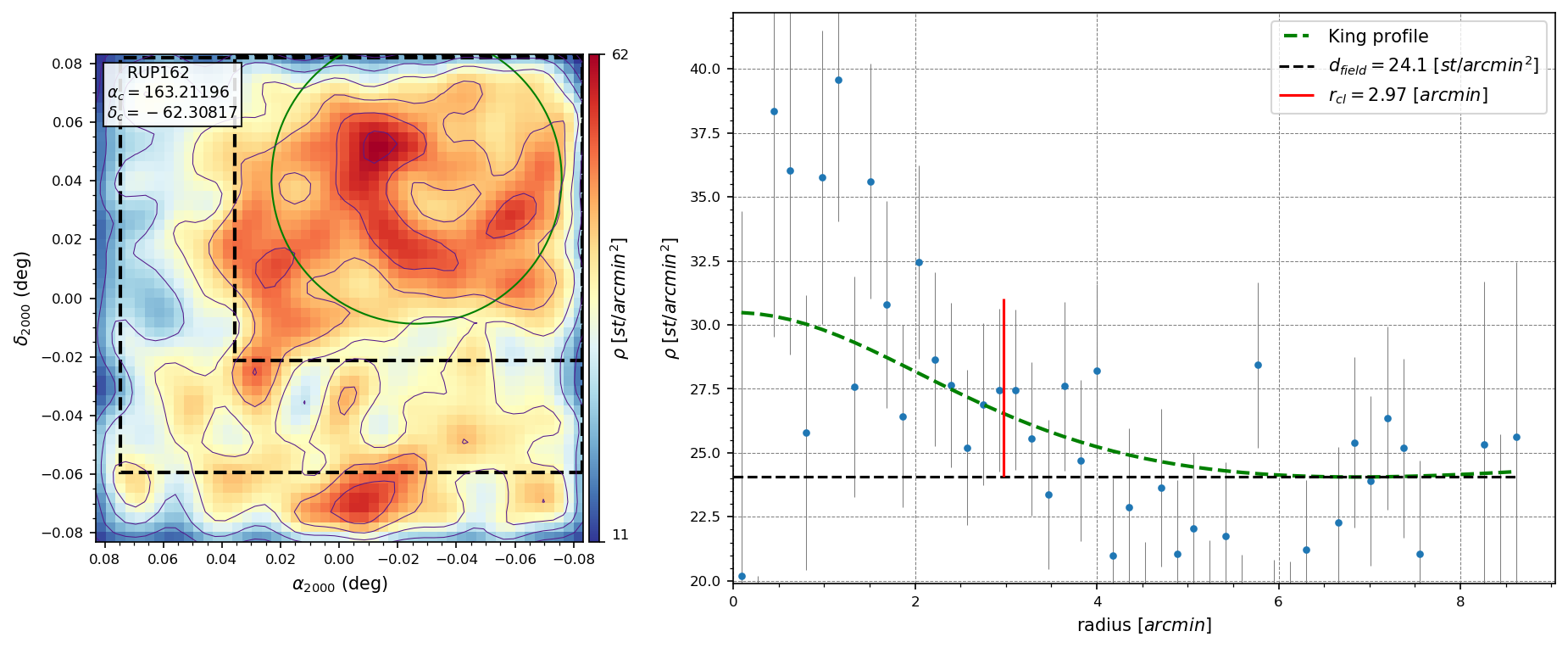}
    \caption{Idem Fig. \ref{fig:struct_vdBH85} for RUP 162.}
    \label{fig48}
\end{figure*}

\texttt{ASteCA} detected an extended and irregular region at the north-west of
the frame in Fig.~\ref{fig48} (where the cluster is supposed to be). Given the
difficulties to set a clear overdensity we decided to focus the attention on
the $\sim3$ arcmin zone encircled in green in Fig. \ref{fig48}, left panel.
The background mean star density is over 20 stars per squared arcmin and, at
the most, the overdensity is just 40 stars at the maximum. This produces
unavoidably a noisy RDP (it is hard to establish a meaningful radius and there
is a quite irregular star distribution across the zone).

\begin{figure*}[ht]
    \centering
    \includegraphics[width=\hsize]{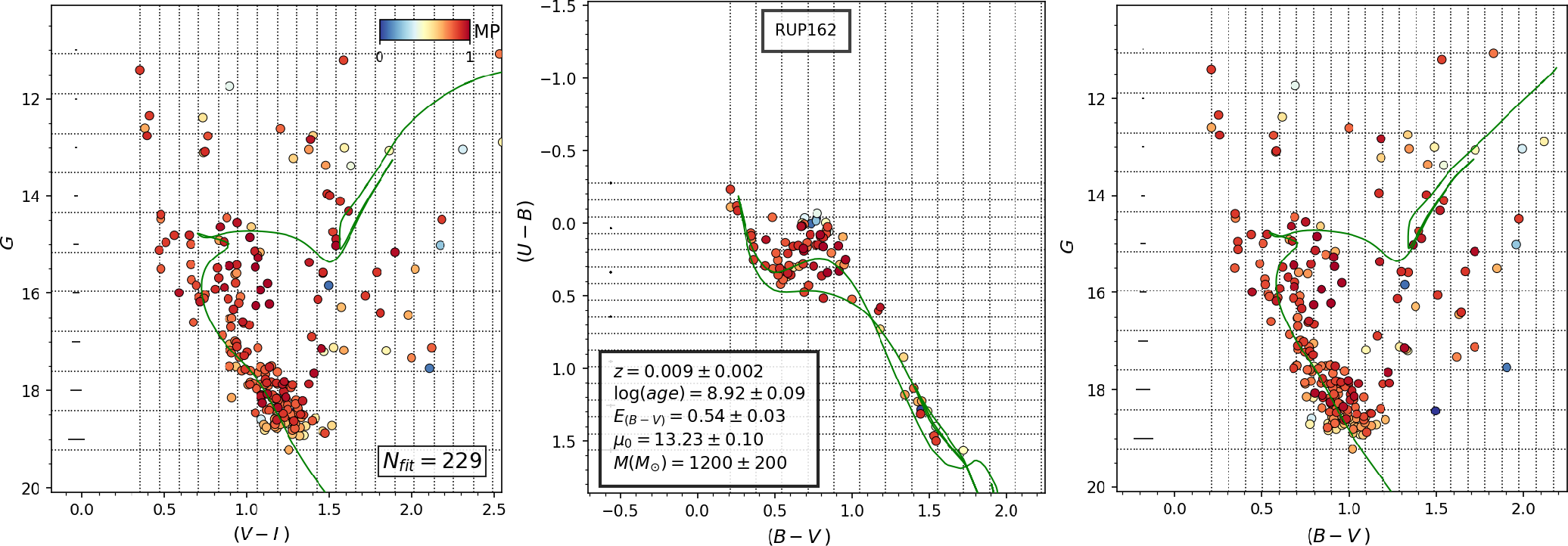}
    \caption{Idem Fig. \ref{fig:fundpars_vdBH85} for RUP 162.}
    \label{fig49}
\end{figure*}

The CMDs and CCD in Fig.~\ref{fig49}, after the removal of field interlopers,
show more than 200 dispersed stars, most of them with large probabilities
assigned. The large scatter in the CMDs and the large MP values assigned even
to stars that are clearly not part of any cluster sequence, point against the
existence of a true cluster in the region.
On the other hand the cleaned CCD, Fig.~\ref{fig49} mid panel, shows a blue
sequence of stars suffering some internal color scatter followed by a tail of
$F$- to $K$-type stars. Therefore this object could be more extended than
supposed. \texttt{ASteCA} found the best fitting with a synthetic cluster with
the following properties:

\begin{itemize}
\item [a)] The color excess affecting the cluster is $E(B-V)=0.54$, well
below the maximum value given by S\&F2011 who estimate $E(B-V)=1.07$.
\item [b)] The absorption free distance modulus is $13.23\pm0.10$ mag
corresponding to a distance of $d=4.43\pm0.20$ kpc.
\end{itemize}

\begin{figure*}[ht]
    \centering
    \includegraphics[width=\hsize]{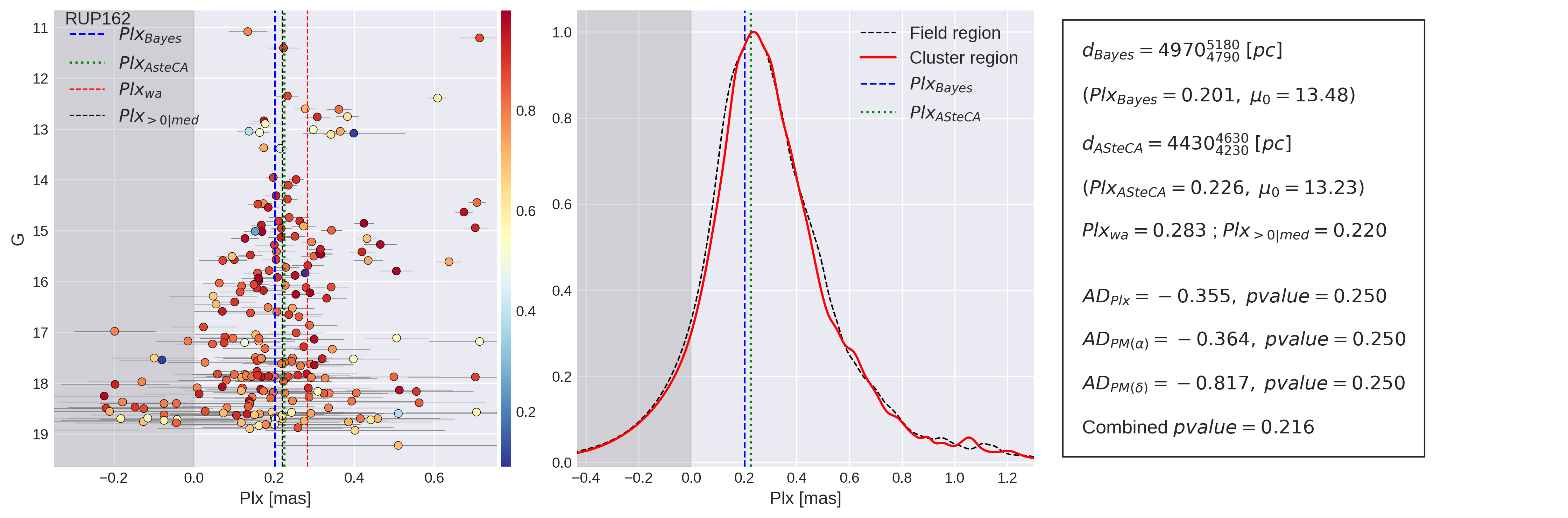}
    \caption{Idem Fig. \ref{fig:plx_bys_vdBH85} for RUP 162.}
    \label{fig50}
\end{figure*}

Anderson-Darling statistical test results are shown in Fig. \ref{fig50}, right
panel. Parallaxes and proper motions $PM(\alpha)$ and $PM(\delta)$ in the
location of RUP 162 and the surrounding field region do not seem to be
different enough from each other as to be efficiently disentangled.\\

Although weak enough the presence of a probable main sequence in the 
panels of Fig. \ref{fig49}, make us cautious leaving some chance for RUP 162 to
be a true cluster about $0.80\pm0.20\times10^9$ years old.
An additional reinforcement as for the hypothetical true entity of this young
object is the existence of a sudden gap along the main sequence at $G=16.5$ mag
and the presence of high probability stars at the red side resembling traces of
a pre-main sequence. Certainly we are just speculating on this fact so that
more and deeper observations are needed to arrive to a concluding result for
RUP 162.


\section{Lynga 15}

\begin{figure*}[ht]
    \centering
    \includegraphics[width=\hsize]{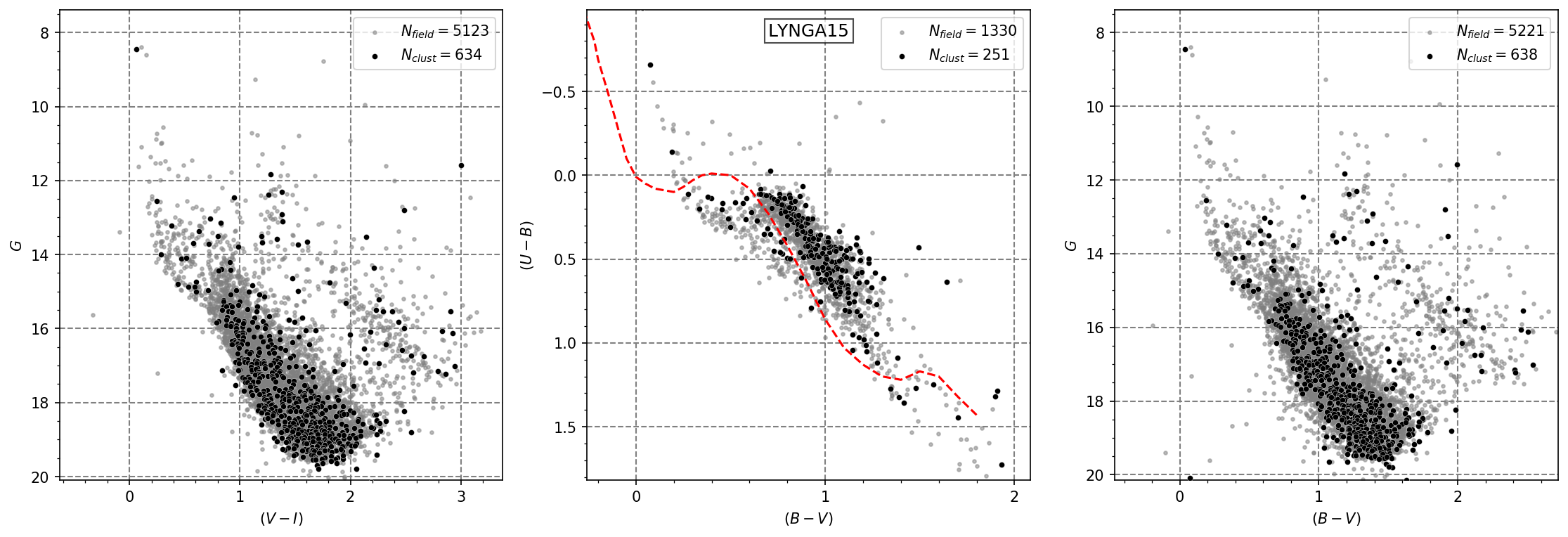}
    \caption{Idem Fig. \ref{fig:photom_vdBH85} for Lynga 15.}
    \label{fig51}
\end{figure*}

This is an intriguing object placed in Centaurus, south west between Crux and
the east border of Carina. More specifically, Lynga 15 is about $1^\circ$
north-east of the star formation region SFR293.64-1.41 \citep{Avedisova_2002}.
Like in many other cases already shown in the $V$ images in Fig. \ref{fig:Vim}
this region does not show, at first glance, any prominent stellar feature
though some stars are bright enough as to attract attention to this place.
However, the overall CMDs and CCD shown in Fig.~\ref{fig51}
are quite surprising since both CMDs depict an extended sequence (from $G=8$
down to $G=15.5$ mag) emerging toward the left side of the main disc population
trace. In the same figure, middle panel, the CCD shows a strip of blue stars
($0.0<(B-V)<0.0$) accompanied by other, probable reddened early type stars,
placed above $(U-B) = 0.0$. The picture seen in the three panels of Fig.
\ref{fig51} induces to think of Lynga 15 as a quite young open cluster.\\

\begin{figure*}[ht]
    \centering
    \includegraphics[width=\hsize]{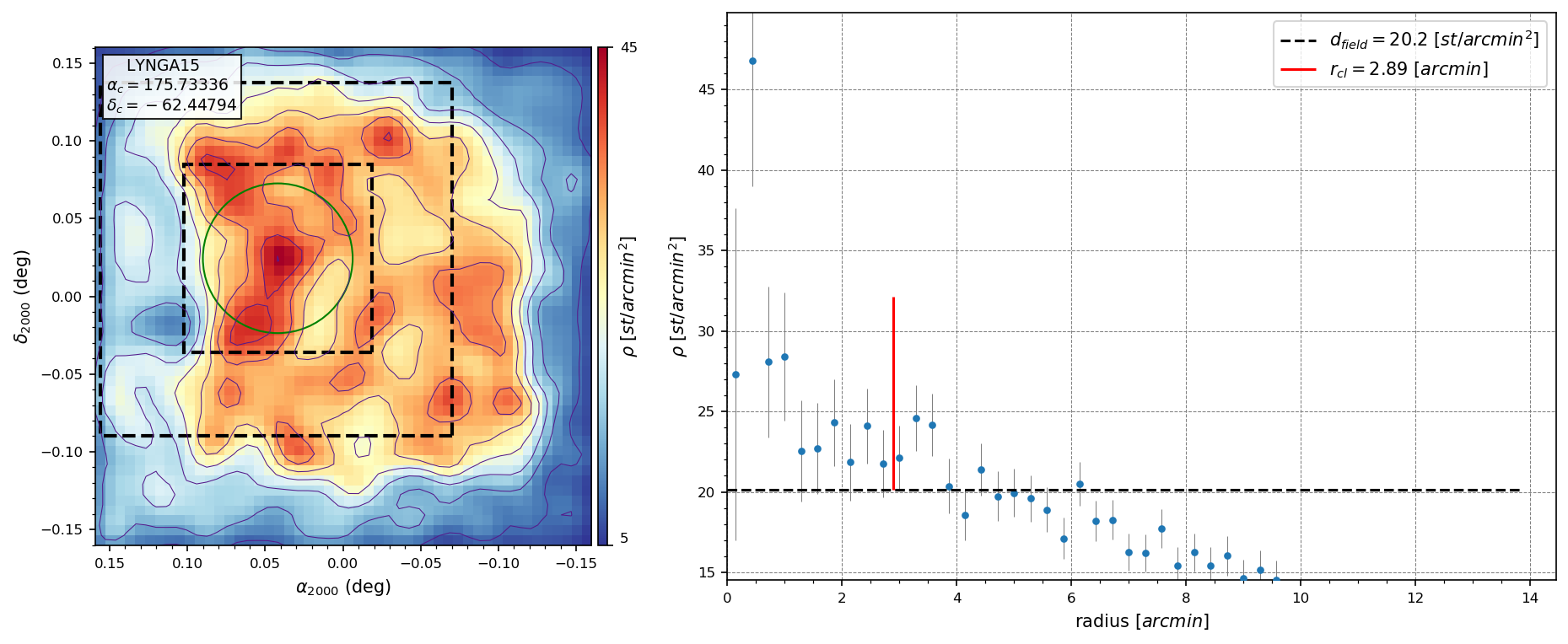}
    \caption{Idem Fig. \ref{fig:struct_vdBH85} for Lynga 15.}
    \label{fig52}
\end{figure*}

\begin{figure*}[ht]
    \centering
    \includegraphics[width=\hsize]{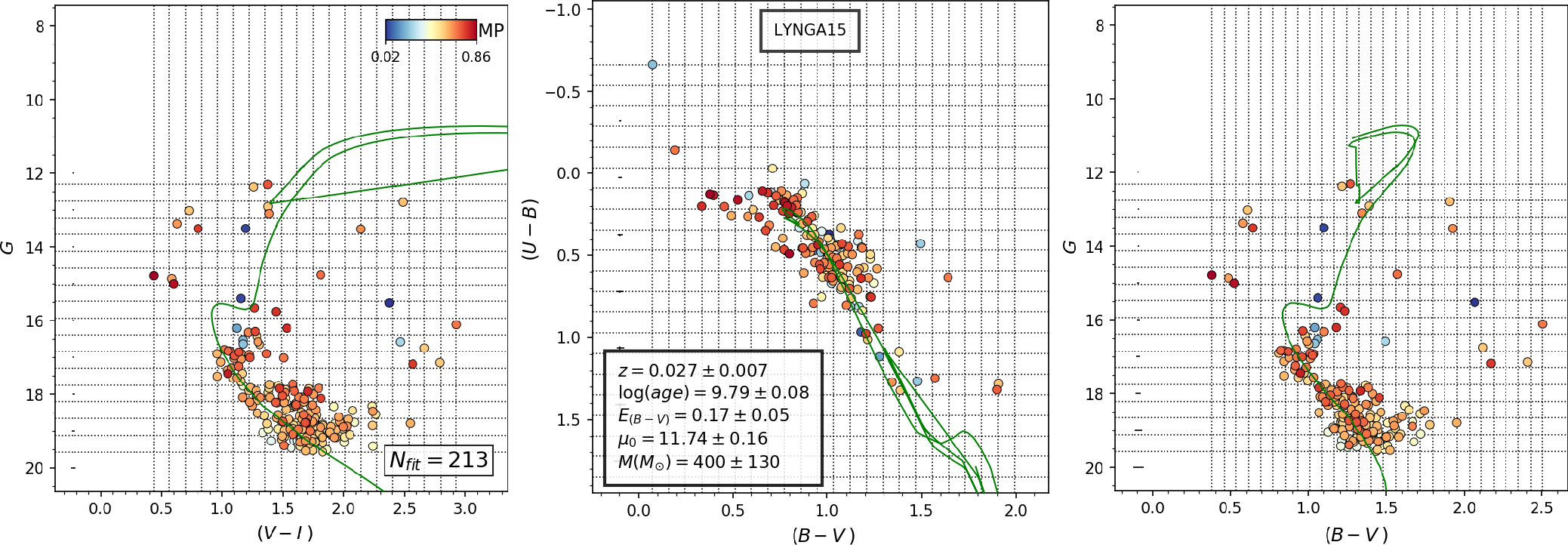}
    \caption{Idem Fig. \ref{fig:fundpars_vdBH85} for Lynga 15.}
    \label{fig53}
\end{figure*}

In turn, \texttt{ASteCA} analysis of the spatial structure found an extended and
irregular stellar density with no indication of a clear overdensity.
The observed frame's density map shows two very distinct stellar
densities, explained by the combination of observations made by two different
telescopes detailed in Sect.~\ref{sec:photo_obs} (same as NGC 4349).
After many attempts to look for the place where the star membership
probabilities reach the highest values we adopted a $\sim2.9$ arcmin radius and
set the potential cluster center in the literature coordinates as indicated in
Fig.~\ref{fig52}, left panel. In this place, the RDP displays a $\sim45$ stars
per squared arcmin peak above the stellar field density, as seen in the right
panel of Fig.~\ref{fig52}.
Even in this position \texttt{ASteCA} yields a conflictive result
since the selected probable members show a large dispersion and, as
seen in Fig.~\ref{fig53} left and right panels, a probable cluster main
sequence mostly composed by lower probability stars appears below approximately
$G=17$ mag.
Above this visual magnitude the main sequence vanishes and we are left with
just a handful of stars with rather large probability values, scattered in
color index and magnitudes. This is, no upper cluster main sequence is
evident in the clean CMDs.
The CCD in the mid panel of Fig. \ref{fig53} contains a few blue stars with no
counterpart in the CMDs. This could be explained this way: all across the
surveyed region there are blue stars (see the overall CCD in Fig. \ref{fig51})
composing a sort of Blue Plume in the respective CMDs and just by chance some
blue stars also appear in the potential cluster region after \texttt{ASteCA}
analysis (mid panel Fig.~\ref{fig53}). It could be possible yet that Lynga 15
is an extended open cluster (even larger than the size of our frame), but the
presence of the huge star gap above $G=17$ mag is unexplainable in a CMD from a
statistical point of view.
In our opinion and from a photometric and spatial point of view Lynga 15 is not
an open cluster. The application of the Anderson-Darling test inform us that
the properties of stars inside the adopted cluster radius and outside of it are
similar, with a probability of $\sim$6\% of mistakenly rejecting the null
hypothesis that both samples arose from the same distribution.\\

We conclude that Lynga 15 is not a true cluster but a superposition of blue
stars at several distances along the line of sight.
This is not odd at all since this object is not far from the galactic equator
so it is probable that blue stars are seen along the direction to this
potential cluster.

\begin{figure*}[ht]
    \centering
    \includegraphics[width=\hsize]{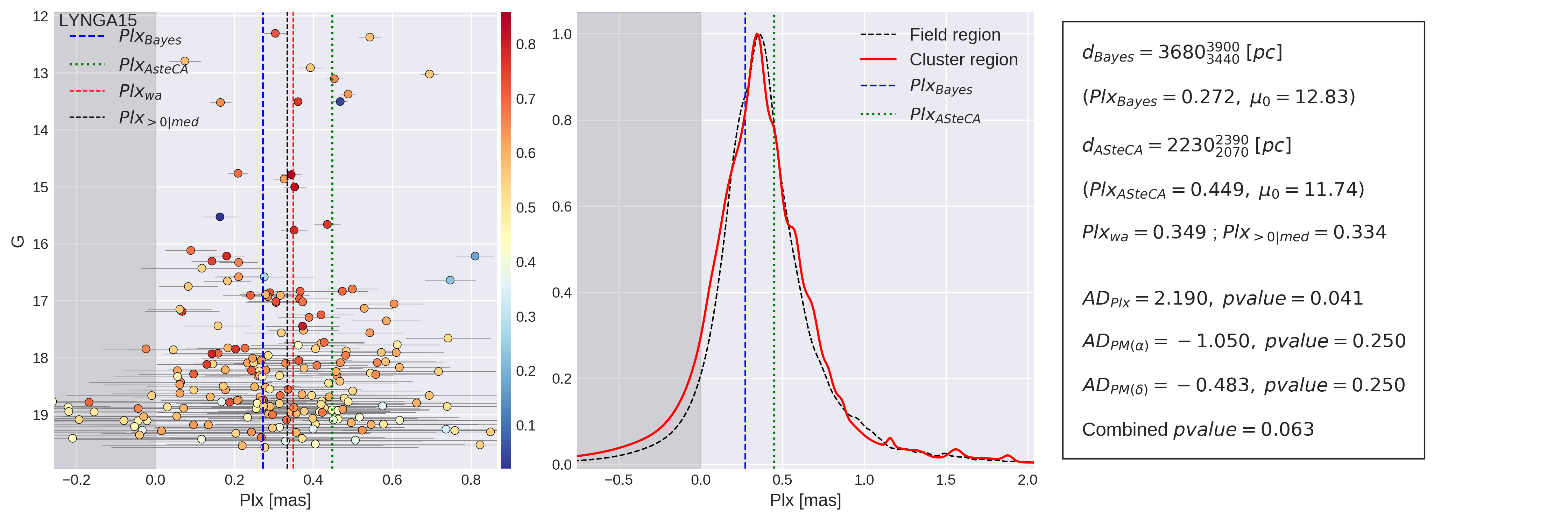}
    \caption{Idem Fig. \ref{fig:plx_bys_vdBH85} for Lynga 15.}
    \label{fig54}
\end{figure*}

\section{Loden 565}

\begin{figure*}[ht]
    \centering
    \includegraphics[width=\hsize]{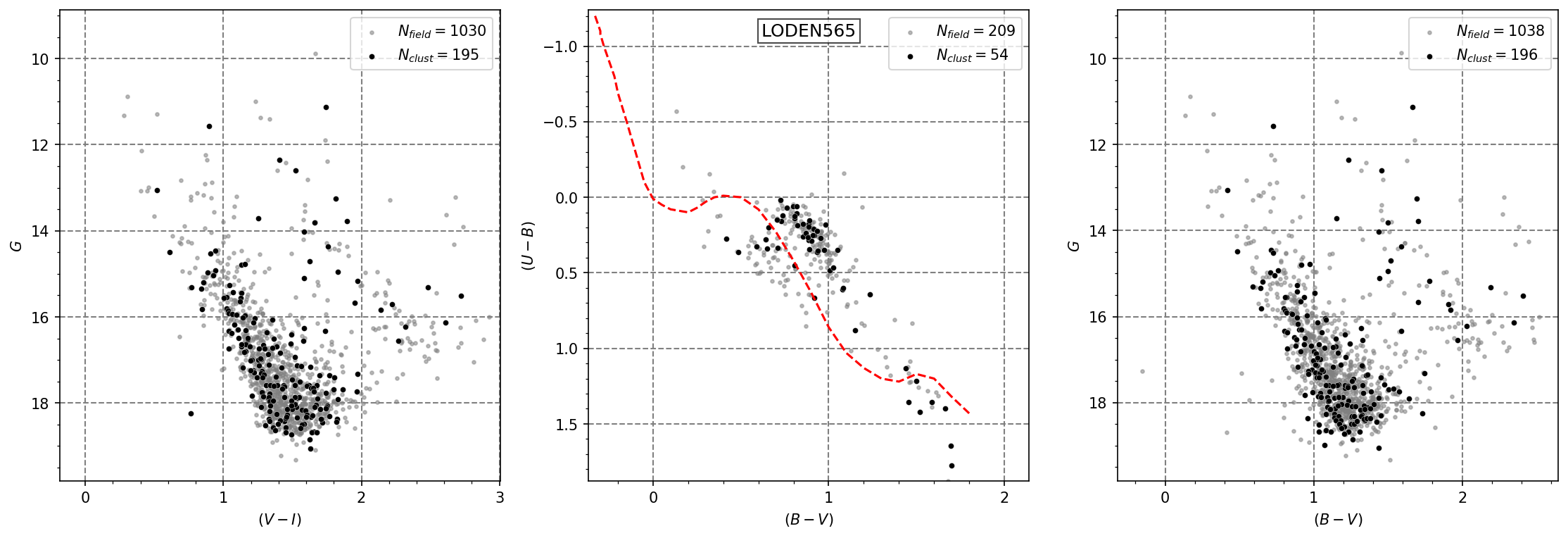}
    \caption{Idem Fig. \ref{fig:photom_vdBH85} for Loden 565.}
    \label{fig55}
\end{figure*}

Placed toward the west side of the Crux constellation the $V$ image in Fig. 
\ref{fig:Vim} of Loden 565 does not show any evident star grouping. Inspection
of the CCD and CMDs in Fig.~\ref{fig55} only suggests the presence of a
dispersed star group down to $G=15-16$ mag approximately. From this magnitude
down the overall CMDs show the common pattern of galactic disc star population
and nothing relevant can be seen in the CCD in the mid panel of
Fig.~\ref{fig55}, but a modest handful of probable slightly reddened late blue
stars for $(B-V)<0.6$.\\

\begin{figure*}[ht]
    \centering
    \includegraphics[width=\hsize]{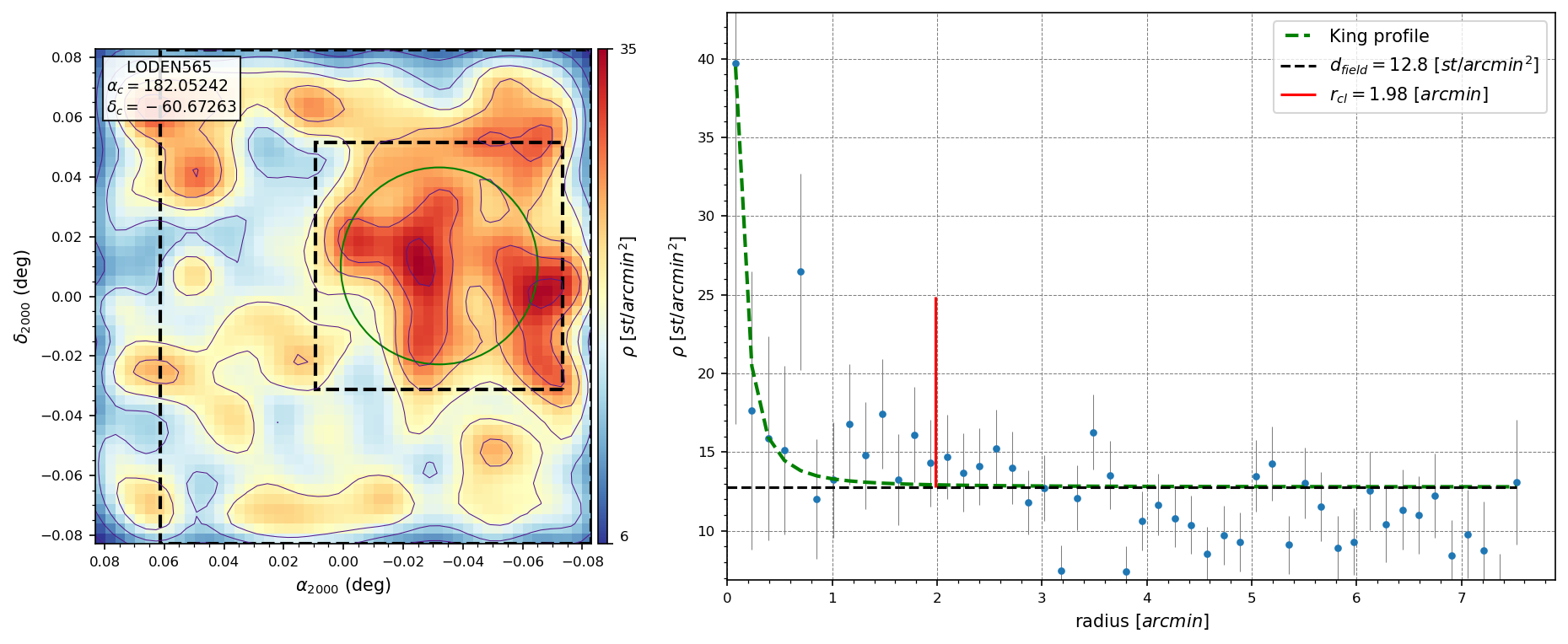}
    \caption{Idem Fig. \ref{fig:struct_vdBH85} for Loden 565.}
    \label{fig56}
\end{figure*}

\begin{figure*}[ht]
    \centering
    \includegraphics[width=\hsize]{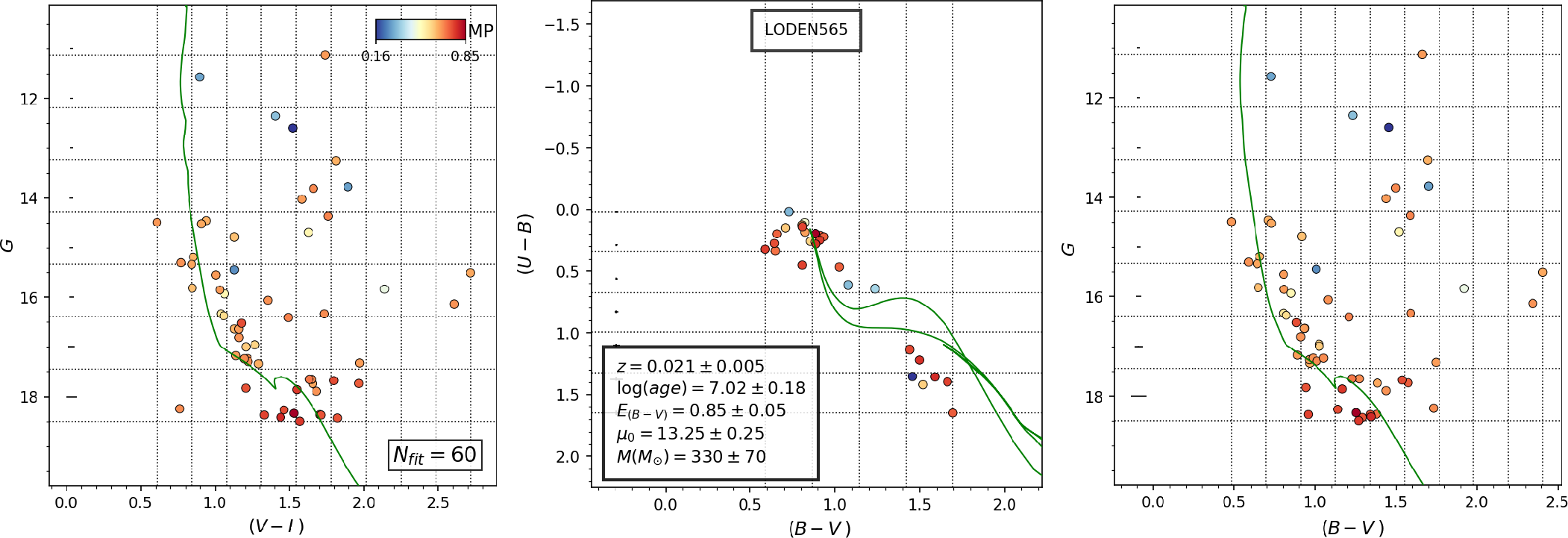}
    \caption{Idem Fig. \ref{fig:fundpars_vdBH85} for Loden 565.}
    \label{fig57}
\end{figure*}

\begin{figure*}[ht]
    \centering
    \includegraphics[width=\hsize]{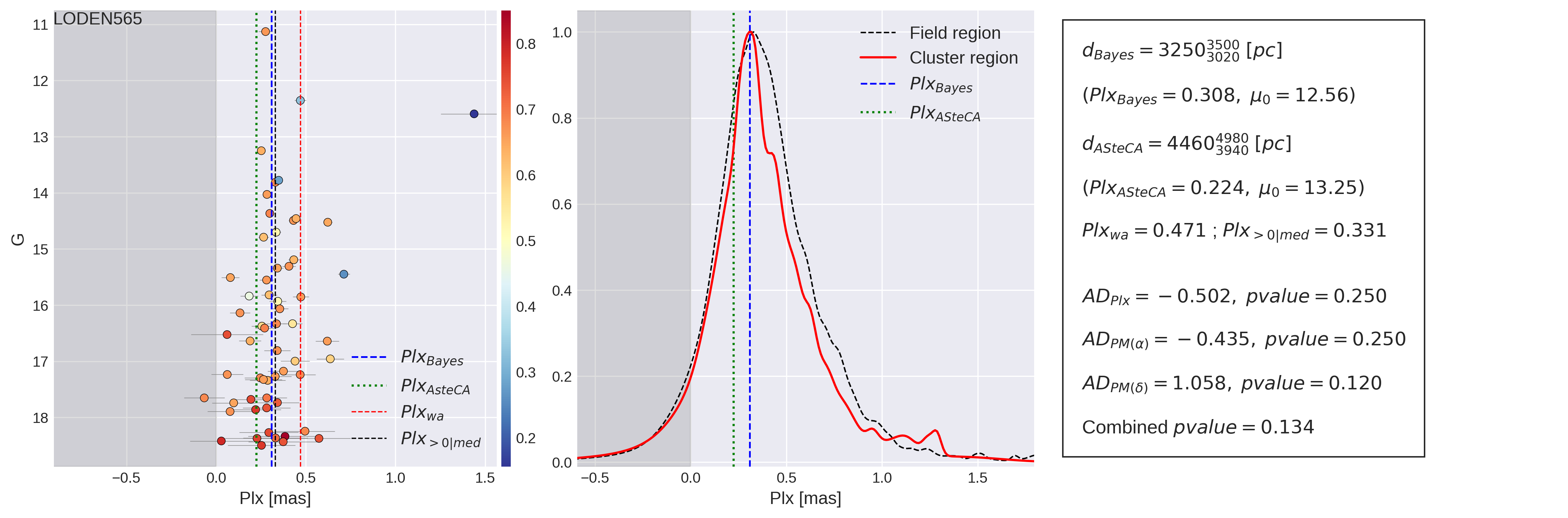}
    \caption{Idem Fig. \ref{fig:plx_bys_vdBH85} for Loden 565.}
    \label{fig58}
\end{figure*}

\texttt{ASteCA} found an irregular overdensity at the north-west corner of the
frame as seen in Fig. \ref{fig56}, left panel. This is the only region across the
entire field where a sudden increase in the star number per area unit is
noticeable showing a $\sim$40 stars per squared arcmin peak at its maximum in
Fig.~\ref{fig56}, right panel.
When looking for membership probabilities only a small number of 60 stars
remain inside the adopted radius with larger probabilities
scattered towards lower magnitudes.
No clear main sequence can be seen present in the CMDs in Fig.~\ref{fig57}.
Notice that none of the stars that
occupy the CCD of Fig. \ref{fig55} (right panel) with $0<(B-V)<0.6$, with some
chances to be reddened early type stars, remain inside the adopted area after 
\texttt{ASteCA}'s membership analysis. The stars that \texttt{ASteCA}
identified inside the adopted radius could be members of an old group but we
conclude that the photometric evidences are not conclusive at all.
More extended and deeper observations are necessary. Previous estimates of the
cluster parameters found for Loden 565 can be found in \cite{Kharchenko_2005}.
These authors concluded that Loden 565 is a moderately young cluster placed at a
distance of $d=0.65$ kpc, affected by a mean reddening $E(B-V)= 0.2$ and
a little older than $10^8$ yrs. The \cite{Kharchenko_2005} atlas shows a
poor fitting to a very sparse available data. In addition, when inspecting the
results from the Anderson-Darling test in the right panel of Fig. \ref{fig58},
it becomes evident that the cluster region is indistinguishable from the
stellar background in terms of parallax and proper motion distributions,
exactly as the clean CCD and CMDs show in Fig.~\ref{fig57}.\\

In conclusion, Loden 565 is more probably a stellar fluctuation.


\section{NGC 4230}

\begin{figure*}[ht]
    \centering
    \includegraphics[width=\hsize]{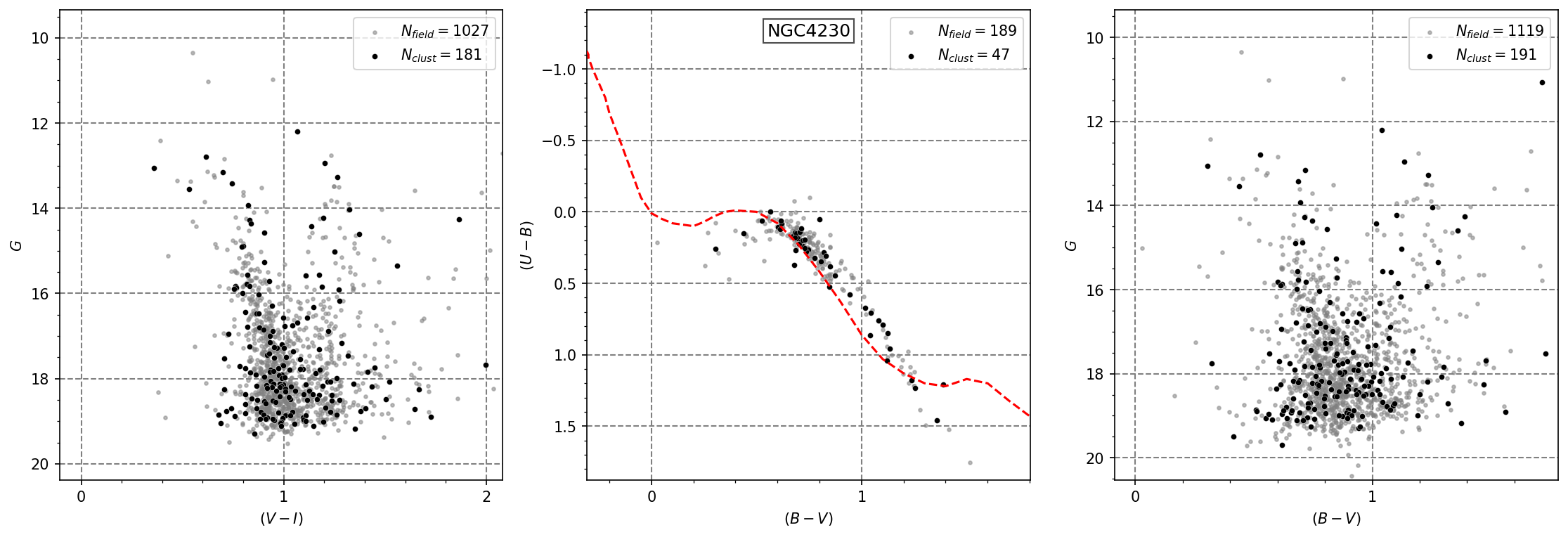}
    \caption{Idem Fig. \ref{fig:photom_vdBH85} for NGC 4230.}
    \label{fig59}
\end{figure*}

This object belongs to the Centaurus region immediately close to the upper
border of Crux. The $V$ image in Fig. \ref{fig:Vim} shows that we are facing
just a modest star grouping near the high proper motion star HD 106826 with 8.8
mag.
Nothing relevant is appreciable in the $V$ image of the inspected zone except
the star already mentioned. A highly scattered and diffuse star distribution
resembling a galactic disc stellar pattern appears in the overall CCD
and CMDs in the panels of Fig. \ref{fig59}.\\

\begin{figure*}[ht]
    \centering
    \includegraphics[width=\hsize]{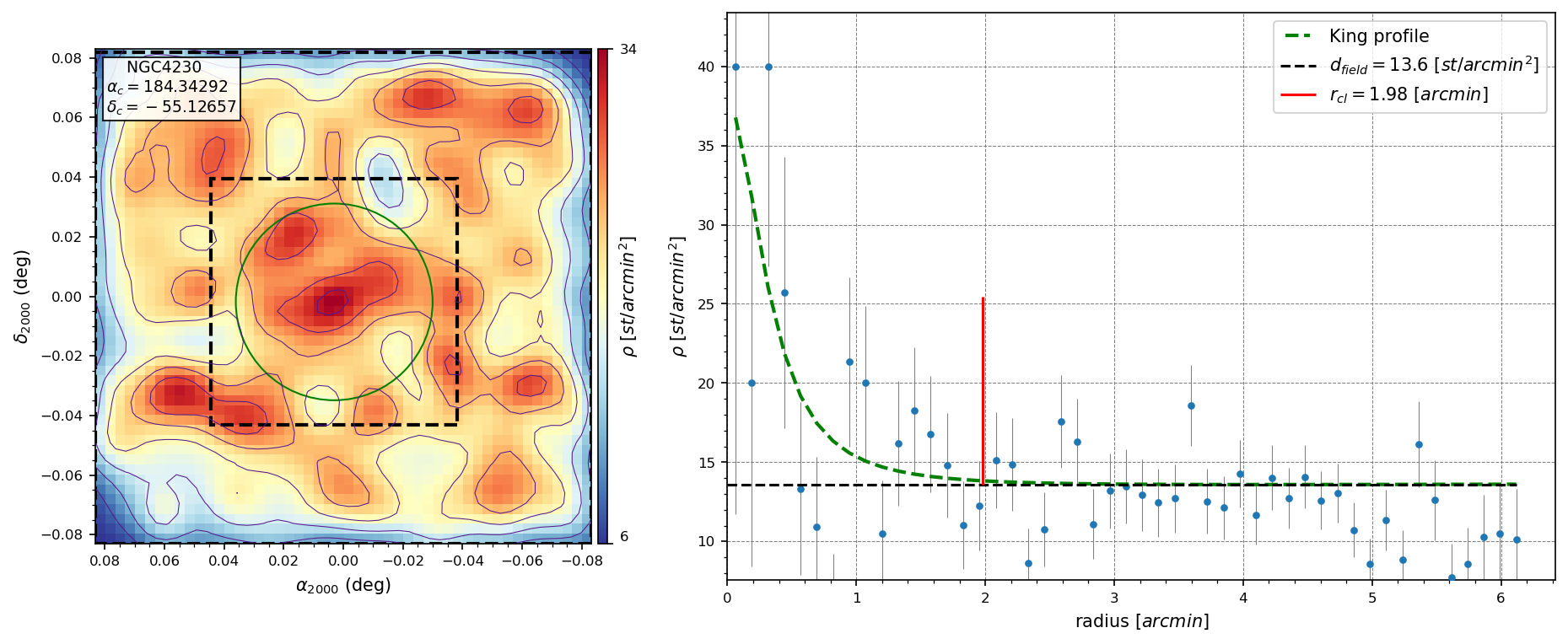}
    \caption{Idem Fig. \ref{fig:struct_vdBH85} for NGC 4230.}
    \label{fig60}
\end{figure*}

\begin{figure*}[ht]
    \centering
    \includegraphics[width=\hsize]{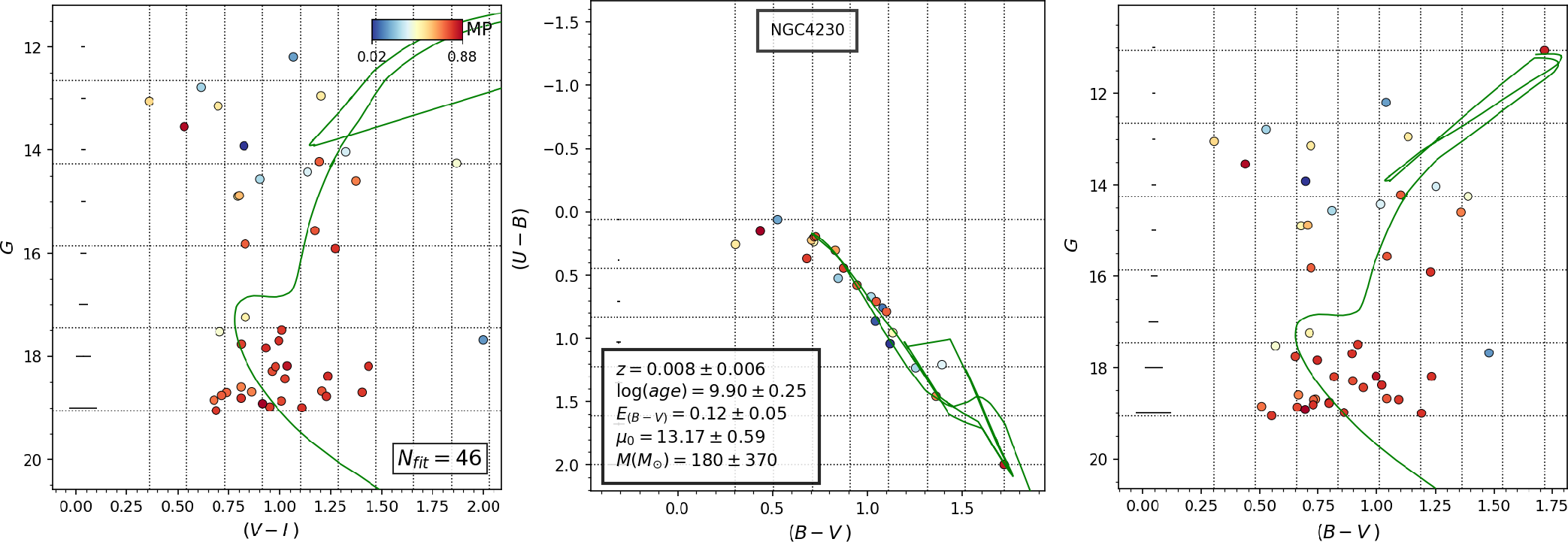}
    \caption{Idem Fig. \ref{fig:fundpars_vdBH85} for NGC 4230.}
    \label{fig61}
\end{figure*}

The spatial inspection performed by \texttt{ASteCA} detected a group of small
stellar overdensities surrounding the central prominence as shown in Fig.
\ref{fig60}, left panel. The peak of the central overdensity shows that the
number of stars per area unit is three times the mean of the background, and the
respective RDP in the right panel of Fig. \ref{fig60} suggests a $\sim$2 arcmin
radius.
However, \texttt{ASteCA} yielded a frustrating result in terms of what
it is expected for a real cluster when analyzing the stellar properties inside
and outside the overdensity.
Only 46 stars remain inside the limits we adopted for NGC 4230. The synthetic
cluster fit is found for the low mass stars with the larger MP values. At this
low number of members and with this large dispersion there is no way to
confidently separate the stellar population into those objects belonging to a 
(putative) real open cluster and those others belonging to the stellar field.
The CCD and CMDs of these stars in Fig.~\ref{fig61} reflect the physical
situation since no main sequence is evident at all. At most, there is a sort of
badly defined giant star sequence whose meaning is dubious because there is no
trace of a main sequence.
The comparison with synthetic clusters performed by \texttt{ASteCA}
fitted mainly a group of stars with low brightness, as shown in the CMDs of
Fig.~\ref{fig61}. This cluster is analyzed in \cite{Tadross_2011} where the
authors find an old 1.7 Gyrs cluster, younger to our result of $\sim$8 Gyrs,
and at a much closer distance (1445 pc versus our result of about 4300 pc)
Therefore the studies do not coincide in the nature of this supposed cluster.\\

\begin{figure*}[ht]
    \centering
    \includegraphics[width=\hsize]{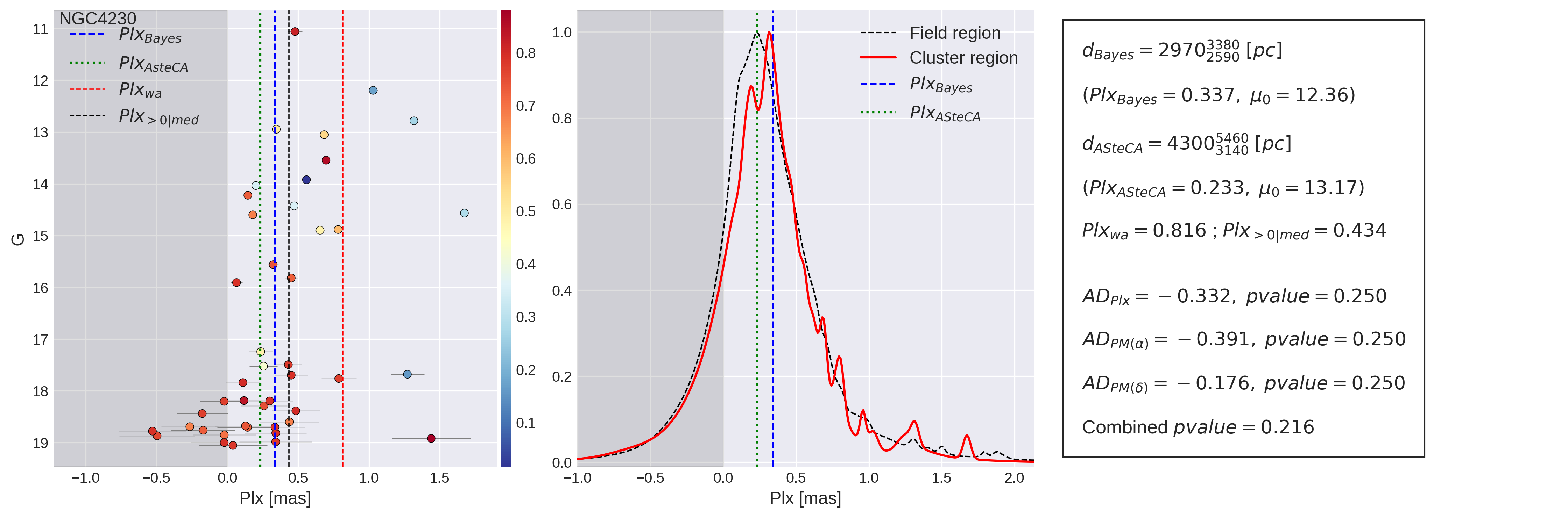}
    \caption{Idem Fig. \ref{fig:plx_bys_vdBH85} for NGC 4230.}
    \label{fig62}
\end{figure*}

Results for the distribution of parallax values and proper motions for the
cluster and field regions are shown in Fig.~\ref{fig62}, right panel.
We see that the Anderson-Darling statistics reveals that the parallax and
proper motions distributions are very similar to stars outside the
cluster region.\\

The lack of a well-defined photometric sequence proper of an open cluster as
demonstrated in Fig. \ref{fig61} together with the results from
the statistical comparison is enough argument to exclude NGC 4230 as a true open
cluster, becoming most probably a random fluctuation of the stellar field.


\end{document}